\documentclass[12pt,a4paper]{article}

\usepackage{a4wide,amsmath,amssymb}
\usepackage{amsfonts,epsfig,epstopdf,psfrag,latexsym}
\usepackage{breqn}
\usepackage{cite,hyperref}

\numberwithin{equation}{section}

\def\beq{\begin{equation}}
\def\eeq{\end{equation}}
\def\be{\begin{equation}}
\def\ee{\end{equation}}
\def\bsp#1\esp{\begin{split}#1\end{split}}

\def \sha{{\,\amalg\hskip -3.6pt\amalg\,}}
\newcommand{\cS}[0]{\mathcal{S}}
\newcommand\lr[1]{{\left({#1}\right)}}

\newcommand{\Gp}{G^+}
\newcommand{\Gm}{G^-}
\newcommand{\Hp}{H^+}
\newcommand{\Hm}{H^-}
\newcommand{\xbar}{\bar{x}}

\begin{document}

\catcode`\@=11
\font\manfnt=manfnt
\def\Watchout{\@ifnextchar [{\W@tchout}{\W@tchout[1]}}
\def\W@tchout[#1]{{\manfnt\@tempcnta#1\relax%
  \@whilenum\@tempcnta>\z@\do{%
    \char"7F\hskip 0.3em\advance\@tempcnta\m@ne}}}
\let\foo\W@tchout
\def\dubious{\@ifnextchar[{\@dubious}{\@dubious[1]}}
\let\enddubious\endlist
\def\@dubious[#1]{%
  \setbox\@tempboxa\hbox{\@W@tchout#1}
  \@tempdima\wd\@tempboxa
  \list{}{\leftmargin\@tempdima}\item[\hbox to 0pt{\hss\@W@tchout#1}]}
\def\@W@tchout#1{\W@tchout[#1]}
\catcode`\@=12

\thispagestyle{empty}

\null\vskip-60pt \hfill
\begin{minipage}[t]{90mm}
HU-Mathematik:~2013-06 HU-EP-13/15 \\
IPPP/13/09 DCPT/13/18  SLAC-PUB-15409 \\
LAPTH-016/13 CERN-PH-TH/2013-058
\end{minipage}

\vskip1.5truecm
\begin{center}
\vskip 0.2truecm

 {\Large\bf
Leading singularities and off-shell conformal integrals}
\vskip 1truecm

\vskip 0.7truecm
{\bf    James Drummond$^{a,b}$, Claude Duhr$^{c,d}$,
  Burkhard Eden$^{e}$,  Paul Heslop$^{f}$, \\
  Jeffrey Pennington$^{g}$,
  Vladimir A. Smirnov$^{e,h}$  \\
}

\vskip 0.3truecm
{\it $^{a}$ CERN, Geneva 23, Switzerland \\\vskip .2truecm
  $^{b}$  LAPTH, CNRS et Universit\'e de Savoie,  F-74941 Annecy-le-Vieux
  Cedex, France \\ 
\vskip .2truecm
$^{c}$ Institut f\"ur Theoretische Physik, ETH Z\"urich, CH-8093, Switzerland\\
\vskip .2truecm
$^{d}$  Institute for Particle Physics Phenomenology, \\
University of Durham,
Durham, DH1 3LE, U.K.\\
\vskip .2truecm
$^{e}$ Institut f\"ur Mathematik, Humboldt-Universit\"at,
Zum gro{\ss}en Windkanal 6,  12489 Berlin \\
\vskip .2truecm
$^{f}$   Mathematics department, Durham University, Durham DH1 3LE,
United Kingdom \\
 \vskip .2truecm
$^{g}$  
SLAC National Accelerator Laboratory, Stanford University,
Stanford, CA 94309, USA \\
\vskip .2truecm 
$^{h}$
Nuclear Physics Institute of Moscow State University, Moscow 119992, Russia                     } \\
\end{center}

\vskip 1 cm 

\centerline{\bf Abstract}
\medskip
\noindent 
{The three-loop four-point function of stress-tensor multiplets in ${\cal N} = 4$ super Yang-Mills theory 
contains two so far unknown, off-shell, conformal integrals, in addition to the known, ladder-type 
integrals.  In this paper we evaluate the unknown integrals, thus obtaining the three-loop correlation 
function analytically. The integrals have the generic structure of rational functions multiplied by 
(multiple) polylogarithms. We use the idea of leading singularities to obtain the rational coefficients, the 
symbol -- with an appropriate ansatz for its structure -- as a means of characterising multiple 
polylogarithms, and the technique of asymptotic expansion of Feynman integrals to obtain the integrals 
in certain limits. The limiting behaviour uniquely fixes the symbols of the integrals, which we then lift to 
find the corresponding polylogarithmic functions. The final formulae are numerically confirmed. The 
techniques we develop
can be applied more generally, and we illustrate this by 
analytically evaluating one of the integrals contributing to the same four-point function at four loops.
This example shows a connection between the leading singularities and the entries of the symbol.}

\vspace{0.8cm}

\begin{center} {\it \noindent In memory of Francis Dolan.}
\end{center}
\newpage

\tableofcontents

\section{Introduction}
\label{sec:Introduce}

The work presented in this paper is motivated by recent progress in planar ${\cal N} = 4$ super Yang-Mills (SYM) theory
in four dimensions, although the methods that we exploit and further develop should be of much wider applicability.

${\cal N}=4$ SYM theory has many striking properties due to its high degree of symmetry; for instance it
is conformally invariant, even as a quantum theory \cite{n4conf}, and the spectrum of anomalous dimensions of
composite operators can be found from an integrable system \cite{dila}. Most strikingly perhaps, it is related to IIB string theory on AdS$_5\times$S$^5$ by the AdS/CFT correspondence \cite{123}.
This is a weak/strong coupling duality in which the same physical system is conveniently described by the field theory picture at weak coupling, while the string theory provides a way of capturing its strong coupling regime. The strong coupling limit of scattering amplitudes in the model has been elaborated in ref.~\cite{aldMald} from a string perspective. The formulae take the form of vacuum expectation values of polygonal Wilson loops with light-like edges.

This duality between amplitudes and Wilson loops remains true at weak coupling \cite{sokaSprad}, extending to the finite terms in ${\cal N} = 4$ SYM previously known relations between the infrared divergences of scattering amplitudes and the ultra-violet divergences of (light-like) Wilson lines in QCD \cite{korRadu}.
Furthermore, it was recently discovered that both sides of this correspondence can be generated from $n$-point correlation functions of stress-tensor multiplets by taking a certain light-cone limit \cite{usRecently}.

The four-point function of stress-tensor multiplets was intensely studied in the early days of the AdS/CFT duality,
in the supergravity approximation \cite{someSugra} as well as at weak coupling. The one-loop
\cite{weak4ptOne} and two-loop \cite{weak4ptTwo} corrections are given by conformal ladder integrals.

A Feynman-graph based three-loop result has never become
available because of the formidable size and complexity of multi-leg
multi-loop computations. Already the two parallel two-loop
calculations \cite{weak4ptTwo} drew heavily upon superconformal
symmetry. However, a formulation on a maximal (`analytic')
superspace~\cite{Galperin:1984av,Hartwell:1994rp} makes it apparent
that the loop corrections to the lowest $x$-space component are given
by a product of a certain polynomial with linear combinations of
conformal integrals,
cf. ref.~\cite{olderAna,Eden:2000qp,Dolan:2001tt,Heslop:2002hp}. Then
in ref.~\cite{hidden,constructing}, using a hidden symmetry permuting
integration variables and external variables, the problem of finding
the three-loop integrand was reduced down to just four unfixed
coefficients without any calculation and furhter down to  only one
overall coefficient after a little further analysis. This single
overall coefficient can then easily be fixed e.g. by
comparing to the MHV four-point three-loop amplitude \cite{ladderTennis} via the correlator/amplitude duality or by requiring
the exponentiation of logarithms in a double OPE limit \cite{hidden}.

Beyond the known ladder and the `tennis court', the off-shell three-loop four-point correlator contains two unknown integrals termed `Easy' and `Hard' in ref.~\cite{hidden}. In this work we embark on an
analytic evaluation of the Easy and Hard integrals postulating that
\begin{itemize}
\item the integrals are sums $\sum_i \, R_i \, F_i$, where $R_i$ are rational functions and $F_i$ are \emph{pure functions}, i.e. $\mathbb{Q}$-linear combinations of logarithms and multiple polylogarithms \cite{Goncharov},
\item the rational functions $R_i$ are given by the so-called \emph{leading singularities} (i.e. residues of global poles) of the integrals \cite{freddy2008},
\item the symbol of each $F_i$ can be pinned down by appropriate constraints and then integrated to a unique transcendental function.
\end{itemize}
The principle of \emph{uniform transcendentality}, innate to the planar ${\cal N} = 4$ SYM theory, implies that the symbols of all the pure functions are tensors of uniform rank six. 
Our strategy will be to make an ansatz for the entries that can appear in the symbols of the pure functions and to write down the most general tensor of uniform rank six of this form. We then impose a set of constraints on this general tensor to pin down the symbols of the pure functions.
First of all, the tensor needs to satisfy the \emph{integrability condition}, a criterion for a general tensor to correspond to the symbol of a transcendental function. Next the symmetries of the integrals induce additional constraints, and finally we equate with single variable expansions corresponding to Euclidean coincidence limits. The latter were elaborated for the Easy and Hard integrals in ref.~\cite{Eden:2012fe,Eden:2012rr} using the method of \emph{asymptotic expansion of Feynman integrals}~\cite{smirnovBook}. 
This expansion technique reduces the original higher-point integrals to two-point integrals, albeit with high exponents of the denominator factors and complicated numerators.

\begin{figure}[htb]
\begin{center}
\includegraphics[width=0.8\textwidth]{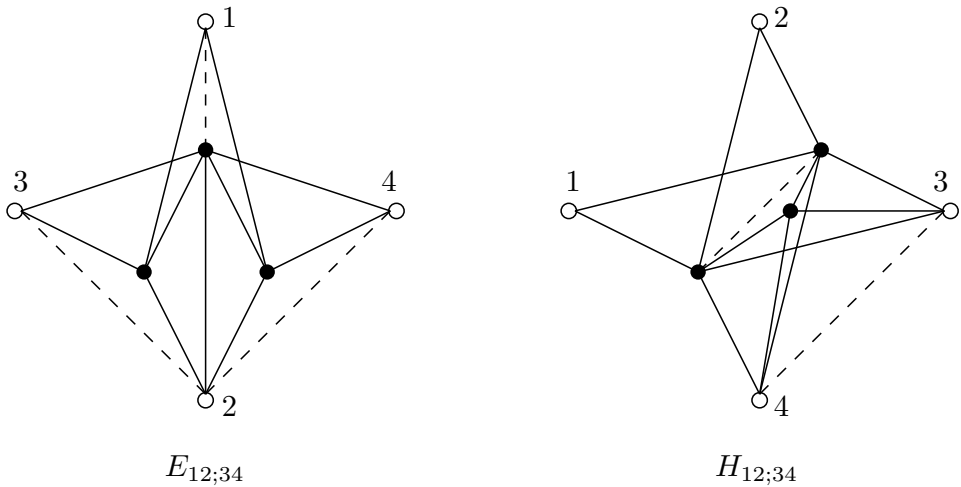}
\caption{\label{fig:EandHpics}The Easy and Hard integrals contributing to the correlator of stress tensor multiplets at three loops.}
\end{center}
\end{figure}

To be specific, up to three loops the off-shell four-point correlator is given by \cite{weak4ptOne,weak4ptTwo,hidden}
\vskip - 0.4 cm
\begin{align} \label{offshc}
  G_{4}(1,2,3,4)= G^{(0)}_{4}  + \frac{2 \, (N_c^2-1)}{(4\pi^2)^{4}} \  R(1,2,3,4)   \   \left[ a  F^{(1)} + a^2 F^{(2)} + a^3 F^{(3)} + O(a^4)  \right] ,
\end{align}
Here $N_c$ denotes the number of colors and $a$ is the 't Hooft coupling. $G_4^{(0)}$ represents the tree-level contribution and $R(1,2,3,4)$ is a universal prefactor, in particular taking into account the different $SU(4)$ flavours which can appear (see ref.~\cite{hidden,constructing} for details). Our focus here is on the loop corrections. These can be written in the compact form (exposing the hidden $S_{4+\ell}$ symmetry) as
\begin{align}
  F^{(\ell)}(x_1,x_2,x_3,x_4)={x_{12}^2 x_{13}^2 x_{14}^2 x_{23}^2 
  x_{24}^2 x_{34}^2\over \ell!\,(\pi^2)^\ell} \int d^4x_5 \dots
  d^4x_{4+\ell} \,  \hat{f}^{(\ell)}(x_1,  \dots , x_{4+\ell})\,,
\end{align}
where
\begin{align}
\hat{f}^{(1)}(x_1,\ldots,x_5) &= {1 \over \prod_{1\leq i<j \leq 5}   x_{ij}^2}\,,\\[3mm]
 \hat{f}^{(2)}(x_1,\ldots,x_6) & = \frac{ {\textstyle \frac1{48}} x_{12}^2x_{34}^2x_{56}^2\ +\ {S_6\ \mathrm{permutations}}}{\prod_{1\leq i<j\leq 6}x_{ij}^2}
\,,\\[3mm]
\hat{f}^{(3)}(x_1,\ldots,x_7) & = \frac{{{\textstyle\frac{1}{20}}}(x_{12}^2)^2( x_{34}^2 x_{45}^2 x_{56}^2 x_{67}^2 x_{73}^2)  \ +\ {S_7\ \mathrm{permutations}}}{\prod_{1\leq i<j\leq 7}x_{ij}^2}\,.
\end{align}
Writing out the sum over permutations in the above expressions, these are written as follows
\begin{align}
F^{(1)} & =  g_{1234}\,,
\\[3mm]
 F^{(2)} & = 
h_{12;34} + h_{34;12} + h_{23;14} +  h_{14;23} 
 \\ & + \, h_{13;24} + h_{24;13}
  + \frac12 \lr{x_{12}^2x_{34}^2+ x_{13}^2 x_{24}^2+x_{14}^2x_{23}^2} 
[ \, g_{1234} ]\, ^2\, , \nonumber \\[3mm]\label{eq:F3}
F^{(3)} & = \big[ \, L_{12;34}+ 5 \mbox{ perms }\big] +
\big[ \, T_{12;34}+ 11 \mbox{ perms }\big]  \\[2mm] 
& + \, \big[ \, E_{12;34}+ 11 \mbox{ perms }\big] +
{\textstyle \frac12}\big[ \, x_{14}^2x_{23}^2 H_{12;34}+ 11 \mbox{ perms }\big] \notag \\[2mm]
 & + \, \big[ \, ({g\times h})_{12;34}+  5 \mbox{ perms }\big] \, ,
  \nonumber 
\end{align}
which involve the following integrals:
\begin{align}\label{eq:int_defs}
& g_{1234}  = \frac{1}{\pi^2}
\int \frac{d^4x_5}{x_{15}^2 x_{25}^2 x_{35}^2 x_{45}^2}  \, , \\
& h_{12;34}  =  \frac{x^2_{34}}{\pi^4}
\int \frac{d^4x_5 \, d^4x_6}{(x_{15}^2 x_{35}^2 x_{45}^2) x_{56}^2
(x_{26}^2 x_{36}^2 x_{46}^2)}\ .  \nonumber
\end{align}
At three-loop order we encounter
\begin{eqnarray}
{(g\times h)}_{12;34} &=&{x_{12}^2 x_{34}^4\over \pi^6}\int 
  \frac{d^4x_5d^4x_6 d^4x_7}{(x_{15}^2  x_{25}^2  x_{35}^2 x_{45}^2) 
 (x_{16}^2  x_{36}^2x_{46}^2) (x_{27}^2  x_{37}^2   x_{47}^2) x_{67}^2} \ ,
\nonumber\\
L_{12;34} & = &  \frac{x^4_{34}}{\pi^6}
\int \frac{d^4x_5 \, d^4x_6 \, d^4x_7}{(x_{15}^2 x_{35}^2 x_{45}^2) x_{56}^2
(x_{36}^2 x_{46}^2) x^2_{67} (x_{27}^2 x_{37}^2 x_{47}^2)} \, , \nonumber \\
T_{12;34}&=&{x_{34}^2\over \pi^6}\int 
  \frac{ d^4x_5 d^4x_6 d^4x_7 \ x_{17}^2 }{(x_{15}^2 x_{35}^2)
(x_{16}^2  x_{46}^2) (x_{37}^2 x_{27}^2  x_{47}^2) x_{56}^2 x_{57}^2 x_{67}^2}\ , \\
E_{12;34} & = &  \frac{x^2_{23} x^2_{24}}{\pi^6}
\int \frac{d^4x_5 \, d^4x_6 \, d^4x_7 \ x^2_{16}}{(x_{15}^2 x_{25}^2 x_{35}^2)
x_{56}^2 (x_{26}^2 x_{36}^2 x^2_{46}) x^2_{67} (x_{17}^2 x_{27}^2 x_{47}^2)}
\, , \nonumber \\
H_{12;34} & = &  \frac{ x_{34}^2 }{\pi^6}
\int \frac{d^4x_5 \, d^4x_6 \, d^4x_7 \ x^2_{57}}{(x_{15}^2 x_{25}^2 x_{35}^2
x^2_{45}) x_{56}^2 (x_{36}^2 x^2_{46}) x^2_{67} (x_{17}^2 x_{27}^2 x^2_{37}
x_{47}^2)} \, . \nonumber
\end{eqnarray}
Here $g,h,L$ are recognised as the one-loop, two-loop and three-loop ladder integrals, respectively, the dual graphs of the off-shell box, double-box and triple-box  integrals.
Off-shell, the `tennis court' integral $T$ can be expressed as the three-loop ladder integral $L$ by using the conformal flip
properties\footnote{Such identities rely on manifest conformal invariance and will be broken by the introduction of most regulators. For instance, the equivalence of $T$ and $L$ is not true for the dimensionally regulated on-shell integrals.}
of a two-loop ladder sub-integral \cite{magic}. The only new integrals are thus $E$ and $H$ (see fig.~\ref{fig:EandHpics}).

Conformal four-point integrals are given by a factor carrying their conformal weight, say, $(x_{13}^2 x_{24}^2)^n$
times some function of the two cross ratios
\begin{equation}
u \, = \, \frac{x_{12}^2 x_{34}^2}{x_{13}^2 x_{24}^2} \, = \, x \, \bar x \, , \qquad
v \, = \, \frac{x_{14}^2 x_{23}^2}{x_{13}^2 x_{24}^2} \, = \, (1-x) (1 - \bar x) \, . \qquad
\label{concrrat}
\end{equation}
Ladder integrals are explicitly known for any number of loops, see ref.~\cite{davyUss} where they 
are very elegantly expressed  
as one-parameter integrals. Integration is simplified by the change of variables 
from the cross-ratios $(u,v)$ to $(x, \bar x)$ as defined in the last equation. The unique rational 
prefactor, $x_{13}^2 x_{24}^2 \, (x - \bar x)$, is common to all cases and can be computed by the 
leading singularity method as we illustrate shortly.
This is multiplied by pure polylogarithm functions which fit with the classification of single-valued 
harmonic polylogarithms (SVHPLs) in ref.~\cite{BrownSVHPLs}. The associated symbols of the 
ladder integrals are then tensors composed of the four letters $\{x, \, \bar x, \, 1 - x, \, 1 - \bar x\}$.

On the other hand, for generic conformal four-point integrals  (of which the Easy and Hard integrals 
are the first examples)  there are no explicit results. Fortunately, in recent years a formalism has 
been developed in the context of scattering amplitudes to find at least the rational prefactors (i.e. 
the leading singularities),
which are given by the residues of the integrals~\cite{freddy2008}. There is one leading singularity 
for each global pole of the integrand and it is obtained by deforming the contour of integration to lie 
on a maximal torus surrounding the pole in question, i.e. by computing the residue at the global 
pole. As an illustration\footnote{The massless box-integral (i.e. the same integral in the limit
$x_{i,i+1}^2 \rightarrow 0$) is discussed in ref.~\cite{NimaandFreddy} in terms of twistor variables 
as the simplest example of a `Schubert problem' in projective geometry. The off-shell case that we 
discuss here was also recently discussed by S.~Caron-Huot (see~\cite{ch}).}, let us apply this 
technique to the massive one-loop box integral $g_{1234}$ defined in eq.~\eqref{eq:int_defs}.
Its leading singularity is obtained by shifting the contour to encircle one of
the global poles of the integrand, where all four terms in the denominator vanish. To find
this let us consider a change of coordinates from $x^\mu_5$ to
$p_i=x_{i5}^2$. The Jacobian for this change of variables is
\begin{equation}
J \, = \, \det \left( { \partial p_i \over \partial x_5^\mu}\right) = \det \left(-2 x_{i5}^\mu \right) \, , \quad
J^2 \, =  \, \det \left(4 x_{i5}\cdot x_{j5}\right) \, = \, 16  \det\left(x_{ij}^2 -x_{i5}^2-x_{j5}^2\right  )\,,
\end{equation}
where the second identity follows by observing that $\det(M)= \sqrt{\det(M M^T)}$.
Using this change of variables the massive box becomes
\begin{equation}\label{eq:1}
g_{1234} \, =  \, \frac{1}{\pi^2} \, \int {d^4 p_i \over p_1 p_2 p_3 p_4 \, J} \ .
\end{equation}
To find its leading singularity we simply  compute the residue around all four poles at $p_i=0$ 
(divided by $2\pi i$). We obtain
\begin{equation}\label{eq:2}
g_{1234} \, \rightarrow \, {1 \over 4\pi^2\lambda_{1234}} \, , \quad  \lambda_{1234} \, = \, \sqrt { \det(x_{ij}^2)_{i,j=1..4}} \, 
= \, x_{13}^2 x_{24}^2 \, (x- \bar x) 
\end{equation}
in full agreement with the analytic result \cite{davyUss}. 

Note that we do not consider explicitly a contour around the branch cut associated with the square root factor J in the denominator of (\ref{eq:1}). Because there is no pole at infinity, the residue theorem guarantees that such a contour is equivalent to the one we already considered. On the other hand, in higher-loop examples, Jacobians from previous integrations cannot be discarded in this manner. In all the examples we consider, these Jacobians always collapse to become simple poles when evaluated on the zero loci of the other denominators and thereby contribute non-trivially to the leading singularity.

The main results of this paper are the analytic evaluations of the Easy and Hard integrals.
Due to Jacobian poles, the Easy integral has three distinct leading singularities, out of which only 
two are algebraically independent, though.  The Hard integral has two distinct leading singularities, 
too. Armed with this information we then attempt to find the pure polylogarithmic functions 
multiplying these rational factors. Our main inputs for this are analytic expressions for the integrals 
in the limit $\bar x \rightarrow 0$ obtained from the results in~\cite{Eden:2012rr}. Matching these 
asymptotic expressions with an ansatz for the symbol of the pure functions we obtain unique 
answers for the pure functions.

The pure functions contributing to the Easy integral are given by SVHPLs, corresponding to a 
symbol with entries drawn from the set $\{x,1-x,\xbar,1-\xbar\}$. In this case there is a very 
straightforward method for obtaining the corresponding function from its asymptotics, by essentially 
lifting HPLs to SVHPLs as we explain in the next section. However, the SVHPLs are not capable of 
meeting all constraints for the pure functions contributing to the Hard integral, so that we need to 
enlarge the set of letters. A natural guess is to include $x - \bar x$ (cf. ref.~\cite{Chavez:2012kn}) 
since it also occurs in the rational factors, and indeed this turns out to be correct. Ultimately, one of 
the pure functions is found to have a four-letter symbol corresponding to SVHPLs, but the symbol of 
the other function contains the new letter: the corresponding function cannot be expressed through 
SVHPLs alone, but it belongs to a more general class of multiple polylogarithms.

Let us stress that the analytic evaluation of the Easy and Hard integrals completes the derivation of 
the three-loop four-point correlator of
stress-tensor multiplets in ${\cal N} = 4$ SYM. The multiple polylogarithms that we find can be 
numerically evaluated to very high precision, which paves the way for tests of future integrable 
system predictions for the four-point function, or for instance for further analyses of the operator 
product expansion.

Finally, since our set of methods has allowed to obtain the analytic result for the Easy and Hard 
integrals in a relatively straightforward way (despite the fact that these are not at all simple to 
evaluate by conventional techniques) we wish to investigate whether this can be repeated to still 
higher orders.  We examine a first relatively simple looking, but non-trivial, four-loop example from 
the list of integrals
contributing to the four-point correlator at that order \cite{constructing}:
\begin{equation}
I^{(4)}_{14;23} = \frac{1}{\pi^8} \, 
\int   \frac{d^4 x_5d^4 x_6 d^4 x_7 d^4 x_8 \ x_{14}^2 x_{24}^2 x_{34}^2}{x_{15}^2 x_{18}^2 x_{25}^2 x_{26}^2 x_{37}^2 x_{38}^2 x_{45}^2 x_{46}^2 x_{47}^2 x_{48}^2 x_{56}^2 x_{67}^2 x_{78}^2 }\,.
 \label{I4int}
\end{equation}
The computation of its unique leading singularity follows the same lines as at three loops.
However, just as for the Hard integral, the alphabet $\{x,1-x,\xbar,1-\xbar\}$ and the corresponding
function space are too restrictive. Interestingly, this integral is related to the Easy integral by a 
differential equation of Laplace type. Solving this equation promotes the denominator factor $1 - u$ 
of the leading singularities of the Easy integral to a new entry
in the symbol of the four-loop integral.
Note that it is at least conceivable that the letter $x - \bar x$ arrives in the symbol of the Hard 
integral due to a similar mechanism, although admittedly not every integral obeys a simple 
differential equation.

\begin{figure}[!t]
\begin{center}
\includegraphics[width=0.35\textwidth]{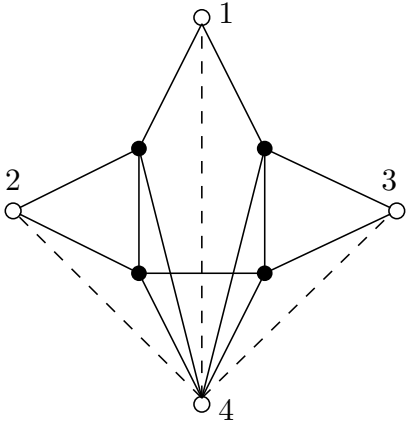}
\caption{\label{fig:I4}The four-loop integral $I^{(4)}_{14;23}$ defined in eq.~\eqref{I4int}.}
\end{center}
\end{figure} 

The paper is organised as follows:
\begin{itemize}
\item In Section~\ref{sec:SVHPLs}, we give definitions of the concepts introduced here: symbols, harmonic polylogarithms, SVHPLs, multiple 
polylogarithms and so on.
\item In Section \ref{sec:asymptotic_expansion}, we comment on the asymptotic expansion of 
Feynman integrals.
\item In Sections \ref{sec:Easy} and \ref{sec:Hard} we derive the leading 
singularities, symbols and ultimately the pure functions corresponding to the Easy and Hard integrals. We also
present numerical data indicating the correctness of our results.
\item In Section~\ref{sec:4loops}, we perform a similar calculation for the four-loop integral, $I^{(4)}$.
\item Finally we draw some conclusions. We include several appendices collecting some formulae 
for the asymptotic expansions of the integrals and alternative ways how to derive the analytic results.
\end{itemize}

\section{Conformal four-point integrals and single-valued polylogarithms}
\label{sec:SVHPLs}

The ladder-type integrals that contribute to the correlator are known. More precisely, if we write
\beq\bsp
g_{13;24} &\,= \frac{1}{x_{13}^2x_{24}^2}\,\Phi^{(1)}(u,v)\,,\\
h_{13;24} &\,= \frac{1}{x_{13}^2x_{24}^2}\,\Phi^{(2)}(u,v)\,,\\
l_{13;24} &\,= \frac{1}{x_{13}^2x_{24}^2}\,\Phi^{(3)}(u,v)\,,
\esp\eeq
then the functions $\Phi^{(L)}(u,v)$ are given by the well-known result~\cite{davyUss},
\beq\bsp\label{eq:UD_function}
\Phi^{(L)}(u,v) &\,= -\frac{1}{L!(L-1)!}\,\int_0^1\frac{d\xi}{v\,\xi^2+(1-u-v)\,\xi+u}\,\log^{L-1}\xi\\
&\,\qquad\times\Big(\log\frac{v}{u}+\log\xi\Big)^{L-1}\,\Big(\log\frac{v}{u}+2\log\xi\Big)\\
&\,=-\frac{1}{x-\xbar}\,f^{(L)}\left(\frac{x}{x-1},\frac{\xbar}{\xbar-1}\right)\,,
\esp\eeq
where the conformal cross ratios are given by eq.~(\ref{concrrat}) and where we defined the pure function
\beq\label{eq:f_def}
f^{(L)}(x,\xbar)=\sum_{r=0}^L\frac{(-1)^{r}(2L-r)!}{r!(L-r)!L!}\log^r(x\bar x)\,\left(\mathrm{Li}_{2L-r}(x)-\mathrm{Li}_{2L-r}(\bar x)\right)\,.
\eeq
At this stage, the variables $(x,\bar x)$ are simply a convenient parametrisation which rationalises 
the two roots of the quadratic polynomial in the denominator of eq.~\eqref{eq:UD_function}. We 
note that $x$ and $\bar x$ are complex conjugate to each other if we work in Euclidean space while 
they are both real in Minkowski signature.

The particular combination of polylogarithms that appears in eq.~\eqref{eq:UD_function} is not 
random, but it has a particular mathematical meaning: in Euclidean space, where $x$ and $\bar x$ 
are complex conjugate to each other, the functions $\Phi^{(L)}$ are single-valued functions of the 
complex variable $x$. In other words, the combination of polylogarithms that appears in the ladder 
integrals is such that they have no branch cuts in the complex $x$ plane. In order to understand the 
reason for this, it is useful to look at the symbols of the ladder integrals. 

\subsection{The symbol}
\label{sec:symbol}

One possible way to define the symbol of a transcendental function is to consider its total differential. 
More precisely, if $F$ is a function whose differential satisfies
\beq
dF = \sum_i F_i\,d\log R_i\,,
\eeq
where the $R_i$ are rational functions,
then we can define the symbol of $F$ recursively by~\cite{Goncharov:2010jf}
\beq\label{eq:GSVV_def}
\cS(F) = \sum_{i}\cS(F_i)\otimes R_i\,.
\eeq
As an example, the symbols of the classical polylogarithms and the ordinary logarithms are given by
\beq
\cS(\mathrm{Li}_{n}(z)) = -(1-z)\otimes\underbrace{z\otimes\ldots\otimes z}_{(n-1)\textrm{ times}} 
{\rm~~and~~} \cS\left(\frac{1}{n!}\ln^nz\right) = \underbrace{z\otimes\ldots\otimes z}_{n\textrm{ times}}\,.
\eeq
In addition the symbol satisfies the following identities,
\beq\bsp
\ldots\otimes (a\cdot b)\otimes\ldots &\,= \ldots\otimes a\otimes\ldots + \ldots\otimes b\otimes\ldots\,,\\
 \ldots\otimes(\pm1)\otimes\ldots &\,= 0\,,\\
 \cS\left(F\,G\right) &\,=  \cS(F)\sha \cS(G)\,,
\esp\eeq
where $\sha$ denotes the shuffle product on tensors. Furthermore, all multiple zeta values are 
mapped to zero by the symbol map. Conversely, an arbitrary tensor
\beq
\sum_{i_1,\ldots,i_n}c_{i_1\ldots i_n}\omega_{i_1}\otimes\ldots\otimes\omega_{i_n}
\eeq
 whose entries are rational functions is the symbol of a function only if the following 
 \emph{integrability condition} is fulfilled,
 \beq \label{intCond}
 \sum_{i_1,\ldots,i_n}c_{i_1\ldots i_n}\,d\log\omega_{i_{k}}\wedge d\log\omega_{i_{k+1}}\,
 \omega_{i_1}\otimes\ldots\otimes\omega_{i_{k-1}}\otimes\omega_{i_{k+2}}\otimes\ldots\otimes\omega_{i_n}=0\,,
\eeq
for all consecutive pairs $(i_k,i_{k+1})$.

The symbol of a function also encodes information about the discontinuities of the function. More 
precisely, the singularities (i.e. the zeroes or infinities) of the first entries of a symbol determine the 
branching points of the function, and the symbol of the discontinuity across the branch cut is 
obtained by dropping this first entry from the symbol. As an example, consider a function $F(x)$ 
whose symbol has the form
\beq
\cS(F(x)) = (a_1-x)\otimes\ldots \otimes (a_n-x)\,,
\eeq
where the $a_i$ are independent of $x$. Then $F(x)$ has a branching point at $x=a_1$, and the 
symbol of the discontinuity across the branch cut is given by
\beq
\cS\left[\mathrm{disc}_{a_1}F(x)\right] = 2\pi i\,(a_2-x)\otimes\ldots \otimes (a_n-x)\,.
\eeq
If $F$ is a Feynman integral, then the branch cuts of $F$ are dictated by Cutkosky's rules. In 
particular, for Feynman integrals without internal masses the branch cuts extend between points 
where one of the Mandelstam invariants becomes zero or infinity. As a consequence, the first 
entries of the symbol of a Feynman integral must necessarily be Mandelstam invariants~
\cite{Gaiotto:2011dt}. In  the case of the four-point position space integrals we are considering in 
this paper, the first entries of the symbol must then be distances between two points, $x_{ij}^2$ for 
$i,j=1\ldots4$. Combined with conformal invariance, this implies that the first entries of the symbols 
of conformally invariant four-point functions can only be cross ratios. As an example, consider the 
symbol of the one-loop four-point function,
\beq
\cS\left[f^{(1)}(x,\xbar)\right] = u\otimes\frac{1-x}{1-\bar x} +v\otimes\frac{\bar x}{x}\,.
\eeq

The first entry condition puts strong constraints on the transcendental functions that can contribute 
to a conformal four-point function. In order to understand this better let us consider a function 
whose symbol can be written in the form
\beq\label{eq:S_Disc}
\cS(F) = u\otimes S_u + v\otimes S_v = (x\bar x)\otimes S_u+[(1-x)(1-\bar x)]\otimes S_v\,,
\eeq
where $S_u$ and $S_v$ are tensors of lower rank. Let us assume we work in Euclidean space 
where $x$ and $\bar x$ are complex conjugate to each other. It then follows from the previous 
discussion that $F$ has potential branching points in the complex $x$ plane at $x\in\{0,1,\infty\}$. 
Let us compute for example the discontinuity of $F$ around $x=0$. Only the first term in eq.~
\eqref{eq:S_Disc} can give rise to a non-zero contribution, and $x$ and $\bar x$ contribute with 
opposite signs. So we find
\beq
\cS\left[\mathrm{disc}_{0}(F)\right] = 2\pi i\,S_u - 2\pi i\,S_u =0\,.
\eeq
The argument for the discontinuities around $x=1$ and $x=\infty$ is similar. We thus conclude that 
$F$ is single-valued in the whole complex $x$ plane. This observation puts strong constraints on 
the pure functions that might appear in the analytical result for a conformal four-point function. In 
particular, the ladder integrals $\Phi^{(L)}$ are related to the single-valued analogues of the 
\emph{classical} polylogarithms,
\beq
D_n(x) = \mathfrak{R}_n\sum_{k=0}^{n-1}\frac{B_k 2^k}{k!}\log^k|x|\,\mathrm{Li}_{n-k}(x)\,,
\eeq
where $\mathfrak{R}_n$ denotes the real part for $n$ odd and the imaginary part otherwise and 
$B_k$ are the Bernoulli numbers. For example, we have
\beq
f^{(1)}(x,\xbar) = 4 i \,D_2(x)\,.
\eeq

\subsection{Single-Valued Harmonic Polylogarithms (SVHPLs)}
\label{sec:harm-polyl}

For more general conformal four-point functions more general classes of polylogarithms may 
appear. The simplest extension of the classical polylogarithms are the so-called harmonic 
polylogarithms (HPLs), defined by the iterated integrals\footnote{In the following we use the word 
\emph{harmonic polylogarithm} in a restricted sense, and only allow for singularities at $x\in\{0,1\}$ 
inside the iterated integrals.}~\cite{Remiddi:1999ew}
\beq
H(a_1,\ldots,a_n;x) = \int_0^xdt \,f_{a_1}(t)\,H(a_2,\ldots,a_n;t)\,,\qquad a_i\in\{0,1\}\,,
\eeq
with
\beq
f_0(x) =\frac{1}{x} {\rm~~and~~}f_1(x) = \frac{1}{1-x}\,.
\eeq
By definition, $H(x) = 1$ and in the case where all the $a_i$ are zero, we use the special definition 
\beq
H(\vec 0_n;x) = \frac{1}{n!}\log^nx\,.
\eeq
The number $n$ of indices of a harmonic polylogarithm is called its \emph{weight}.
Note that the harmonic polylogarithms contain the classical polylogarithms as special cases,
\beq
H(\vec 0_{n-1},1;x) = \mathrm{Li}_n(x)\,.
\eeq
In ref.~\cite{Drummond:2012bg} it was shown that infinite classes of generalised ladder integrals 
can be expressed in terms of single-valued combinations of HPLs. Single-valued analogues of 
HPLs were studied in detail in ref.~\cite{BrownSVHPLs}, and an explicit construction valid for all 
weights was presented. Here it suffices to say that for every harmonic polylogarithm of the form 
$H(\vec a;x)$ there is a function $\mathcal{L}_{\vec a}(x)$ with essentially the same properties as 
the ordinary harmonic polylogarithms, but in addition it is single-valued in the whole complex $x$ 
plane. We will refer to these functions as \emph{single-valued harmonic polylogarithms} (SVHPLs). 
Explicitly, the functions $\mathcal{L}_{\vec a}(x)$ can be expressed as 
\beq
\mathcal{L}_{\vec a}(x) = \sum_{i,j}c_{ij}\,H(\vec a_i;x)\,H(\vec a_j;\bar x)\,,
\eeq
where the coefficients $c_{ij}$ are polynomials of multiple $\zeta$ values such that all branch cuts cancel.

There are two natural symmetry groups acting on the space of SVHPLs. The first symmetry group 
acts by complex conjugation, i.e., it exchanges $x$ and $\bar x$. The conformal four-point functions 
we are considering are real, and thus eigenfunctions under complex conjugation, while the SVHPLs 
defined in ref.~\cite{BrownSVHPLs} in general are not. It is therefore convenient to diagonalise the 
action of this symmetry by defining
\beq\bsp
L_{\vec a}(x) &\,= \frac{1}{2}\left[\mathcal{L}_{\vec a}(x)-(-1)^{|\vec a|}\mathcal{L}_{\vec a}(\bar x)\right]\,,\\
\overline{L}_{\vec a}(x) &\,= \frac{1}{2}\left[\mathcal{L}_{\vec a}(x)+(-1)^{|\vec a|}\mathcal{L}_{\vec a}(\bar x)\right]\,,
\esp\eeq
where $|\vec a|$ denotes the weight of $\mathcal{L}_{\vec a}(x)$.
Note that we have apparently doubled the number of functions, so not all the functions $L_{\vec a}(x)$ 
and $\overline{L}_{\vec a}(x)$ can be independent. Indeed, one can observe that
\beq
\overline{L}_{\vec a}(x) = \textrm{[product of lower weight SVHPLs of the form } L_{\vec a}(x) \textrm{ ]}\,.
\eeq
The functions $\overline{L}_{\vec a}(x)$ can thus always be rewritten as linear combinations of 
products of SVHPLs of lower weights. In other words, the multiplicative span of the functions $L_{\vec a}(x)$ 
and multiple zeta values spans the whole algebra of SVHPLs.
As an example, in this basis the ladder integrals take the very compact form
\beq
f^{(L)}(x,\xbar) = (-1)^{L+1}\,{2}\,L_{\scriptsize\underbrace{\small 0,\ldots,0}_{L-1},0,1,\scriptsize\underbrace{ 0,\ldots,0}_{L-1}}(x)\,.
\eeq
While we present most of our result in terms of the $L_{\vec a}(x)$, we occassionally find it convenient to employ the $\overline{L}_{\vec a}(x)$ and the $\mathcal{L}_{\vec a}(x)$ to obtain more compact expressions.

The second symmetry group is the group $S_3$ which acts via the transformations of the argument
\begin{eqnarray}\label{eq:S3}
x \to x\,, & x\to 1- x\,,& x\to 1/(1-x)\,,\\
\nonumber x\to 1/ x\,, & x\to 1-1/x\,,& x\to  x/( x-1)\,.
\end{eqnarray}
This action of $S_3$ permutes the three singularities $\{0,1,\infty\}$ in the integral representations 
of the harmonic polylogarithms. In addition, this action has also a physical interpretation. The 
different cross ratios one can form out of four points $x_i$ are parametrised by the group 
$S_4/(\mathbb{Z}_2\times\mathbb{Z}_2)\simeq S_3$. The action~\eqref{eq:S3} is the 
representation of this group on the cross ratios in the parametrisation~\eqref{concrrat}. 

\subsection{The $\bar x \rightarrow 0$  limit of SVHPLs}
\label{sec:asympt-limit-svhpls}
We will be using knowledge of the asymptotic expansions of integrals in the limit $\bar x \rightarrow 0$ 
in order to constrain, and even determine, the integrals themselves. If the function lives in the 
space of SVHPLs there is a very direct and simple way to obtain the full function from its asymptotic 
expansion.

This direct procedure relies on the close relation between the series expansion of SVHPLs around 
$\bar x=0$ and ordinary HPLs.  In the case where SVHPLs are analytic at $(x,\bar x)=0$ (i.e. when the 
corresponding word 
ends in a `1') then
\begin{align}
  \lim_{\bar x \rightarrow 0} {\cal L}_w(x) &= H_w (x) \ .
\end{align}
Similar results exist in the case where ${\cal L}_w(x)$ is not analytic at the origin. In that case the 
limit does strictly speaking not exist, but we can, nevertheless, represent the function in a 
neighbourhood of the origin as a polynomial in $\log u$, whose coefficients are analytic functions. More 
precisely, using the shuffle algebra properties of SVHPLs, we have a unique decomposition
\beq
{\cal L}_w(x) = \sum_{p,w'} a_{p,w'} \log^{p}u  \, \mathcal{L}_{w'}(x)\,,
\eeq
where $a_{p,w'}$ are integer numbers and $\mathcal{L}_{w'}(x)$ are analytic at the origin $(x,\bar x)=0$.

Conversely, if we are given a function $f(x,\bar x)$ that around $\bar x=0$ admits the asymptotic expansion
\beq\label{eq:f_expansion}
f(x,\bar x) = \sum_{p,w} a_{p,w} \log^{p}u  \, H_{w}(x) + \mathcal{O}(\bar x)\,,
\eeq
where the $a_{p,w}$ are independent of $(x,\bar x)$ and $w$ are words made out of the letters 0 and 1 
ending in a 1, there is a unique function $f_{\textrm{SVHPL}}(x,\bar x)$ which is a linear combination of products of
SVHPLs that has the same asymptotic expansion around $\bar x=0$ as $f(x,\bar x)$. Moreover, this 
function is simply obtained by replacing the HPLs in eq.~\eqref{eq:f_expansion} by their single-valued 
analogues,
\begin{align}
  f_{\textrm{SVHPL}}(x,\bar x) = \sum_{p,w} a_{p,w} \log^{p}u \,  {\cal L}_{w}(x)\ . 
\end{align}
In other words, $f(x,\bar x)$ and $f_{\textrm{SVHPL}}(x,\bar x)$ agree in the limit $\bar x\to0$ up to power-suppressed terms.

It is often the case that we find simpler expressions by expanding out all products, i.e. by not explicitly writing the powers of logarithms of $u$. 
More precisely, replacing $\log u$ by $\log x+\log\bar x$ in eq.~\eqref{eq:f_expansion} and using the shuffle product for HPLs, we can write eq.~\eqref{eq:f_expansion} in the form
\beq\label{eq:f_expansion_2}
f(x,\bar x) = \sum_{w} a_{w} \, H_{w}(x) + \log\bar x\,P(x,\log\bar x) + \mathcal{O}(\bar x)\,,
\eeq
where $P(x,\log\bar x)$ is a polynomial in $\log\bar x$ whose coefficients are HPLs in $x$. From the 
previous discussion we know that there is a linear combination of SVHPLs that agrees with $f(x,\bar x)$ 
up to power-suppressed terms. In fact, this function is independent of the actual form of the polynomial 
$P$, and is completely determined by the first term in the left-hand side of eq.~\eqref{eq:f_expansion_2},
\begin{align}
 f_{\textrm{SVHPL}}(x,\bar x) &= \sum_{w} a_{w} \, {\cal L}_{w}(x)\,.
\end{align}

So far we have only described how we can always construct a linear combination of SVHPLs that agrees with a given function in the limit $\bar x\to0$ up to power-suppressed terms. The inverse is obviously not true, and we will encounter such a situation for the Hard integral. In such a case we need to enlarge the space of functions to include more general classes of multiple polylogarithms.
Indeed, while SVHPLs have symbols whose entries are all drawn from the set $\{x,\bar x,1-x,1-\bar x\}$, it was observed in ref.~\cite{Chavez:2012kn} that the symbols of three-mass three-point functions (which are related to conformal four-point functions upon sending a point to infinity) in dimensional regularisation involve functions whose symbols also contain the entry $x-\bar x$. Function of this type cannot be expressed in terms of HPLs alone, but they require more general classes of multiple polylogarithms, defined recursively by $G(x)=1$ and,
\beq
G(a_1,\ldots,a_n;x) = \int_0^x\frac{dt}{t-a_1}\,G(a_2,\ldots,a_{n};t) \, , \quad G(\vec 0_p;x) = \frac{\log^p(x)}{p!} \, ,
\eeq
where $a_i\in\mathbb{C}$. We will encounter such functions in later sections when constructing the analytic results for the Easy and Hard integrals.


\section{The short-distance limit}
\label{sec:asymptotic_expansion}

In this section we sketch how the method of `asymptotic expansion of Feynman integrals' can
deliver asymptotic series for the $\bar x \to 0$ limit of the Easy and the Hard integral. These expansions
contain enough information about the integrals to eventually fix ans\"atze for the full expressions.

In ref.~\cite{Eden:2012fe,Eden:2012rr} asymptotic expansions were derived for both the Easy and 
Hard integrals in the limits where 
one of the cross ratios, say $u$, tends to zero. 
The limit $u\rightarrow 0,v\rightarrow 1$ can be described as a short-distance limit, $x_2\rightarrow x_1$. 
Let us assume that we have got rid of the coordinate $x_4$ by sending it to infinity and that we are 
dealing with a function
of three coordinates, $x_1,x_2,x_3$, one of which, say $x_1$, can be set to zero.
The short-distance limit we are interested in then corresponds to $x_2\rightarrow 0$, so that the 
coordinate $x_2$ is small (soft) and the coordinate $x_3$ is large (hard).
This is understood in the Euclidean sense, i.e. $x_2$ tends to zero precisely when each of its 
component tends to zero.
One can formalise this by multiplying $x_2$ by a parameter $\rho$ and then considering the limit
$\rho \rightarrow 0$ upon which $u \sim \rho^2, v - 1 \sim \rho$. 

 For a Euclidean limit in momentum space, one can apply the well-known formulae
for the corresponding asymptotic expansion written in graph-theoretical language (see ref.~\cite{smirnovBook} for a review).
One can also write down similar formulae in position space. In practice, it is often more efficient
to apply the prescriptions of the strategy of expansion by regions~\cite{smirnovBook,Beneke:1997zp}
 (see also Chapter~9 of ref.~\cite{smirnovNewBook} for a recent review),
which are equivalent to the graph-theoretical prescriptions in the case of Euclidean limits.
The situation is even simpler in position space where we work with propagators $1/x_{ij}^2$.
It turns out that in order to reveal all the regions contributing to the asymptotic expansion of a 
position-space Feynman integral
it is sufficient to consider each of the integration coordinates $x_i$ either soft (i.e. of order $x_2$) or 
hard (i.e. of order $x_3$).
Ignoring vanishing contributions, which correspond to integrals without scale, one obtains
a set of regions relevant to the given limit.
One can reveal this set of regions automatically, using  the code described in refs.~\cite{Pak:2010pt,Jantzen:2012mw}.
  
The most complicated contributions in the expansion correspond to regions where the internal 
coordinates are either all hard or soft. For the Easy and Hard integrals, this gives three-loop 
two-point integrals with numerators.
In ref.~\cite{Eden:2012fe}, these integrals were evaluated by treating three numerators as extra 
propagators with negative exponents, so
that the number of the indices in the given family of integrals was increased from nine to twelve.
The integrals were then reduced to master integrals using integration-by-parts (IBP) identities using 
the {\tt c++} version of the code {\tt FIRE} \cite{Smirnov:2008iw}. While this procedure is not 
optimal, it turned out to be sufficient for the computation in ref.~\cite{Eden:2012fe}.
In ref.~\cite{Eden:2012rr}, a more efficient way was chosen: performing a tensor decomposition
and reducing the problem to evaluating integrals with nine indices by the well-known {\tt MINCER} 
program~\cite{Gorishnii:1989gt},
which is very fast because it is based on a hand solution of the IBP relations for this specific family 
of integrals.
This strategy has given the possibility to evaluate much more terms of the asymptotic expansion.

It turns out that the expansion we consider includes, within dimensional regularisation,
the variable $u$ raised to powers involving an amount proportional to $\epsilon=(4-d)/2$.
A characteristic feature of asymptotic expansions is that individual contributions may exhibit poles. 
Since the conformal integrals we are dealing with are finite in four dimensions,
the poles necessarily cancel, leaving behind some logarithms. The resulting expansions contain 
powers and logarithms
of $u$ times polynomials in $v-1$. Instead of the variable $v$, we turn to the variables $(x,\xbar)$ 
defined in eq.~\eqref{concrrat}.
Note that it is easy to see that in terms of these variables the limit $u\rightarrow 0,v\rightarrow 1$
corresponds to both $x$ and  $\bar{x}$ becoming small.

In fact, we only need the leading power term with respect to $u$ and {\em all} the terms with 
respect to $x$.
The results of ref.~\cite{Eden:2012rr} were presented in terms of infinite sums involving harmonic 
numbers, i.e., for each inequivalent 
permutation of the external points, it was shown that one can write
\beq
I(u,v) = \sum_{k=0}^3\,\log^ku\,f_k(x)+\mathcal{O}(u)\,,
\eeq
where $I(u,v)$ denotes either the Easy or the Hard integral, and $v=1-x+\mathcal{O}(\bar x)$. The 
coefficients $f_k(x)$ were expressed as 
combinations of terms of the form
\beq\label{eq:sample_sum}
\sum_{s=1}^\infty\frac{x^{s-1}}{s^i}\,S_{\vec\jmath}(s) {\rm~~or~~} \sum_{s=1}^\infty\frac{x^{s-1}}{(1+s)^i}\,S_{\vec \jmath}(s)\,,
\eeq
where $S_{\vec \jmath}(s)$ are nested harmonic sums~\cite{Vermaseren:1998uu},
\beq
S_{i}(s) = \sum_{n=1}^s\frac{1}{n^i} {\rm~~and~~} S_{i\vec\jmath}(s)=\sum_{n=1}^s\frac{S_{\vec \jmath}(n)}{n^i}\,.
\eeq

To arrive at such explicit results for the coefficients $f_k(x)$ a kind of experimental mathematics
suggested in ref.~\cite{Fleischer:1998nb} was applied:
the evaluation of the first terms in the expansion in $x$ gave a hint about the possible dependence of the coefficient at the $n$-th
power of $x$. Then an ansatz in the form of a linear combination of nested sums was constructed and
the coefficients in this ansatz were fixed by the information about the first terms. Finally, the validity of
the ansatz was confirmed using information about the next terms. The complete $x$-expansion was thus inferred from the
leading terms.

For the purpose of this paper, it is more convenient to work with polylogarithmic functions in $x$ 
rather than harmonic sums. Indeed, sums of the type~\eqref{eq:sample_sum} can easily be 
performed in terms of harmonic polylogarithms using the algorithms described 
in ref.~\cite{Moch:2001zr}. We note, however, that during the summation process, sums of the 
type~\eqref{eq:sample_sum} with $i=0$ are generated. Sums of this type are strictly speaking not 
covered by the algorithms of ref.~\cite{Moch:2001zr}, but we can easily reduce them to the case $i\neq0$ using the following procedure,
\beq\bsp
\sum_{s=1}^\infty x^{s-1}\,S_{i\vec\jmath}(s) &\, = \frac{1}{x}\sum_{s=1}^\infty\,x^s\,\sum_{n=1}^{s}\,{1\over n_1^i}\,S_{\vec \jmath}(n) = \frac{1}{x}\sum_{s=0}^\infty\,x^s\,\sum_{n=0}^{s}\,{1\over n^i}\,S_{\vec \jmath}(n)\,,
\esp\eeq
where the last step follows from $S_{\vec \jmath}(0) =0$. Reshuffling the sum by letting $s=n_1+n$,
we obtain the following relation which is a special case of eq.~(96) in ref.~\cite{Fleischer:1998nb}:
\beq\bsp
\sum_{s=1}^\infty x^{s-1}\,S_{i\vec\jmath}(s)&\, 
=\frac{1}{x}\sum_{n_1=0}^\infty\,x^{n_1}\,\sum_{n=0}^{\infty}{\,x^{n}\over n^i}\,S_{\vec \jmath}(n) 
= \frac{1}{1-x}\,\sum_{s=1}^{\infty}{\,x^{s-1}\over s^i}\,S_{\vec \jmath}(s)\,.
\esp\eeq
The last sum is now again of the type~\eqref{eq:sample_sum} and can be dealt with 
using the algorithms of ref.~\cite{Moch:2001zr}.

Performing all the sums that appear in the results of ref.~\cite{Eden:2012rr}, we find for example
\begin{eqnarray}
x_{13}^2\,x_{24}^2\,E_{14;23} & = & \frac{\log u}{x}\Big(H_{2,2,1}-H_{2,1,2}+H_{1,3,1}+2 H_{1,2,1,1}-H_{1,1,3}-2 H_{1,1,1,2}\\
\nonumber&-&6 \zeta _3 H_2-6 \zeta _3 H_{1,1} \label{aeE1423}\Big) -\frac{2}{x} \Big(2 \zeta _ 3 H_{2,1}-4 \zeta _ 3 H_{1,2}+4 \zeta _ 3 \
H_{1,1,1}+H_{3,2,1}\\
\nonumber&-&H_{3,1,2}+H_{2,3,1}-H_{2,1,3}+2 H_{1,4,1}+2 \
H_{1,3,1,1}+2 H_{1,2,2,1}-2 H_{1,1,4}\\
\nonumber&-&2 H_{1,1,2,2}-2 H_{1,1,1,3}-6 \
\zeta _ 3 H_ 3\Big)+\mathcal{O}(u)\,,\\
\nonumber\phantom{A}&&\\
x_{13}^4\,x_{24}^4\,H_{12;34} & = & \frac{4\log u}{x^2} \Big(H_{1,1,2,1}-H_{1,1,1,2}-6 \zeta_3 H_{1,1}\Big) \label{aeH1234}-\frac{2}{x^2}\Big(4 H_{2,1,2,1}-4 \
H_{2,1,1,2}\\
\nonumber&+&4 H_{1,1,3,1}-H_{1,1,2,1,1}-4 H_{1,1,1,3}+H_{1,1,1,2,1}-24 \zeta_3 H_{2,1}+6 \zeta_3 H_{1,1,1}\Big)\\
\nonumber&+&\mathcal{O}(u)\,,
\end{eqnarray}
where we used the compressed notation, e.g., $H_{2,1,1,2}\equiv H(0,1,1,1,0,1;x)$.
The results for the other orientations are rather lengthy, so we do not show them here, but we collect them in 
Appendix~\ref{app:asymptotics}. Let us however comment about the structure of the functions $f_k(x)$ that appear in the expansions. 
The functions $f_k(x)$ can always be written in the form
\beq
f_k(x) = \sum_{l}R_{k,l}(x)\times[\textrm{HPLs in }x]\,,
\eeq
where $R_{k,l}(x)$ may represent any of the following rational functions
\beq
\frac{1}{x^2}\,,\qquad\frac{1}{x}\,,\qquad \frac{1}{x(1-x)}\,.
\eeq
We note that the last rational function only enters the asymptotic expansion of $H_{13;24}$.

The aim of this paper is to compute the Easy and Hard integrals by writing for each integral an ansatz
of the form
\beq
\sum_iR_i(x,\bar x) P_i(x,\bar x)\,,
\eeq
and to fix the coefficients that appear in the ansatz by matching the limit $\bar{x}\to0$ to the asymptotic expansions 
presented in this section. In the previous section we argued that a natural space of functions for the polylogarithmic 
part $P_i(x,\bar x)$ are functions that are single-valued in the complex $x$ plane in Euclidean space. We however still 
need to determine the rational prefactors $R_i(x,\bar x)$, which are not constrained by single-valuedness. 

A natural ansatz would consist in using the same rational prefactors as those appearing in the ladder type integrals. 
For ladder type integrals we have
\beq
R_i^{\textrm{ladder}}(x,\bar x) = \frac{1}{(x-\bar x)^\alpha}\,,\qquad \alpha\in\mathbb{N}\,,
\eeq
plus all possible transformations of this function obtained from the action of the $S_3$ symmetry~\eqref{eq:S3}. 
Then in the limit $u\to0$ we obtain
\beq
\lim_{\bar x\to0}R_i^{\textrm{ladder}}(x,\bar x) = \frac{1}{x^\alpha}\,.
\eeq
We see that the rational prefactors that appear in the ladder-type integrals can only give rise to rational prefactors 
in the asymptotic expansions with are pure powers of $x$, and so they can never account for the rational function $1/(x(1-x))$ 
that appears in the asymptotic expansion of $H_{13;24}$. We thus need to consider more general prefactors than those 
appearing in the ladder-type integrals. This issue will be addressed in the next sections.

\section{The Easy integral}
\label{sec:Easy}

\subsection{Residues of the Easy integral}

The Easy integral is defined as
\begin{equation} \label{eq:7}
  E_{12;34}   =   \frac{x^2_{23} x^2_{24}}{\pi^6}
\int \frac{d^4x_5 \, d^4x_6 \, d^4x_7 \ x^2_{16}}{(x_{15}^2 x_{25}^2 x_{35}^2)
x_{56}^2 (x_{26}^2 x_{36}^2 x^2_{46}) x^2_{67} (x_{17}^2 x_{27}^2
x_{47}^2)}
 \, .
\end{equation}
To find all its leading singularities we order the integrations as follows
\begin{equation}
  E_{12;34}  =   \frac{x_{23}^2 x_{24}^2
  }{\pi^6}
\left[ \int \frac{ d^4x_6 \ x^2_{16}}{x_{26}^2 x_{36}^2 x^2_{46}} 
\left( \int \frac{ d^4x_5 } {x_{15}^2 x_{25}^2 x_{35}^2
x_{56}^2 }
\right) \left(\int \frac{  d^4x_7 } {x_{17}^2 x_{27}^2
x_{47}^2  x^2_{67} }
\right)\right] \, .
\end{equation}

First the $x_7$ and $x_5$ integrations: they are both the same as the
massive box computed in the Introduction and thus
give leading singularities (see eq.~\eqref{eq:2})
\begin{equation}
\pm{1 \over
 4\,\lambda_{1236}} \qquad \pm{1 \over 4\,
 \lambda_{1246}}\ ,
\end{equation} respectively.  So we can move
directly to the final $x_6$ integration
\begin{equation}\label{eq:6}
\frac{1}{16\,\pi^6}  \int\frac{ d^4x_6\ x^2_{16}}{x_{26}^2 x_{36}^2 x^2_{46}
  \lambda_{1236}  \lambda_{1246}} .
\end{equation}

Here there are five factors in the denominator and we want to take the residues when four of them vanish
to compute the leading singularity, so there are various choices to
consider. The simplest option is to cut the three propagators
$1/x_{i6}^2$. Then on this cut we have $\lambda_{1236|\mathrm{cut}}=\pm x_{16}^2 x_{23}^2$
and $\lambda_{1246|\mathrm{cut}}=\pm x_{16}^2 x_{24}^2$, where the vertical line indicates the value on the cut, and the integral reduces to
the massive box. This simplification of the $\lambda$ factors is similar to the phenomenon of 
composite leading singularities \cite{Buchbinder:2005wp}.
 Thus cutting either of the two $\lambda$s will result in\footnote{With a slight abuse of language, in 
the following we use the word `cut' to designate that we look at the zeroes of a certain denominator factor.}
\begin{align}
\text{  leading  singularity \#1 of } E_{12;34} = \pm  \frac{1}{64\,\pi^6  \lambda_{1234} }  \ .
\end{align}

The only other possibility is cutting both $\lambda$'s.
There are then three possibilities, firstly we could cut $x_{26}^2$ and $x_{36}^2$ as well as the two 
$\lambda's$. On this cut $\lambda_{1236}$ reduces to $\pm x_{16}^2x_{23}^2$ and one obtains 
residue $\#1$ again. Similarly in the second case where we cut $x_{26}^2$,  $x_{46}^2$ and the 
two $\lambda$s.

So finally we consider the case where we cut $x_{36}^2$,  $x_{46}^2$ and the two $\lambda$'s. In 
this case $\lambda_{1236|\mathrm{cut}}=\pm (x^2_{16}x^2_{23} - x^2_{13}x^2_{26})$
and $\lambda_{1246|\mathrm{cut}}=\pm (x^2_{16}x^2_{24} - x^2_{14}x^2_{26})$. Notice
that setting $\lambda_{1236}=\lambda_{1246}=0$ means setting $x_{16}^2=x_{26}^2=0$.
We then need to compute the Jacobian associated with cutting $x_{36}^2, x_{46}^2, \lambda_{1236},
\lambda_{1246}$
\beq\bsp
\det&\left.\left({\partial(x_{36}^2, x_{46}^2, \lambda_{1236},
    \lambda_{1246}) \over \partial x_6^\mu}  \right)\right|_{\mathrm{cut}}  \\
    &= \pm  16 \det\left.\Big(x^\mu_{36},\ x^\mu_{46},\ x^\mu_{16}x_{23}^2 - x^2_{13}x^\mu_{26} ,\ x^\mu_{16}x^2_{24} - x^2_{14}x^\mu_{26}\Big)\right|_{\mathrm{cut}} \\
&= \pm   16 \det\left.(x^\mu_{36},x^\mu_{46},x^\mu_{16} ,x^\mu_{26})(x_{23}^2 x^2_{14} - x^2_{24}x^2_{13})\right|_{\mathrm{cut}} \\
&=\pm 4 \lambda_{1234}(x_{23}^2 x^2_{14} - x^2_{24}x^2_{13})\ ,
\esp\eeq
The  result of the $x_6$ integral~(\ref{eq:6}) is
\begin{equation}
\frac{1}{64\,\pi^6} \left.\frac{  x^2_{16}}{ x^2_{26} \lambda_{1234}   (x_{23}^2 x^2_{14} - x^2_{24}x^2_{13})}\right|_{\mathrm{cut}}
  \end{equation}

  At this point there is a subtlety, since on the cut we have simultaneously $x^2_{16}x^2_{23} - x^2_{13}x^2_{26}=x^2_{16}x^2_{24} - x^2_{14}x^2_{26}=0$, i.e. $x^2_{16}=x^2_{26}=0$  
  and so $x^2_{16} \over  x^2_{26}$ is undefined. More specifically, the integral depends on 
  whether we take $x^2_{16}x^2_{23} - x^2_{13}x^2_{26}=0$ first or $x^2_{16}x^2_{24} - x^2_{14}x^2_{26}=0$ first. 
So we get two possibilities (after multiplying by the external factors $x^2_{23}x^2_{24}$ in eq.~\eqref{eq:7}) :
  \begin{align}
    \text{  leading  singularity \#2 of } E_{12;34} &= \pm  \frac{ x^2_{13}x^2_{24}}{64\,\pi^6\,  \lambda_{1234}   (x_{23}^2 x^2_{14} - x^2_{24}x^2_{13})}\\
    \text{  leading  singularity \#3 of } E_{12;34}    &=\pm
    \frac{ x^2_{14}x^2_{23}} {64\,\pi^6\, \lambda_{1234}   (x_{23}^2
      x^2_{14} - x^2_{24}x^2_{13})}\ .
\end{align}

We conclude that the Easy integral
takes the `leading singularity times pure function' form\footnote{A similar form of the Easy leading singularities, as well as those of the Hard integral discussed in the next section, was independently obtained by S.~Caron-Huot \cite{Simonunpublished}.}
\begin{align}
  E_{12;34}= {1 \over x_{13}^2 x_{24}^2}\left[ {E^{(a)}(x,\bar
    x) \over x-\bar x } + {E^{(b)}(x,\bar
    x) \over (x-\bar x)(v-1)}+ {v\, E^{(c)}(x,\bar
    x)  \over (x-\bar x)(v-1)}    \right]\ .
\end{align}
We note that the $x_3 \leftrightarrow x_4$ symmetry relates
$E^{(b)}$ and $E^{(c)}$. Furthermore, putting everything over a common
denominator it is easy to see that $E^{(a)}$ can be absorbed
into the other two functions. We conclude that there is in fact only
one independent function, and the Easy integral can be written in
terms of a single pure function $E(x,\bar x)$ as
\begin{align} \label{e1234}
  E_{12;34}= {1 \over x_{13}^2 x_{24}^2\,(x-\bar x)(v-1)}\left[ E(x,\bar
    x) + v \, E\left(\frac{x}{x-1},\frac{\bar
    x}{\bar x-1}\right)    \right]\ .
\end{align}
The function $E(x,\bar x)$ is antisymmetric under the interchange of
$x,\bar x$
\begin{align}
  E(\bar x,x) = - E(x, \bar x)\,,
\end{align}
to ensure that $E_{12;34}$ is a symmetric function of $x, \bar x$, but
it possesses no other symmetry.

The other two orientations of the Easy integral are then found by
permuting various points and are given by
\begin{align}
    E_{13;24}&= {1 \over x_{13}^2 x_{24}^2\,(x-\bar x)(u-v)}\left[
      u \, E \left({1\over x},{1\over \bar
    x}\right)+ v\, E\left({1\over 1-x},{1 \over 1-\bar
    x}\right)    \right]\,,\\
    E_{14;23}&= {1 \over x_{13}^2 x_{24}^2\,(x-\bar x)(1-u)}\left[
       E({1- x},{1-\bar    x})+u \, E\left({1-{1\over x}},{1 -{1\over\bar x}}\right)     \right]\ .
\end{align}
It is thus enough to have an expression for $E(x,\xbar)$ to determine all possible orientations of the 
Easy integral. The functional form of $E(x,\xbar)$ will be the purpose of the rest of this section.

\subsection{The symbol of $E(x,\bar x)$}

In this subsection we determine the symbol of $E(x,\bar x)$, and in the next section we describe its uplift to a function. This strategy
seems over-complicated in the case at hand, because $E(x,\bar x)$ can in fact directly be obtained
in terms of SVHPLs of weight six from its asymptotic expansion using the method described in
Section~\ref{sec:asympt-limit-svhpls}. 
The two-step derivation (symbol and subsequent uplift)  is included mainly for pedagogical purposes because it
equally applies to the Hard integral and our four-loop example, where the functions are not writeable in terms of SVHPLs
only so that a direct method yet has to be found.

Returning to the Easy integral, we start by writing down the most general tensor of rank six that
\begin{itemize}
\item has all its entries drawn from the set $\{x,1-x,\xbar,1-\xbar\}$,
\item satisfies the first entry condition, i.e. the first factors in each tensor are either $x\xbar$ or $(1-x)(1-\xbar)$,
\item is odd under an exchange of $x$ and $\xbar$.
\end{itemize}
This results in a tensor that depends on $2\cdot4^5/2 = 1024$ free coefficients (which we assume 
to be rational numbers). Imposing the integrability condition~\eqref{intCond} reduces the number of 
free coefficients to 28, which is the number of SVHPLs of weight six that are odd under an 
exchange of $x$ and $\xbar$. The remaining free coefficients can be fixed by matching to the limit 
$u \to 0, v \to 1$, or equivalently $\xbar \to 0$.

In order to take the limit, we drop every term in the symbol containing an entry $1-\bar x$  and
we replace $\bar x \rightarrow u/x$, upon which the singularity is hidden
in $u$. As a result, every permutation of our ansatz yields a symbol composed of the three letters 
$\{u, x, 1-x \}$. This tensor can immediately be matched to the symbol of the asymptotic expansion 
of the Easy integral discussed in Section~\ref{sec:asymptotic_expansion}.
Explicitly, the limits
\begin{eqnarray}
x_{13}^2 x_{24}^2 \, E_{12;34} & \rightarrow & -\frac{1}{x^2} \left[ \lim_{\bar x \rightarrow 0} \, E(x,\bar x) + \lim_{\bar x \rightarrow 0} E\left(\frac{x}{x-1},\frac{ \bar x }{\bar x - 1} \right) \right] \nonumber \\ 
& & + \frac{1}{x} \lim_{\bar x \rightarrow 0}  E\left(\frac{x}{x-1}, \frac{\bar x}{\bar x - 1} \right)\label{eq:8}      \\[2mm]
x_{13}^2 x_{24}^2 \, E_{13;24} & \rightarrow & - \frac{1}{x} \, \lim_{\bar x \rightarrow 0} \, E\left(\frac{1}{1-x}, \, \frac{1}{1 - \bar x}\right) \\[2mm]
x_{13}^2 x_{24}^2 \, E_{14;23} & \rightarrow & \phantom{-} \frac{1}{x} \, \lim_{\bar x \rightarrow 0} \, E(1-x,\, 1- \bar x)
\end{eqnarray}
can be matched with the asymptotic expansions recast as HPLs. All three conditions are consistent 
with our ansatz; each of them on its own suffices to determine all remaining constants. 
The resulting symbol is a linear combination of 1024 tensors with entries drawn from the set 
$\{x,1-x,\xbar,1-\xbar\}$ and with coefficients $\{ \pm 1, \, \pm 2\}$. 

Note that the uniqueness of the uplift procedure for SVHPLs given in Section~\ref{sec:asympt-limit-svhpls} implies
that each asymptotic limit is sufficient to fix the symbol.

\subsection{The analytic result for $E(x,\bar x)$: uplifting from the symbol}

In this section we determine the function $E(x,\bar x)$ defined in eq.~\eqref{e1234} starting from its symbol.
As the symbol has  all its entries drawn from the set $\{x,1-x,\bar x,1-\bar x\}$, the function $E(x,\bar x)$ can be 
expressed in terms of the SVHPLs classified in \cite{BrownSVHPLs}. Additional single-valued terms\footnote{In principle we cannot exclude at this stage more 
complicated functions of weight less than six multiplied by zeta values.} 
proportional to zeta values
can be fixed by again appealing to the asymptotic expansion of the integral.

We start by writing down an ansatz for $E(x,\bar x)$ as a linear combination 
of weight six  of SVHPLs 
that is odd under exchange of $x$ and $\bar x$. Note that we have some freedom
w.r.t. the basis for our ansatz. In the following we choose 
basis elements containing a single factor of the form $L_{\vec a}(x)$. This ensures that all the terms 
are linearly independent.

Next we fix the free coefficients in our ansatz by requiring its symbol to agree 
with that of $E(x,\bar x)$ determined in the previous section. As we had
started from SVHPLs with the correct symmetries and weight, all coefficients are 
fixed in a unique way. 
We arrive at the 
following expression for $E(x,\bar x)$:
\beq\bsp\label{eq:E_result}
E(x,\bar x) &\,=
4 L_{2,4}-4 L_{4,2}-2 L_{1,3,2}+2 L_{2,1,3}-2 L_{3,1,2}+4 L_{3,2,0}\\
&\,-2 L_{2,2,1,0}+8 L_{3,1,0,0}+2 L_{3,1,1,0}-2 L_{2,1,1,1,0}
\,
\esp\eeq
For clarity, we suppressed the argument of the $L$ functions and we employed the compressed
notation for HPLs, e.g., $L_{3,2,1}\equiv L_{0,0,1,0,1,1}(x,\bar x)$.
The asymptotic limits of the last expression correctly reproduce the terms proportional to zeta values in eq.~(\ref{aeE1423}) and the formulae in Appendix~\ref{app:asymptotics}. 



\subsection{The analytic result for $E(x,\bar x)$: the direct approach}
\label{sec:analytic-result-ex}


Here we quickly give the direct method for obtaining $E(x,\bar x)$ 
explicitly from its asymptotics via the method outlined in Section~\ref{sec:asympt-limit-svhpls}.

The asymptotic value of the Easy integral in the permutation $E_{12;34}$ is given in Appendix~\ref{app:asymptotics}. Comparing
eq.~(\ref{app1eq:1}) with eq.~(\ref{eq:8})
and further writing $\log u =\log x + \log \bar x$ and expanding out products of 
functions we find for the asymptotic value of $E(x,\bar x)$:
\beq\bsp
 E(x,\bar x) &\,= 4 \zeta_3 H_{2,1}+2 H_{2,4}-2
   H_{4,2}+H_{1,2,3}-H_{1,3,2}-2
   H_{1,4,0}+H_{2,1,3}-H_{3,1,2}\\
   &\,+2
   H_{3,2,0}-H_{1,3,1,0}+H_{2,1,2,0}-2
   H_{2,2,0,0}-H_{2,2,1,0}+H_{3,1,1,0}+2
   H_{1,2,0,0,0}\\
   &\,+H_{1,2,1,0,0}-H_{2,1,1,0,0}-20 \zeta_5 H_1+8
   \zeta_3 H_3+2 \zeta_3 H_{1,2}\\
   &\, + \ \log\bar x\,P(x,\log\bar x) + \mathcal{O}(\bar x)\,, 
\esp\eeq
where $P$ is a polynomial in $\log\bar x$ with coefficients that are HPLs in $x$.
From the discussion in Section~\ref{sec:asympt-limit-svhpls} we know that there is a unique combination of SVHPLs with this precise asymptotic behavior, and so we find a natural ansatz for $E(x,\bar x)$,
\begin{align}
  E(x,\bar x) &\,= 4 \zeta_3 {\cal L}_{2,1}+2 {\cal L}_{2,4}-2
   {\cal L}_{4,2}+{\cal L}_{1,2,3}-{\cal L}_{1,3,2}-2
   {\cal L}_{1,4,0}+{\cal L}_{2,1,3}-{\cal L}_{3,1,2}+2
   {\cal L}_{3,2,0} \nonumber \\
   &\,-{\cal L}_{1,3,1,0}+{\cal L}_{2,1,2,0}-2
   {\cal L}_{2,2,0,0}-{\cal L}_{2,2,1,0}+{\cal L}_{3,1,1,0}+2
   {\cal L}_{1,2,0,0,0}+{\cal L}_{1,2,1,0,0} \nonumber \\
   &\,-{\cal L}_{2,1,1,0,0}-20 \zeta_5 {\cal L}_1+8
   \zeta_3 {\cal L}_3+2 \zeta_3 {\cal L}_{1,2}\ . \label{eq:E_result}
\end{align}
We have 
lifted this function from its asymptotics in just one  limit $\bar x \rightarrow 0$ while we also know
two other limits of this function given in eq.~(\ref{aeE1423}) and Appendix A. Remarkably, eq.~(\ref{eq:E_result}) is automatically consistent with these two limits, giving a strong indication that it is indeed the right function.
Furthermore, eq.~\eqref{eq:E_result} can then in turn be rewritten in a way that makes the antisymmetry under exchange of $x$ and $\bar x$ manifest, and we recover eq.~(\ref{eq:E_result}).
Note also that antisymmetry in $x \leftrightarrow \bar x$ was not 
input anywhere, and the fact that the resulting function is indeed antisymmetric is a non-trivial consistency check.

As an aside we also note here that the form of $E(x,\bar x)$, expressed in the particular basis of SVHPLs we chose to work with, is very simple, having only coefficients 
$\pm 1$ or $\pm2$ for the polylogarithms of weight six.
Indeed other orientations of $E$ have even simpler forms, for instance
\beq\bsp
 E(1/x,1/\bar x)&\,= \mathcal{L}_{2,4}-\mathcal{L}_{3,3}-\mathcal{L}_{1,2,3}
   +\mathcal{L}_{1,3,2}-\mathcal{L}_{1,4,0}-\mathcal{L}_{2,1,3}+\mathcal{L}_
   {3,1,2}-\mathcal{L}_{4,0,0}+\mathcal{L}_{4,1,0}\\
   &\,+\mathcal{L}_{1,3,0,0}+\mathcal{L}_{1,3,1,0}-\mathcal{L}_{2,1,2,0}+\mathcal{L}_{2,2,1,0}+\mathcal{L
   }_{3,0,0,0}-\mathcal{L}_{3,1,1,0}-\mathcal{L}_{1,2,1,0,0}\\
   &\,-\mathcal{L}_{2,
   1,0,0,0}+\mathcal{L}_{2,1,1,0,0}+8 \zeta_3 \mathcal{L}_3-2 \zeta_3 \mathcal{L}_{1,2}-6 \zeta_3 \mathcal{L}_{2,0}-4 \zeta_3
   \mathcal{L}_{2,1}\,,
\esp\eeq
with all coefficients of the weight six SVHPLs being $\pm 1$,  or in the manifestly antisymmetric 
form with all weight six SVHPLs with coefficient $+1$
\begin{dmath}
E(1/x,1/\bar x)=
  L_{2,4}+L_{1,3,2}+L_{3,1,2}+L_{4,1,0}+L_{1,3,0,0}+L_{1,3,1,0}+L_{
   2,2,1,0}+L_{3,0,0,0}+L_{2,1,1,0,0}+6\zeta_3 L_3 -2 \zeta_3 L_{2,1} .  
\end{dmath}

\subsection{Numerical consistency tests for $E$}
We have determined the analytic result for the Easy integral relying
on the knowledge of its residues, symbol and asymptotic expansions. In order to check the 
correctness of the result, we evaluated $E_{14;23}$ numerically\footnote{All polylogarithms 
appearing in this paper have been evaluated numerically using the 
{\sc GiNaC}~\cite{Bauer:2000cp} and {\tt HPL}~\cite{Maitre:2005uu} packages.} and 
compared it to a direct numerical evaluation of the coordinate space integral using {\tt FIESTA} 
\cite{Smirnov:2008py,Smirnov:2009pb}.

To be specific, we evaluate the conformally-invariant function $x_{13}^2 x_{24}^2\,E_{14;23}$. 
Applying a conformal transformation to send $x_4$ to infinity, the integral takes the simplified form,
\begin{equation}
\lim_{x_4\to \infty} x_{13}^2 x_{24}^2\,E_{14;23} =\frac{1}{\pi^6} \int \frac{d^4x_5 d^4x_6 d^4x_7 \; x_{13}^2 x_{16}^2}{(x_{15}^2x_{25}^2) x_{56}^2 (x_{26}^2 x_{36}^2) x_{67}^2 (x_{17}^2 x_{37}^2)}\, ,
\end{equation}
with only 8 propagators. We use the remaining freedom to fix $x_{13}^2=1$ so that $u=x_{12}^2$ 
and $v=x_{23}^2$. Other numerical values for $x_{13}^2$ are possible, of course, but we found 
that this choice yields relatively stable numerics.

After Feynman parameterisation, the integral is only seven-dimensional and can be evaluated with 
off-the-shelf software. We generate the integrand with {\tt FIESTA} and perform the numerical 
integration with a stand-alone version of  {\tt CIntegrate}. 
Using the algorithm Divonne\footnote{Experience shows that Divonne outperforms other algorithms of the Cuba library for problems roughly this size.}, 
we obtain roughly five digits of precision after five million function evaluations.

In total, we checked 40 different pairs of values for the cross ratios and we found very good 
agreement in all cases. A sample of the numerical checks is shown in Table~\ref{tab:easy}. Note that 
$\delta$ denotes the relative error between the analytic result and the number obtained by {\tt FIESTA},
\beq
\delta = \left|\frac{N_{analytic} - N_{FIESTA}}{N_{analytic} + N_{FIESTA}}\right|\,.
\eeq

\begin{table}[!t]
\begin{center}
\begin{tabular}{ccccc}
\hline\hline
$u$&$v$& Analytic & {\tt FIESTA} & $\delta$\\
\hline
0.1 & 0.2 & 82.3552 & 82.3553 & 6.6e-7 \\
 0.2 & 0.3 & 57.0467 & 57.0468 & 3.2e-8 \\
 0.3 & 0.1 & 90.3540 & 90.3539 & 5.9e-8 \\
 0.4 & 0.5 & 37.1108 & 37.1108 & 1.9e-8 \\
 0.5 & 0.6 & 31.9626 & 31.9626 & 1.9e-8 \\
 0.6 & 0.2 & 54.2881 & 54.2881 & 6.9e-8 \\
 0.7 & 0.3 & 42.6519 & 42.6519 & 4.4e-8 \\
 0.8 & 0.9 & 23.0199 & 23.0199 & 1.7e-8 \\
 0.9 & 0.5 & 30.8195 & 30.8195 & 2.4e-8\\
  \hline\hline
 \end{tabular}
 \caption{\label{tab:easy}Numerical comparison of the analytic result for $x_{13}^2 x_{24}^2\,E_{14;23}$ against {\tt FIESTA} for several values of the conformal cross ratios.}
  \end{center}
 \end{table}


\section{The Hard integral}
\label{sec:Hard}


\subsection{Residues of the Hard integral}

To find all the leading singularities we consider each integration
sequentially as follows
\begin{equation}
  H_{12;34}  =   \frac{x_{34}^2
  }{\pi^6}
  \left\{ \int \frac{ d^4x_6 }{x_{16}^2 x_{26}^2 x_{36}^2
x^2_{46}} 
\left[ \int\frac{ d^4x_5 \ x^2_{56}}{x_{15}^2 x_{25}^2 x_{35}^2
x^2_{45}} 
\left( \int\frac{ d^4x_7 } { x_{37}^2 x^2_{47} x_{57}^2 x^2_{67}}
\right)\right] \right\} \, .
\end{equation}
Let us start with the $x_7$ integration,
\begin{equation}
  \int\frac{ d^4x_7 }{x_{37}^2 x^2_{47} x_{57}^2 x^2_{67}}.
\end{equation}
This is simply the off-shell box considered in Section~\ref{sec:Introduce},
and so its leading singularities are (see eq.~(\ref{eq:2}))
\begin{equation}
\pm{1 \over4\,
 \lambda_{3456}}\ .
\end{equation}
Next we turn to the $x_5$ integration, which now takes the form
\begin{equation}\label{eq:3}
 \int \frac{  d^4x_5 \ x^2_{56} } {\, x_{15}^2 x^2_{25} x_{35}^2
   x^2_{45} \lambda_{3456}} \ .
\end{equation}
There are five factors in the denominator, and we want to cut four of them
to compute the leading singularity. The simplest option is to cut the four propagators
$1/x_{i5}^2$. Doing so would yield a new Jacobian factor
$1/\lambda_{1234}$ (exactly as in the previous subsection) and
freeze $\lambda_{3456|\mathrm{cut}}=\pm x_{56}^2 x_{34}^2$. This latter factor
simply cancels the numerator and we are left with the final $x_6$
integration being that of the box in the Introduction. Putting
everything together, the leading singularity for this choice is
\begin{align}
\text{  leading  singularity \#1 of } H_{12;34} = \pm
\frac{1}{64\pi^6\,\lambda_{1234}^2}  \ .
\end{align}

Returning to the $x_5$ integration, eq.~(\ref{eq:3}), we must consider
the possibility of cutting $\lambda_{3456}$ and three other
propagators. Cutting $x_{35}^2$ and $x_{45}^2$ immediately freezes
$\lambda_{3456|\mathrm{cut}}=\pm x_{56}^2 x_{34}^2$ which is canceled by the
numerator. Thus it is not  possible to cut these two propagators
and $\lambda_{3456}$. However, cutting $x_{15}^2, x_{25}^2,
x_{35}^2$ and $\lambda_{3456}$ is possible (the only other
possibility, i.e. cutting 
$x_{15}^2, x_{25}^2, x_{45}^2$ and $\lambda_{3456}$, gives the same
result by  by invariance of the integral under exchange of $x_3$ and $x_4$).
Indeed one finds that when $x_{35}^2=0$,
\begin{equation}
\lambda_{3456} = \pm (x^2_{45} x^2_{36}-x^2_{56}x^2_{34}) \,.
\end{equation}
To compute the leading singularity associated with this pole we need to compute
the Jacobian
\begin{equation}
J= \det \left( {\partial( x_{15}^2,x_{25}^2,x_{35}^2,\lambda_{3456})
     \over \partial x_{5}^\mu} \right)\,,
\end{equation}
As in the box case, it is useful to consider the
square of $J$ (on the cut),
\begin{align}\label{eq:5}
  J^2= 16 \det\left(
  \begin{array}{cc}
    x_{ij}^2 & -2 x_i \cdot \partial \lambda_{3456}/\partial x_5 \\
    -2 x_i \cdot \partial \lambda_{3456}/\partial x_5 & (\partial
    \lambda_{3456}/\partial x_5)^2
\end{array}
\right)\ .
\end{align}
The result of the $x_5$ integration is then simply
\begin{align}\label{eq:4}
 \left. {  x_{56}^2 \over J x_{45}^2}\right|_{\mathrm{cut}} = \left.{ x_{36}^2 \over J x_{34}^2}\right|_{\mathrm{cut}}\,,
\end{align}
where the second equality follows since $x_{56}^2$ and $x_{45}^2$ are
to be evaluated on the cut (indicated by the vertical line) for which $x^2_{45} x^2_{36}-x^2_{56}
x^2_{34}=0$.
Finally we need to turn to the remaining $x_6$ integral. We are simply left with
\begin{align}
\left.\frac{1}{16\pi^6}\, \int \frac{ d^4x_6}{\, x_{16}^2 x^2_{26} x_{46}^2 J }\right|_{\mathrm{cut}}  \,,
\end{align}
where we note that the $x_{36}^2$ propagator term has canceled with
the numerator in eq.~(\ref{eq:4}). So we have no choice left for the
quadruple cut as there are only four poles. In fact on the other cut
of the three propagators we find $J_{|\mathrm{cut}}= 4 (x_{14}^2 x_{23}^2 - x_{13}^2
x_{24}^2) x_{36}^2$, and so this brings back the propagator $x_{36}^2$.

Computing the Jacobian associated with this final integration thus  yields the final result for the leading singularity,
\begin{align}
  \text{  leading  singularity \#2 of } H_{12;34} = \pm  \frac{1}{64\pi^6\, (x_{14}^2 x_{23}^2 - x_{13}^2 x_{24}^2)
  \lambda_{1234} }  \ .
\end{align}

We conclude that the Hard integral can be written as these leading
singularities times pure functions, i.e. it has the form
\begin{align}\label{eq:H1234}
{H_{12;34}} ={1 \over x_{13}^4 x_{24}^4}  \left[ { H^{(a)}(x,\bar x) \over
 (x-\bar x)^2} + { H^{(b)}(x,\bar x) \over  (v-1)(x-\bar x)}\right]\,,
\end{align}
where $H^{(a),(b)}$ are pure polylogarithmic functions. The pure 
functions must furthermore satisfy the following properties
\begin{align}
  H^{(a)}(x,\bar x) &= H^{(a)}(\bar x, x)\,,  &
  \qquad
  H^{(b)}(x,\bar x)& = -H^{(b)}(\bar x, x)\,, \label{eq60residues}\\
  H^{(a)}(x,\bar x) &= H^{(a)}( x/ (x-1), \bar x/(\bar x-1))\,,  & \qquad
  H^{(b)}(x,\bar x) &= H^{(b)}( x/ (x-1), \bar x/ (\bar x-1))\,,\nonumber
\end{align}
in order that $H_{12;34}$ be symmetric in $x,\bar x$ and under the
permutation $x_1 \leftrightarrow x_2$. Furthermore we would expect
that $H^{(a)}(x,x)=0$ in order to cancel the pole at $x-\bar x$. In
fact it will turn out in this section that even without imposing this
condition by hand we will arrive at a unique result
which nevertheless has this particular property. 

By swapping the points around we automatically get
\begin{align}\label{eq:9}
{H_{13;24}} &={ 1 \over x_{13}^4 x_{24}^4}  \left[ { H^{(a)}(1/x,1/\bar x) \over
 (x-\bar x)^2} + { H^{(b)}(1/x,1/\bar x) \over  (u-v)(x-\bar x)}\right]\,,\\ 
{H_{14;23}} &={ 1 \over x_{13}^4 x_{24}^4}  \left[ { H^{(a)}(1-x,1-\bar x) \over
 (x-\bar x)^2} + { H^{(b)}(1-x,1-\bar x) \over  (1-u)(x-\bar
 x)}\right] \ .
\end{align}

\subsection{The symbols of $H^{(a)}(x,\bar x)$ and $H^{(b)}(x,\bar x)$}
In order to determine the pure functions contributing to the Hard integral, we proceed just like for 
the Easy integral and first determine the symbol.
For the Hard integral we have to start from two ans\"{a}tze for the symbols 
${\cal S}[H^{(a)}(x,\bar x)]$ and ${\cal S}[H^{(b)}(x,\bar x)]$.
While both pure functions are invariant under the exchange $x_1 \leftrightarrow x_2$, ${\cal S}[H^{(a)}]$ 
must be symmetric under the exchange of $x,\bar x$ and ${\cal S}[H^{(b)}]$ has to be 
antisymmetric, cf. eq.~(\ref{eq60residues}). Going through exactly the same steps as for $E$ we 
find that the single-variable limits of the symbols \emph{cannot} be matched against the data from 
the asymptotic expansions using only entries from the set $\{x,1-x,\xbar,1-\xbar\}$.
We thus need to enlarge the ansatz. 

Previously, the letter $x-\bar x\sim \lambda_{1234}$ has been encountered in 
ref.~\cite{Chavez:2012kn,Schnetz:2013hqa} in a similar context. 
We therefore consider all possible integrable symbols made from the letters 
$\{x,1-x,\bar x, 1-\bar x,x - \bar x\}$ which obey the initial entry condition (\ref{eq:S_Disc}). 
In the case of the Easy integral, the integrability condition only implied that terms depending on 
both $x$ and $\bar x$ come from products of single-variable functions. Here, on the other hand, the 
condition is more non-trivial since, for example,
\beq\bsp
 d\log \frac{x}{\bar x} \wedge d\log(x-\bar x) &\,=  d\log x \wedge d\log\bar x \, , \\
 d\log\frac{1-x}{1-\bar x} \wedge d\log(x-\bar x) &\,= d\log(1-x) \wedge d\log(1-\bar x) \, .
\esp\eeq
We summarise the dimensions of the spaces of such symbols, split according to parity under 
exchange of $x$ and $\bar x$, in Table~\ref{tab:x-xbar_dimensions}.
\begin{table}[!h]
\begin{center}
\begin{tabular}{ccc}
\hline\hline
Weight & Even & Odd\\
\hline
1 & 2 & 0 \\
2 & 3 & 1 \\
3 & 6 & 3 \\
4 & 12 & 9 \\
5 & 28 & 24 \\
6 & 69 & 65 \\
 \hline\hline
 \end{tabular}
 \caption{\label{tab:x-xbar_dimensions}Dimensions of the spaces of integrable symbols with entries drawn from the set $\{x,1-x,\bar x, 1-\bar x,x - \bar x\}$ and split according to the parity under 
exchange of $x$ and $\bar x$.}
  \end{center}
 \end{table}

Given our ansatz for the symbols of the functions we are looking for, we then match against the 
twist two asymptotics as described previously. We find a unique solution for the symbols of both 
$H^{(a)}$ and $H^{(b)}$ compatible with all asymptotic limits. Interestingly, the limit of $H_{13;24}$ 
leaves one undetermined parameter in ${\cal S}[H^{(a)}]$, which we may fix by appealing to 
another limit. 
In the resulting symbols, the letter $x - \bar x$ occurs only in the last two entries of $\mathcal{S}[H^{(a)}]$ 
while it is absent from $\mathcal{S}[H^{(b)}]$. Although we did not impose this as a constraint, ${\cal S}[H^{(a)}]$
 goes to zero when $x \rightarrow \bar x$, which is necessary since the integral cannot have a pole at $x=\bar x$.

\subsection{The analytic results for $H^{(a)}(x,\bar x)$ and $H^{(b)}(x,\bar x)$}
In this section we integrate the symbol of the Hard integral to a function, i.e. we determine the full 
answers for the \emph{functions} $H^{(a)}(x,\xbar)$ and $H^{(b)}(x,\xbar)$ that contribute to the 
Hard integral $H_{12;34}$.

In the previous section we already argued that the symbol of $H^{(b)}(x,\xbar)$ has all its entries
 drawn form the set $\{x,1-x,\xbar,1-\xbar\}$, and so it is reasonable to assume that $H^{(b)}(x,\xbar)$ 
 can be expressed in terms of SVHPLs only. We may therefore proceed  by lifting directly from the 
 asymptotic form as we did in Section~\ref{sec:analytic-result-ex} for the Easy integral.
By comparing the form of $H_{13;24}$, eq.~(\ref{eq:9}), with its asymptotic value~(\ref{eq:2}) we 
can read off the asymptotic form of $H(1/x,1/\bar x)$. Writing $\log u$ as $\log x + \log \bar x$, 
expanding out all the functions and neglecting $\log \bar x$ terms, we can the lift directly to the full 
function by simply converting HPLs to SVHPLs. In this way we arrive at
\beq\bsp\label{eq:Hb_result}
H^{(b)}(1/x,1/\bar x)&\,=  2 \mathcal{L}_{2,4}-2 \mathcal{L}_{3,3}-2
   \mathcal{L}_{1,1,4}-2 \mathcal{L}_{1,4,0}+2 \mathcal{L}_{1,4,1}-2
   \mathcal{L}_{2,3,1}+2 \mathcal{L}_{3,1,2}\\
   &\,-2 \mathcal{L}_{4,0,0}+2
   \mathcal{L}_{4,1,0}+2 \mathcal{L}_{1,1,1,3}+2 \mathcal{L}_{1,1,3,0}+2
   \mathcal{L}_{1,3,0,0}-2 \mathcal{L}_{1,3,1,1}\\
   &\,-2 \mathcal{L}_{2,1,1,2}+2
   \mathcal{L}_{2,1,2,1}+2 \mathcal{L}_{3,0,0,0}-2 \mathcal{L}_{3,1,1,0}-2
   \mathcal{L}_{1,1,1,2,1}-2 \mathcal{L}_{1,1,2,1,0}\\
   &\,+2
   \mathcal{L}_{1,1,2,1,1}-2 \mathcal{L}_{1,2,1,0,0}+2
   \mathcal{L}_{1,2,1,1,0}-2 \mathcal{L}_{2,1,0,0,0}+2
   \mathcal{L}_{2,1,1,0,0}\\
   &\,+16 \zeta_3 \mathcal{L}_3 -16 \zeta_3 \mathcal{L}_{2,1}\ .
\esp\eeq
Other orientations although still quite simple do not all share the property that they only have coefficients $\pm 2$. Using the basis of SVHPLs that makes the parity under exchange of $x$ and $\bar x$ explicit, we can write the last equation in the equivalent form
\begin{align}
H^{(b)}(x,\bar x)&\,=
16 L_{2,4}-16 L_{4,2}-8 L_{1,3,2}-8 L_{1,4,1}+8 L_{2,1,3}-8 L_{2,2,2}+8 L_{2,3,1}-8 L_{3,1,2} \nonumber \\
&\,+16 L_{3,2,0}+8 L_{3,2,1}-8 L_{4,1,1}+4 L_{1,2,2,1}-8 L_{1,3,1,1}-4 L_{2,1,1,2}+8 L_{2,1,2,1} \nonumber \\
&\,-8 L_{2,2,1,0}-4 L_{2,2,1,1}+8 L_{3,1,1,0}-4 L_{1,1,2,1,1}-24 L_{2,1,1,1,0}\,.
\end{align}

Next, we turn to the function $H^{(a)}(x,\xbar)$. As the symbol of $H^{(a)}(x,\xbar)$ contains the 
entry $x-\xbar$, it cannot be expressed through SVHPLs only. Single-valued functions whose 
symbols have entries drawn form the set $\{x,1-x,\xbar,1-\xbar,x-\xbar\}$ have been studied up to 
weight four in ref.~\cite{Chavez:2012kn}, and a basis for the corresponding space of functions was 
constructed. The resulting single-valued functions are combinations of logarithms of $x$ and $\xbar$ 
and multiple polylogarithms $G(a_1,\ldots,a_n;1)$, with $a_i\in \{0,1/x,1/\xbar\}$. Note that the 
harmonic polylogarithms form a subalgebra of this class of functions, because we have, e.g., 
\beq
G\left(0,\frac{1}{x},\frac{1}{x};1\right) = H(0,1,1;x)\,.
\eeq
This class of single-valued functions thus provides a natural extension of the SVHPLs we have 
encountered so far. In the following we show how we can integrate the symbol of $H^{(a)}(x,\xbar)$ 
in terms of these functions. The basic idea is the same as for the case of the SVHPLs: we would 
like to write down the most general linear combination of multiple polylogarithms of this type and fix 
their coefficients by matching to the symbol and the asymptotic expansion of $H^{(a)}(x,\xbar)$. 
Unlike the SVHPL case, however, some of the steps are technically more involved, and we 
therefore discuss these points in detail.

Let us denote by $\mathcal{G}$ the algebra generated by $\log x$ and $\log\xbar$ and by multiple 
polylogarithms $G(a_1,\ldots,a_n;1)$, with $a_i\in \{0,1/x,1/\xbar\}$, with coefficients that are 
polynomials in multiple zeta values. Note that without loss of generality we may assume that $a_n\neq0$. 
In the following we denote by $\mathcal{G}^\pm$ the linear subspaces of $\mathcal{G}$ of the 
functions that are respectively even and odd under an exchange of $x$ and $\xbar$. Our first goal 
will be to construct a basis for the algebra $\mathcal{G}$, as well as for its even and odd 
subspaces. As we know the generators of the algebra $\mathcal{G}$, we automatically know a 
basis for the underlying vector space for every weight. It is however often desirable to choose a 
basis that ``recycles'' as much as possible information from lower weights, i.e. we would like to 
choose a basis that explicitly includes all possible products of lower weight basis elements. Such a 
basis can always easily be constructed: indeed, a theorem by Radford~\cite{Radford} states that 
every shuffle algebra is isomorphic to the polynomial algebra constructed out of its Lyndon words. 
In our case, we immediately obtain a basis for $\mathcal{G}$ by taking products of $\log x$ and 
$\log\xbar$ and $G(a_1,\ldots,a_n;1)$, where $(a_1,\ldots,a_n)$ is a Lyndon word in the three 
letters $\{0,1/x,1/\xbar\}$. Next, we can easily construct a basis for the eigenspaces 
$\mathcal{G}^\pm$ by decomposing each (indecomposable) basis function into its even and odd 
parts. In the following we use the shorthands
\beq
G^\pm_{m_1,\ldots,m_k}(x_1,\ldots,x_k) = \frac{1}{2}G\Big(\underbrace{0,\ldots,0}_{m_1-1},\frac{1}{x_1},\ldots, \underbrace{0,\ldots,0}_{m_k-1},\frac{1}{x_k};1\Big) \pm (x\leftrightarrow \xbar)\,.
\eeq
In doing so we have seemingly doubled the number of basis functions, and so not all the 
eigenfunctions corresponding to Lyndon words can be independent. Indeed, we have for example
\beq\label{eq:G11_relation}
G_{1,1}^+(x,\xbar) = \frac{1}{2}G_{1}^+(x)^2 -\frac{1}{2}G_{1}^-(x)^2\,.
\eeq
It is easy to check this relation by computing the symbol of both sides of the equation. Similar 
relations can be obtained without much effort for higher weight functions. The resulting linearly 
independent set of functions are the desired bases for the eigenspaces. We can now immediately 
write down the most general linear combination of elements of weight six in $\mathcal{G}^+$  and 
determine the coefficients by matching to the symbol of $H^{(a)}(x,\xbar)$. As we are working with 
a basis, all the coefficients are fixed uniquely.

At this stage we have determined a function in $\mathcal{G}^+$ whose symbol matches the symbol 
of $H^{(a)}(x,\xbar)$. We have however not yet fixed the terms proportional to zeta values. We start 
by parametrising these terms by writing down all possible products of zeta values and basis 
functions in $\mathcal{G}^+$. Some of the free parameters can immediately be fixed by requiring 
the function to vanish for $x=\xbar$ and by matching to the asymptotic expansion. Note that our 
basis makes it particularly easy to compute the leading term in the limit $\xbar\to0$, because
\beq
\lim_{\xbar\to0}G^\pm_{\vec m}(\ldots,\xbar,\ldots) = 0\,.
\eeq
In other words, the small $u$ limit can easily be approached by dropping all terms which involve 
(non-trivial) basis functions that depend on $\xbar$. The remaining terms only depend on $\log\xbar
$ and harmonic polylogarithms in $x$.
However, unlike for SVHPLs, matching to the asymptotic expansions does not fix uniquely the 
terms proportional to zeta values. The reason for this is that, while in the SVHPL case we could rely 
on our knowledge of a basis for the single-valued subspace of harmonic polylogarithms, in the 
present case we have been working with a basis for the full space, and so the function we obtain 
might still contain non-trivial discontinuities. In the remainder of this section we discuss how on can 
fix this ambiguity. 

In ref.~\cite{Chavez:2012kn} a criterion was given that allows one to determine whether a given 
function is single-valued. In order to understand the criterion, let us consider the algebra $
\overline{\mathcal{G}}$ generated by multiple polylogarithms $G(a_1,\ldots,a_n;a_{n+1})$, with 
$a_i\in \{0,1/x,1/\xbar\}$ and $a_{n+1}\in\{0,1,1/x,1/\xbar\}$, with coefficients that are polynomials in 
multiple zeta values. Note that $\overline{\mathcal{G}}$ contains $\mathcal{G}$ as a subalgebra. 
The reason to consider the larger algebra $\overline{\mathcal{G}}$ is that $\overline{\mathcal{G}}$ 
carries a Hopf algebra structure\footnote{Note that we consider a slightly extended version of the 
Hopf algebra considered in ref.~\cite{Goncharov:2001} that allows us to include consistently 
multiple zeta values of even weight, see ref.~\cite{Brown:2011ik,Duhr:2012fh}.}~\cite{Goncharov:2001}, 
i.e. $\overline{\mathcal{G}}$ can be equipped with a coproduct $\Delta:
\overline{\mathcal{G}}\to\overline{\mathcal{G}}\otimes\overline{\mathcal{G}}$. Consider now the 
subspace $\mathcal{G}_{SV}$ of $\overline{\mathcal{G}}$ consisting of single-valued functions. It is 
easy to see that $\mathcal{G}_{SV}$ is a subalgebra of $\overline{\mathcal{G}}$. However, it is not 
a sub-Hopf algebra, but rather $\mathcal{G}_{SV}$ is a $\overline{\mathcal{G}}$-comodule, i.e. $
\Delta:\mathcal{G}_{SV}\to\mathcal{G}_{SV}\otimes\overline{\mathcal{G}}$. In other words, when 
acting with the coproduct on a single-valued function, the first factor in the coproduct must itself be 
single-valued. As a simple example, we have
\beq
\Delta(L_2) = \frac{1}{2}L_0\otimes\log\frac{1-x}{1-\xbar} + \frac{1}{2}L_1\otimes\log\frac{\xbar}{x}\,.
\eeq
Note that this is a natural extension of the first entry condition discussed in Section~\ref{sec:SVHPLs}.
This criterion can now be used to recursively fix the remaining ambiguities to obtain a single-valued 
function. In particular, in ref.~\cite{Chavez:2012kn} an explicit basis up to weight four was 
constructed for $\mathcal{G}_{SV}$. We extended this construction and obtained a complete basis 
at weight five, and we refer to ref.~\cite{Chavez:2012kn} about the construction of the basis. All the 
remaining ambiguities can then easily be fixed by requiring that after acting with the coproduct, the 
first factor can be decomposed into the basis of $\mathcal{G}_{SV}$ up to weight five. We then 
finally arrive at
\beq\bsp\label{eq:Ha_result}
H^{(a)}(x,\xbar) &\,= \mathcal{H}(x,\xbar) 
-\frac{28}{3} \zeta _3 L_{1,2}+164 \zeta _3 L_{2,0}+\frac{136}{3} \zeta _3 L_{2,1}-\frac{160}{3} L_3 L_{2,1}-66 L_0 L_{1,4}\\
&\,-\frac{148}{3} L_0 L_{2,3}+\frac{64}{3} L_2 L_{3,1}+\frac{52}{3} L_0 L_{3,2}+16 L_1 L_{3,2}+36 L_0 L_{4,1}+64 L_1 L_{4,1}\\
&\,+\frac{70}{3} L_0 L_{1,2,2}+24 L_0 L_{1,3,1}+\frac{26}{3} L_1 L_{1,3,1}-8 L_2 L_{2,1,1}+64 L_0 L_{2,1,2}\\
&\,-\frac{58}{3} L_0 L_{2,2,0}-4 L_0 L_{2,2,1}+\frac{50}{3} L_1 L_{2,2,1}-12 L_0 L_{3,1,0}-\frac{88}{3} L_0 L_{3,1,1}\\
&\,+18 L_1 L_{3,1,1}-\frac{32}{3} L_0 L_{1,1,2,1}-18 L_0 L_{1,2,1,1}+\frac{166}{3} L_0 L_{2,1,1,0}-8 L_0 L_{2,1,1,1}\\
&\,+328 \zeta _3 L_3+32 L_3{}^2-64 L_2 L_4
\,.
\esp\eeq
The function $\mathcal{H}(x,\bar x)$ is a single-valued combination of multiple polylogarithms that cannot be expressed through SVHPLs alone,
\begin{align}
\mathcal{H}&(x,\xbar) =-128 \Gp_{4,\bar{2}}-512 \Gp_{5,\bar{1}}-64 \Gp_{3,1,\bar{2}}+64 \Gp_{3,\bar{1},2}-64 \Gp_{3,\bar{1},\bar{2}}-128 \Gp_{3,\bar{2},\bar{1}} \nonumber \\
&\,+64 \Gp_{4,1,\bar{1}}-64 \Gp_{4,\bar{1},1}-448 \Gp_{4,\bar{1},\bar{1}}+64 \Gp_{2,\bar{1},2,1}+64 \Gp_{2,\bar{1},\bar{2},\bar{1}}+64 \Gp_{2,2,1,\bar{1}}+64 \Gp_{2,2,\bar{1},1}  \nonumber \\
&\,-64 \Gp_{2,2,\bar{1},\bar{1}}+128 \Gp_{2,\bar{2},1,1}+128 \Gp_{2,\bar{2},\bar{1},\bar{1}}+256 \Gp_{3,1,1,\bar{1}}+128 \Gp_{3,1,\bar{1},1}-128 \Gp_{3,1,\bar{1},\bar{1}} \nonumber \\
&\,+192 \Gp_{3,\bar{1},1,1}-64 \Gp_{3,\bar{1},1,\bar{1}}-64 \Gp_{3,\bar{1},\bar{1},1}+192 \Gp_{3,\bar{1},\bar{1},\bar{1}}+128 \Hp_{2,4}-128 \Hp_{4,2} \nonumber \\
&\,+\frac{640}{3} \Hp_{2,1,3}-\frac{64}{3} \Hp_{2,3,1}-\frac{256}{3} \Hp_{3,1,2}+64 \Hp_{2,1,1,2}-64 \Hp_{2,2,1,1}+64 L_0 \Gp_{3,\bar{2}} 
\end{align}
\begin{align}
&\,+256 L_0 \Gp_{4,\bar{1}}+32 L_0 \Gp_{2,1,\bar{2}}+64 L_0 \Gp_{2,2,\bar{1}}+96 L_0 \Gp_{3,1,\bar{1}}+32 L_0 \Gp_{3,\bar{1},1}+96 L_0 \Gp_{3,\bar{1},\bar{1}} \nonumber \\
&\,-64 L_0 \Gp_{2,1,1,\bar{1}}+64 L_0 \Gp_{2,\bar{1},\bar{1},\bar{1}}-32 L_1 \Gp_{3,\bar{2}}-128 L_1 \Gp_{4,\bar{1}}-16 L_1 \Gp_{2,1,\bar{2}} \nonumber \\
&\,-32 L_1 \Gp_{2,2,\bar{1}}-80 L_1 \Gp_{3,1,\bar{1}}-16 L_1 \Gp_{3,\bar{1},1}-16 L_1 \Gp_{3,\bar{1},\bar{1}}-64 L_2 \Gm_{2,\bar{1},\bar{1}}+64 L_4 \Gm_{1,\bar{1}} \nonumber \\
&\,+32 L_{2,2} \Gm_{1,\bar{1}}-\frac{32}{3} \Hp_2 \Hp_{2,2}-64 \Hp_2 \Hp_{2,1,1}-128 \Hp_2 \Hp_4-64 \Hm_1 L_0 \Gm_{2,\bar{1},\bar{1}} \nonumber \\
&\,-32 L_0^2 \Gp_{3,\bar{1}}-32 L_0^2 \Gp_{2,\bar{1},\bar{1}}+32 L_0^2 \Gp_{1,1,1,\bar{1}}+32 L_1 L_0 \Gp_{3,\bar{1}}+16 L_1 L_0 \Gp_{2,1,\bar{1}} \nonumber \\
&\,+16 L_1 L_0 \Gp_{2,\bar{1},\bar{1}}-\frac{80}{3} \Hm_1 L_0 L_{2,2}-48 \Hm_1 L_0 L_{2,1,1}+12 \Hm_1 L_1 L_{2,2}+16 L_0^2 \Hp_{2,2} \nonumber \\
&\,+32 L_0^2 \Hp_{2,1,1}-64 \Hm_1 L_4 L_0+16 \Hm_1 L_1 L_4+64 L_3 \Gp_{1,1,\bar{1}}-\frac{640}{3} \Hm_3 \Hm_{2,1}
\nonumber \\
&\,+64 (\Hm_{2,1})^2+128 (\Hm_3)^2+32 L_0 L_2 \Gm_{2,\bar{1}}-32 L_0 L_2 \Gm_{1,1,\bar{1}}-16 L_1 L_2 \Gm_{2,\bar{1}}
\nonumber \\
&\,+\frac{16}{3} L_0 L_2 \Hm_{2,1}+16 \Hm_1 L_2 L_{2,1}-\frac{112}{3} \Hp_2 L_0 L_{2,1}-8 \Hp_2 L_1 L_{2,1}-32 \Hm_3 L_0 L_2  \nonumber \\
&\,-48 \Hm_1 L_3 L_2+32 \Hp_2 L_0 L_3+16 \Hp_2 L_1 L_3+32 \Hm_1 L_0^2 \Gm_{2,\bar{1}}-16 \Hm_1 L_1 L_0 \Gm_{2,\bar{1}}\nonumber \\
&\,+\frac{16}{3} L_0^3 \Gp_{2,\bar{1}}+\frac{16}{3} L_0^3 \Gp_{1,1,\bar{1}}-8 L_1 L_0^2 \Gp_{2,\bar{1}}-8 L_1 L_0^2 \Gp_{1,1,\bar{1}}+\frac{16}{3} \Hm_1 L_0^2 \Hm_{2,1} \nonumber \\
&\,-16 (\Hm_1)^2 L_0 L_{2,1}-32 \Hm_1 \Hm_3 L_0^2+\frac{8}{3} (\Hm_1)^2 L_3 L_0-12 (\Hm_1)^2 L_1 L_3+28 \Hp_2 L_2^2
\nonumber \\
&\,+\frac{368 (\Hp_2)^3}{9}-16 L_0^2 L_2 \Gm_{1,\bar{1}}-8 L_0 L_1 L_2 \Gm_{1,\bar{1}}+\frac{56}{3} \Hm_1 \Hp_2 L_0 L_2-8 \Hm_1 \Hp_2 L_1 L_2 \nonumber \\
&\,+8 (\Hm_1)^2 L_2^2+8 (\Hp_2)^2 L_0^2+8 (\Hp_2)^2 L_0 L_1+\frac{28}{3} (\Hm_1)^2 \Hp_2 L_0^2-4 (\Hm_1)^2 \Hp_2 L_0 L_1 \nonumber \\
&\,-96 \Hm_2 (\Hm_1)^3 L_0+\frac{160}{3} (\Hm_1)^3 L_0 L_2+\frac{52}{3} \Hm_1 L_0^3 L_2+4 \Hm_1 L_0 L_1^2 L_2
\nonumber \\
&\,+4 \Hm_1 L_0^2 L_1 L_2+\Hp_2 L_0 L_1^3+\frac{2}{3} \Hp_2 L_0^2 L_1^2-8 \Hp_2 L_0^3 L_1+\frac{148}{3} (\Hm_1)^4 L_0^2 \nonumber \\
&\,+\frac{10}{3} (\Hm_1)^2 L_0^4+5 (\Hm_1)^2 L_0^2 L_1^2-\frac{10}{3} (\Hm_1)^2 L_0^3 L_1-128 \zeta _3 \Gp_{2,\bar{1}}-128 \zeta _3 \Gp_{1,1,\bar{1}} \nonumber \\
&\,+\frac{16}{3} \zeta _3 (\Hm_1)^2 L_0+24 \zeta _3 (\Hm_1)^2 L_1+\frac{64}{3} \zeta _3 \Hm_1 L_2\,, \nonumber
\end{align}
where we used the obvious shorthand
\beq\bsp
H^\pm_{\vec m} &\,\equiv \frac{1}{2}H_{\vec m}(x)\pm (x\leftrightarrow\xbar)\,.
\esp\eeq
and similarly for $G^\pm_{\vec m}$. In addition, for $G^\pm_{\vec m}$ the position of $\xbar$ is indicated by the bars in the indices, e.g.,
\beq
G^\pm_{1,\bar{2},\bar{3}}\equiv G^\pm_{1,2,3}(x,\xbar,\xbar)\,.
\eeq
Note that we have expressed $\mathcal{H}(x,\bar x)$ entirely using the basis of $\mathcal{G}^+$ 
constructed at the beginning of this section. As a consequence, all the terms are linearly 
independent and there can be no cancellations among different terms.

\subsection{Numerical consistency checks for $H$}
In the previous section we have determined the analytic result for the Hard integral. In order to 
check that our method indeed produced the correct result for the integral, we have compared our 
expression numerically against {\tt FIESTA}.
Specifically, we evaluate the conformally-invariant function $x_{13}^4 x_{24}^4\,H_{13;24}$. 
Applying a conformal transformation to send $x_4$ to infinity, the integral takes the simplified form,
\begin{equation}
\lim_{x_4\to \infty} x_{13}^4 x_{24}^4\,H_{13;24} = \frac{1}{\pi^6}\int \frac{d^4x_5 d^4x_6 d^4x_7 \; x_{13}^4 x_{57}^2}{(x_{15}^2x_{25}^2 x_{35}^2) x_{56}^2 (x_{36}^2) x_{67}^2 (x_{17}^2 x_{27}^2 x_{37}^2)}\, ,
\end{equation}
with 9 propagators. As we did for $E_{14;23}$, we use the remaining freedom to fix $x_{13}^2=1$ 
so that $u=x_{12}^2$ and $v=x_{23}^2$, and perform the numerical evaluation using the same 
setup. We compare at 40 different values, and find excellent agreement in all cases. A small sample 
of the numerical checks is shown in Table~\ref{tab:hard}.

\begin{table}[!t]
\begin{center}
\begin{tabular}{ccccc}
\hline\hline
$u$&$v$& Analytic & {\tt FIESTA} & $\delta$\\
\hline
0.1 & 0.2 & 269.239 & 269.236 & 6.4e-6 \\
 0.2 & 0.3 & 136.518 & 136.518 & 1.9e-6 \\
 0.3 & 0.1 & 204.231 & 204.230 & 1.3e-6 \\
 0.4 & 0.5 & 61.2506 & 61.2505 & 5.0e-7 \\
 0.5 & 0.6 & 46.1929 & 46.1928 & 3.5e-7 \\
 0.6 & 0.2 & 82.7081 & 82.7080 & 7.4e-7 \\
 0.7 & 0.3 & 57.5219 & 57.5219 & 4.7e-7 \\
 0.8 & 0.9 & 24.6343 & 24.6343 & 2.0e-7 \\
 0.9 & 0.5 & 34.1212 & 34.1212 & 2.6e-7\\
 \hline\hline
 \end{tabular}
 \caption{\label{tab:hard}Numerical comparison of the analytic result for $x_{13}^4 x_{24}^4\, H_{13;24}$ against {\tt FIESTA} for several values of the conformal cross ratios.}
  \end{center}
 \end{table}


\section{The analytic result for the three-loop correlator}
\label{sec:correlator}
In the previous sections we computed the Easy and Hard integrals analytically. Using 
eq.~\eqref{eq:F3}, we can therefore immediately write down the analytic answer for the three-loop 
correlator of four stress tensor multiplets. We find
\beq\bsp
x_{13}^2&\,x_{24}^2\,F_3 =\frac{6}{x-\xbar}\left[f^{(3)}(x) + f^{(3)}\left(1-\frac{1}{x}\right)+ f^{(3)}\left(\frac{1}{1-x}\right)\right]\\
&\,+\frac{2}{(x-\xbar)^2}\,f^{(1)}(x)\,\left[v\,f^{(2)}(x) + f^{(2)}\left(1-\frac{1}{x}\right)+ u\, f^{(2)}\left(\frac{1}{1-x}\right)\right]
\qquad \phantom{w}
\esp\eeq
\begin{align}
&\,+\frac{4}{x-\xbar}\,\left[\frac{1}{v-1}\,E(x) + \frac{v}{v-1}\,E\left(\frac{x}{x-1}\right)
+\frac{1}{1-u}\,E(1-x) \right. \nonumber \\
&\,\left.\,\,\qquad\qquad+ \frac{u}{1-u}\,E\left(1-\frac{1}{x}\right)
+\frac{u}{u-v}\,E\left(\frac{1}{x}\right) + \frac{v}{u-v}\,E\left(\frac{1}{1-x}\right)\right] \nonumber \\
&\,+\frac{1}{(x-\xbar)^2}\,\left[(1+v)\,H^{(a)}(x) + (1+u)\,H^{(a)}\left(1-x\right)+(u+v)\,H^{(a)}\left(\frac{1}{x}\right)\right] \nonumber \\
&\,+\frac{1}{x-\xbar}\,\left[\frac{v+1}{v-1}\,H^{(b)}(x) + \frac{1+u}{1-u}\,H^{(b)}\left(1-x\right)+\frac{u+v}{u-v}\,H^{(b)}\left(\frac{1}{x}\right)\right]\,. \nonumber
\end{align}
The pure functions appearing in the correlator are defined in 
eqs.~\eqref{eq:f_def},~\eqref{eq:E_result},~\eqref{eq:Hb_result} and~\eqref{eq:Ha_result}. For 
clarity, we suppressed the dependence of the pure functions on $\xbar$, i.e. we write $f^{(L)}(x)
\equiv f^{(L)}(x,\xbar)$ and so on. All the pure functions can be expressed in terms of SVHPLs,  
except for $H^{(a)}$ which contains functions whose symbols involve $x-\xbar$ as an entry. We 
checked that these contributions do not cancel in the sum over all contributions to the correlator.


\section{A four-loop example}
\label{sec:4loops}

In this section we will discuss a four-loop integral to illustrate how our techniques can be applied at higher 
loops.  The example we consider contributes to the four-loop four-point function of stress-tensor multiplets in ${\cal N}=4$ SYM. Specifically, we consider the Euclidean, conformal, four-loop integral,
 
\begin{equation}
I^{(4)}_{14;23} \,= \frac{1}{ \pi^8}\int   \frac{d^4 x_5d^4 x_6 d^4 x_7 d^4 x_8 x_{14}^2 x_{24}^2 x_{34}^2}{x_{15}^2 x_{18}^2 x_{25}^2 x_{26}^2 x_{37}^2 x_{38}^2 x_{45}^2 x_{46}^2 x_{47}^2 x_{48}^2 x_{56}^2 x_{67}^2 x_{78}^2 }
\, = \frac{1}{x_{13}^2 x_{24}^2} f(u,v)\,,
\label{I4}
\end{equation}
where the cross ratios $u$ and $v$ are defined by eq.~\eqref{concrrat}. As we will demonstrate in the following sections, this integral obeys a second-order differential equation whose solution is uniquely specified by imposing single-valued behaviour, similar to the generalised ladders considered in ref.~\cite{Drummond:2012bg}. 

The four-loop contribution to the stress-tensor four-point function in ${\cal N}=4$ SYM contains some integrals that do not obviously obey
any such differential equations, and with the effort presented here we also wanted to learn to what extent the two-step procedure of deriving symbols and subsequently uplifting them to functions can be repeated for those cases. Our results are encouraging: the main technical obstable is obtaining sufficient data from the asymptotic expansions;  we show that this step is indeed feasible, at least for $I^{(4)}$, and present the results in Section~\ref{sec:asyexp}. Ultimately we find it simpler to evaluate $I^{(4)}$ by solving a differential equation, and in this case the asymptotic expansions provide stringent consistency checks. 


\subsection{Asymptotic expansions}
\label{sec:asyexp}
Let us first consider the limits of the four-loop integral~\eqref{I4} and its point permutations for $x_{12}, x_{34} \rightarrow 0$. We derive expressions for its asymptotic expansion in the limit where $u \to 0, v \to 1$ similar to those for the Easy and Hard integrals obtained in Section~\ref{sec:asymptotic_expansion}. 
The logarithmic terms can be fully determined, while the non-logarithmic part of the expansion requires four-loop IBP techniques that
allow us to reach spin 15. This contains enough information to fix the $\zeta_n \log^0(u)$ terms (important for beyond-the-symbol contributions) while the purely rational part of the
asymptotic series remains partially undetermined. However, our experience with Easy and Hard
has shown that each of the three coincidence limits is (almost) sufficient to pin down the various symbols. Inverting the integrals from one orientation to another ties non-logarithmic terms in one expansion to logarithmic ones in another, so that we do in fact command over much more data than it superficially seems. It is also conceivable to take into account more than the lowest order in $u$.

We start by investigating the asymptotic expansion of the integral $I^{(4)}_{14;23}$ whose
coincidence limit $x_{12}, x_{34} \rightarrow 0$ diverges as
$\log^2u$. There are three contributing regions: while in the first two regions the original integral factors into a 
product of two two-loop integrals or a one-loop integral and a trivial three-loop integral, the third part
corresponds to the four-loop `hard' region in which the original integral is simply expanded in the small distances.
The coefficients of the logarithmically divergent terms in the asymptotic expansion, i.e. the coefficients 
of $\log^2u$ and $\log u$, can be worked out from the first two regions alone. It is easy to reach high powers 
in $x$ and we obtain a safe match onto harmonic series of the type~\eqref{eq:sample_sum} with $i>1$.
Similar to the case of the Easy and Hard integrals discussed in 
Section~\ref{sec:asymptotic_expansion}, we can sum up the harmonic sums in terms of HPLs. Note 
that the absence of harmonic sums with $i=1$ implies the absence of HPLs of the form $H_{1,\ldots}(x)$.

In the hard region, we have explicitly worked out the contribution from spin zero through eight, i.e., up
 to and including terms of $\mathcal{O}(x^8)$.  By what has been said above about the form 
 of the series, this amount of data is sufficient to pin down the terms involving zeta values, while we cannot
 hope to fix the purely rational part where the dimension of the ansatz is larger than the number of constraints we can obtain.
The linear combination displayed below was found from the limit $\bar x\to0$ of the symbol of the four-loop integral
 derived in subsequent sections. Its expansion around $x=0$ reproduces the asymptotic expansion of the integral
 up to $O(x^8)$.
 We find
\begin{align}
& x_{13}^2\,x_{24}^2\,I^{(4)}_{14;23} = \label{I1423} \\
& \frac{1}{2 \, x} \,  \log^2u \, \Bigl[  H_{2, 1, 3}  -  H_{2, 3, 1}  + 
H_{3, 1, 2}  -  H_{3, 2, 1}  + 2  H_{2, 1, 1, 2}  - 2  H_{2, 2, 1, 1}  +
\zeta_3 (6  H_{3}  + 6  H_{2, 1} ) \Bigr] \, + \nonumber \\ &
\frac{1}{x} \, \log u \, \Bigl[ -4  H_{2, 1, 4}  + 4  H_{2, 4, 1}  - 3  H_{3, 1, 3}  +  3  H_{3, 3, 1}  - 3  H_{4, 1, 2}  + 3  H_{4, 2, 1}  -  4  H_{2, 1, 1, 3}  - 4  H_{2, 1, 2, 2} \nonumber \\ 
& + 4  H_{2, 2, 2, 1}  +  4  H_{2, 3, 1, 1}  - 2  H_{3, 1, 1, 2}  + 2  H_{3, 2, 1, 1}  + 
 \zeta_3 (-18  H_{4}  - 8  H_{2, 2}  - 2  H_{3, 1}  + 8  H_{2, 1, 1})  \Bigr]  +
\nonumber \\
& \frac{1}{x} \, \Bigl[
10  H_{2, 1, 5}  + 2  H_{2, 2, 4}  - 2  H_{2, 3, 3}  -  10  H_{2, 5, 1}  + 8  H_{3, 1, 4}  - 8  H_{3, 4, 1}  +  6  H_{4, 1, 3}  - 6  H_{4, 3, 1} \nonumber 
\end{align}
\begin{align}
& + 6  H_{5, 1, 2}  -  6  H_{5, 2, 1}  + 8  H_{2, 1, 1, 4}  + 6  H_{2, 1, 2, 3}  +  8  H_{2, 1, 3, 2}  - 2  H_{2, 1, 4, 1}  + 2  H_{2, 2, 2, 2}  -  4  H_{2, 2, 3, 1}  \nonumber \\[1mm] 
& - 4  H_{2, 3, 1, 2}  - 10  H_{2, 3, 2, 1}  -  4  H_{2, 4, 1, 1}  + 4  H_{3, 1, 1, 3}  + 6  H_{3, 1, 2, 2}  -  6  H_{3, 2, 2, 1}  - 4  H_{3, 3, 1, 1}  + 4  H_{2, 1, 1, 2, 2} 
\nonumber \\[1mm] & - 4  H_{2, 1, 2, 2, 1}  - 4  H_{2, 2, 1, 1, 2} 
+ 4  H_{2, 2, 2, 1, 1}  +
\zeta_3 ( 36 H_{5}  + 8 H_{2, 3}  + 12 H_{3, 2}  - 12 H_{4, 1}  -  4 H_{2, 1, 2} \nonumber \\ & - 16 H_{2, 2, 1}  - 8 H_{3, 1, 1}) +
\zeta_5 (10 H_{3}  + 10 H_{2, 1} ) \Bigr] +\mathcal{O}(u)\,.\nonumber
\end{align}
Next we turn to the asymptotic expansion of the orientation $I^{(4)}_{12;34}$. Here
the Euclidean coincidence limit $x_{12}, x_{34} \to 0$ is finite, and thus the only region we need to
analyse is the four-loop hard region, for which we have determined the asymptotic expansion up to 
and including terms of $\mathcal{O}(x^{15})$. Just like for the non-logarithmic part in the asymptotic expansion 
of $I^{(4)}_{14;23}$, eq.~\eqref{I1423}, we have fixed the terms proportional to zeta values by matching
an ansatz in terms of HPLs onto this data, and
once again, the terms not containing zeta values are taken from the relevant limit of the symbol. We find 
\begin{align}
& x_{13}^2 \,x_{24}^2\,I^{(4)}_{12;34} = \\
& \frac{1}{x} \, \Bigl[
4  H_{1, 3, 4}  - 4  H_{1, 5, 2}  + 2  H_{1, 1, 2, 4}  -  2  H_{1, 1, 4, 2}  + 2  H_{1, 2, 1, 4}  - 2  H_{1, 2, 3, 2}  +  2  H_{1, 3, 1, 3}  + 2  H_{1, 3, 3, 1}  \nonumber \\ 
& - 2  H_{1, 4, 1, 2}  -  2  H_{1, 5, 1, 1}  +  H_{1, 1, 2, 1, 3}  +  H_{1, 1, 2, 3, 1}  -   H_{1, 1, 3, 1, 2}  -  H_{1, 1, 3, 2, 1}  +  H_{1, 2, 1, 1, 3}  +   H_{1, 2, 1, 3, 1}  \nonumber \\ 
& -  H_{1, 2, 2, 1, 2}  +  H_{1, 2, 2, 2, 1}  -  2  H_{1, 2, 3, 1, 1}  +  H_{1, 3, 1, 2, 1}  -  H_{1, 3, 2, 1, 1}  +   H_{1, 2, 1, 1, 2, 1}  -  H_{1, 2, 1, 2, 1, 1}  + \nonumber \\ 
&
\zeta_3 (8  H_{1, 1, 3}  - 8  H_{1, 2, 2}  + 4  H_{1, 1, 2, 1}  -  4  H_{1, 2, 1, 1} ) +
70 \, \zeta_7  H_{1}  \Bigr] + \mathcal{O}(u)\,. \nonumber
\end{align}
The expansion around $x=0$ of this expression reproduces the
asymptotic expansion of the integral up to $\mathcal{O}(x^{15})$.

The most complicated integrals appearing in the asymptotic expansion of $I^{(4)}_{14;23}, I^{(4)}_{12;34}$
are four-loop two-point dimensionally regularised (in position space) integrals which belong to the 
family of integrals contributing to the evaluation of the five-loop
contribution to the Konishi anomalous dimension \cite{Eden:2012fe},
\begin{align} \notag
G(a_1,\ldots,a_{14})= \int &
\frac{d^{d} x_6 d^{d} x_7 d^{d} x_8d^{d} x_9}{
 (x_{16}^2)^{a_1} (x_{17}^2)^{a_2}(x_{18}^2)^{a_3}(x_{19}^2)^{a_4}
(x_{6}^2)^{a_5} (x_{7}^2)^{a_6}(x_{8}^2)^{a_7}}
\\
& \times
\frac{1}{(x_{9}^2)^{a_8}  (x_{67}^2)^{a_9} (x_{68}^2)^{a_{10}}(x_{69}^2)^{a_{11}} (x_{78}^2)^{a_{12}}(x_{79}^2)^{a_{13}} (x_{89}^2)^{a_{14}}}\,,
\end{align}
with various integer indices $a_1,\ldots,a_{14}$ and $d=4-2\epsilon$.

The complexity of the IBP reduction to master integrals is determined, in a first approximation, by 
the number of positive indices
and the maximal deviation
from the
corner point of a sector,
which has indices equal to $0$ or $1$ for non-positive and positive indices, correspondingly.
This deviation can be characterised by the number  $\sum_{i\in \nu_{+}} (a_i-1)-\sum_{i\in \nu_{-}} a_i$
where $\nu_{\pm}$ are sets of positive (negative) indices.
So the most complicated (for an IBP reduction) integrals appearing in the contribution of spin $s$ to the asymptotic
expansion in the short-distance limit have nine positive indices and the deviation from the corner point is
equal to $2s-2$. It was possible to get results up to spin 15.
 
As in ref.~\cite{Eden:2012fe} the IBP reduction
was performed by the {\tt c++} version of the code {\tt FIRE} \cite{Smirnov:2008iw}.
The master integrals of this family either reduce, via a dual transformation, to the corresponding momentum
space master integrals \cite{Baikov:2010hf,Lee:2011jt} or can be taken from ref.~\cite{Eden:2012fe}.
To arrive at contributions corresponding to higher spin values, 
{\tt FIRE} was combined with a recently developed alternative code to solve IBP relations
{\tt LiteRed} \cite{Lee:2012cn} based on the algebraic properties of IBP relations revealed in ref.~\cite{Lee:2008tj}. (See ref.~\cite{AS_VS} where this combination was
presented within the {\tt Mathematica} version of {\tt FIRE}.)


\subsection{A differential equation}
\label{sec:diffeq}
We can use the magic identity \cite{magic} on the two-loop ladder subintegral
\be
I^{(2)}(x_1,x_2,x_4,x_7) = h_{14;27}=\frac{1}{\pi^4} \int \frac{d^4 x_5 d^4 x_{6} x_{24}^2}{x_{15}^2 x_{25}^2 x_{26}^2 x_{45}^2 x_{46}^2 x_{67}^2 x_{56}^2} \,.
\ee
The magic identity reads
\be
I^{(2)}(x_1,x_2,x_4,x_7) = I^{(2)}(x_2,x_1,x_7,x_4)\,,
\ee
and using it on the four-loop integral we find
\begin{align}
I^{(4)}_{14;23} &= \frac{1}{\pi^4} \int \frac{d^4 x_7 d^4 x_8}{x_{18}^2 x_{37}^2 x_{38}^2 x_{47}^2 x_{48}^2 x_{78}^2} I^{(2)}(x_1,x_2,x_4,x_7)\notag \\ 
&=  \frac{1}{\pi^4} \int \frac{d^4 x_7 d^4 x_8}{x_{18}^2 x_{37}^2 x_{38}^2 x_{47}^2 x_{48}^2 x_{78}^2} I^{(2)}(x_2,x_1,x_7,x_4) \notag \\
&= \frac{1}{\pi^8}\int   \frac{d^4 x_5d^4 x_6 d^4 x_7 d^4 x_8 x_{17}^2 x_{14}^2 x_{34}^2}{x_{18}^2 x_{37}^2 x_{38}^2 x_{47}^2 x_{48}^2 x_{78}^2 x_{25}^2 x_{15}^2 x_{16}^2 x_{75}^2 x_{76}^2 x_{64}^2 x_{56}^2}\,.
\label{4loopmagic}
\end{align}

The resulting integral (\ref{4loopmagic}) is `boxable', i.e. we may apply the Laplace operator at the 
point $x_2$. The only propagator which depends on $x_2$ is the one connected to the point $x_5$ and we have
\be
\Box_2 \frac{1}{x_{25}^2} = -4 \pi^2 \delta^4(x_{25})\,.
\ee
The effect of the Laplace operator is therefore to reduce the loop order by one \cite{magic}.
Thus on the full integral $I^{(4)}$ we have
\begin{align}
\Box_2 I^{(4)}_{14;23} &= - \frac{4}{\pi^6} 
\int 
\frac{d^4 x_6 d^4 x_7 d^4 x_8 x_{17}^2 x_{14}^2 x_{34}^2}{x_{18}^2 x_{37}^2 x_{38}^2 x_{47}^2 x_{48}^2 x_{78}^2 x_{12}^2 x_{16}^2 x_{72}^2 x_{76}^2 x_{64}^2 x_{26}^2} \notag \\
&= - 4 \frac{x_{14}^2}{x_{12}^2 x_{24}^2} E_{14;23}\,,
\label{4loopdiffeqbox}
\end{align}
where we have recognised the Easy integral, 
\be
E_{14;23} = \frac{1}{\pi^6} \int \frac{d^4 x_6 d^4 x_7 d^4 x_8 x_{34}^2 x_{24}^2 x_{17}^2}{x_{18}^2 x_{37}^2 x_{38}^2 x_{47}^2 x_{48}^2 x_{78}^2  x_{16}^2 x_{72}^2 x_{76}^2 x_{64}^2 x_{26}^2} = \frac{1}{x_{13}^2 x_{24}^2} f_E(u,v)\,.
\label{Easy}
\ee

The differential equation (\ref{4loopdiffeqbox}) becomes an equation for the function $f$,
\be
\Box_2 \frac{1}{x_{13}^2 x_{24}^2} f(u,v) = - 4\frac{x_{14}^2}{x_{12}^2 x_{13}^2 x_{24}^4} f_E(u,v)\,.
\ee
Applying the chain rule we obtain the following equation in terms of $u$ and $v$,
\be
\Delta^{(2)} f(u,v) = - \frac{4}{u} f_E(u,v)\,,
\ee
where
\be
\Delta^{(2)} = 4[2(\partial_u + \partial_v) + u \partial_u^2 + v \partial_v^2 - (1-u-v)\partial_u \partial_v]\,.
\ee
In terms of ($x$,$\bar x$) 
we have
\be
x \bar x \partial_x \partial_{\bar x} \hat{f}(x,\bar x) = - \hat{f}_E(x,\bar x) \,,
\ee
where
\be
\hat{f}(x,\bar x) = -(x-\bar x) f(u,v) 
\label{deffhat}
\ee
and similarly for $\hat{f}_E$. Note that $\hat{f}(x,\bar x) = -\hat{f}(\bar x, x)$.
Now we recall that the function $f_E(u,v)$ defined by eq.~(\ref{Easy}) in the orientation 
$E_{14;23}$ is of the form
\be
f_E(u,v) = \frac{1}{(x-\bar x)(1-x\bar x)}\Bigl[E(1-x,1-\bar x) + x \bar x E\Bigl(1-\frac{1}{x},1-\frac{1}{\bar x}\Bigr)\Bigr]\,.
\ee
Hence we find the following equation for $\hat f$,
\be
(1-x \bar x)x \bar x \partial_x \partial_{\bar x} \hat{f}(x,\bar x) =  -\Bigl[E(1-x,1-\bar x) + x \bar x E\Bigl(1-\frac{1}{x},1-\frac{1}{\bar x}\Bigr)\Bigr]\,.
\label{diffeq}
\ee
Without examining the equation in great detail we can immediately make the following observations about $\hat{f}$.
\begin{itemize}
\item The function $\hat f$ is a pure function of weight eight. From eq. (\ref{deffhat}) the only leading 
singularity of the four-loop integral $I^{(4)}$ is therefore of the $1/(x-\bar x)$ type, just as for the ladders.
\item The final entries of the symbol of $\hat{f}(x,\bar x)$ can be written as functions only of $x$ or 
of $\bar x$, but not both together. This follows because the right-hand side of eq.~(\ref{diffeq}) 
contains only functions of weight six, whereas there would be a contribution of weight seven if the 
final entries could not be separated into functions of $x$ or $\bar x$ separately.
\item The factor $(1-x\bar x)$ on the left-hand side implies that the next-to-final entries in the 
symbol of $\hat{f}(x,\bar x)$ contain the letter $(1-x\bar x)$.
\end{itemize}


In ref.~\cite{Drummond:2012bg}, slightly simpler, but very similar, equations were analysed for a 
class of generalised ladder integrals.
The analysis of ref.~\cite{Drummond:2012bg} can be adapted to the case of the four-loop integral 
$I^{(4)}$ and, as in ref.~\cite{Drummond:2012bg}, the solution to the equation (\ref{diffeq}) is 
uniquely determined by imposing single-valued behaviour on $\hat{f}$. 

First of all we note that any expression of the form $h(x) - h(\bar x)$ obeys the homogeneous 
equation and antisymmetry under the exchange of $x$ and $\bar x$ and hence can be added to 
any solution of eq.~(\ref{diffeq}). However, the conditions of single-valuedness,
\be
[{\rm disc}_x - {\rm disc}_{\bar x}]\hat{f}(x,\bar x) = 0\,, \qquad [{\rm disc}_{1-x} - {\rm disc}_{1-\bar x}]\hat{f}(x,\bar x)=0\,,
\label{SVconds}
\ee
and that 0 and 1 are the only singular points, fix this ambiguity. 

Let us see how the ambiguity is fixed. Imagine that we have a single-valued solution and we try to 
add $h(x) - h(\bar x)$ to it so that it remains a single-valued solution. Then the conditions 
(\ref{SVconds}) on the discontinuites tell us that $h$ can have no branch cuts at $x=0$ or $x=1$. 
Since these are the only places that the integral has any singularities, we conclude it has no branch 
cuts at all. Since the only singularities of the integral are logarithmic branch points, $h$ has no 
singularities at all and the only allowed possibility is that $h$ is constant, which drops out of the 
combination $h(x) - h(\bar x)$. Thus there is indeed a unique single-valued solution to 
eq.~(\ref{diffeq}). The argument we have just outlined is identical to the one used in ref.~
\cite{Drummond:2012bg} to solve for the generalised ladders.

A very direct way of obtaining the symbol of the single-valued solution to eq.~(\ref{diffeq}) is to 
make an ansatz of weight eight from the five letters
\be
\left\{ x,1-x,\bar x , 1-\bar x, 1 - x \bar x\right\},
\label{newletters}
\ee
and impose integrability and the initial entry condition. Then imposing that the differential equation 
is satisfied directly at symbol level leads to a unique answer.

\subsection{An integral solution}

Now let us look at the differential equation (\ref{diffeq})  in detail and construct the single-valued 
solution. It will be convenient to organise the right-hand side of the differential equation (\ref{diffeq}) 
according to symmetry under $x \leftrightarrow 1/x$. We define
\begin{align}
E_+(x,\bar x) &= \frac{1}{2}\Bigl[ E(1-x,1-\bar x) + E\Bigl(1-\frac{1}{x},1-\frac{1}{\bar x}\Bigr)\Bigr]\,, \notag \\
E_-(x,\bar x) &= \frac{1}{2}\Bigl[ E(1-x,1-\bar x) - E\Bigl(1-\frac{1}{x},1-\frac{1}{\bar x}\Bigr)\Bigr]\,.
\end{align}
Then the differential equation reads
\be
(1-x \bar x) x \bar x \partial_x \partial_{\bar x} \hat{f}(x,\bar x) = -(1-x\bar x)E_-(x,\bar x) - (1+x\bar x)E_+(x,\bar x)\,.
\label{diffeqHpm}
\ee
We may now split the equation (\ref{diffeqHpm}) into two parts
\begin{align}
\label{diffeqfa}
x\bar x \partial_x \partial_{\bar x} f_a (x,\bar x) &= - E_-(x,\bar x)\,, \\
\label{diffeqfb}
(1-x\bar x) x \bar x \partial_x \partial_{\bar x} f_b(x,\bar x) &= - (1 +x \bar x)E_+(x,\bar x)\,.
\end{align}
Note that we may take both $f_a$ and $f_b$ to be antisymmetrc under $x \leftrightarrow 1/x$. 

The equation (\ref{diffeqfa}) is of exactly the same form as the equations considered in ref.~\cite{Drummond:2012bg}.
Following the prescription given in ref.~\cite{Drummond:2012bg}, Section 6.1, it is a simple matter to find 
a single-valued solution to the equation (\ref{diffeqfa}) in terms of single-valued polylogs. We find
\begin{align}
\label{fa}
f_a(x,\bar x) 
=& 
L_{3, 4, 0} - 2 L_{4, 3, 0} + L_{5, 2, 0} + 2 L_{3, 2, 2, 0} - 
 2 L_{4, 1, 2, 0} - L_{4, 2, 0, 0} - 2 L_{4, 2, 1, 0} + 
 L_{5, 0, 0, 0} \nonumber  \\
 & + 2 L_{5, 1, 0, 0} + 2 L_{5, 1, 1, 0} + 
 2 L_{4, 1, 1, 0, 0}
 -4 \zeta_3  \bigl(\bar L_{5} - 2 \bar L_{3, 2} + 2 \bar L_{4, 0} + 3 \bar L_{4, 1} \bigr)
%
\end{align}

We now treat the equation (\ref{diffeqfb}) for $f_b$. Let us split it into two parts so that $f_b(x,\bar x) = f_1(x,\bar x) + f_2(x,\bar x)$,
\begin{align}
(1-x\bar x)x\bar x \partial_x \partial_{\bar x} f_1(x,\bar x) &= -E_+(x,\bar x)\,, \notag \\
\label{diffeqf1f2}
(1-x\bar x) \partial_x \partial_{\bar x} f_2(x,\bar x) &=- E_+(x,\bar x)\,.
\end{align}
We may write integral solutions
\be
f_1(x,\bar x)= -\int_1^x \frac{dt}{t} \int_1^{\bar x} \frac{d\bar t}{\bar t} \frac{E_+(t,\bar t)}{1-t \bar t} 
\ee
and
\be
f_2(x,\bar x) = -f_1(1/x, 1/ \bar x)=  -\int_1^x dt \int_1^{\bar x} d\bar t \frac{E_+(t,\bar t)}{1-t \bar t}
\label{f2}
\ee
which obey the equations (\ref{diffeqf1f2}).

It follows that the full function $\hat f$ is given by
\be
\hat{f}(x,\bar x) = f_a(x,\bar x)+ f_1(x,\bar x) + f_2(x,\bar x) + h(x) - h(\bar x)
\ee
for some holomorphic function $h$. We note that $\hat{f}(x,1) -f_a(x,1)= h(x) - h(1)$.

Now we examine the function $f_2$ in more detail. Writing $E_+(t,\bar t) = \sum_i H_{w_i}(t) H_{w^\prime_i}(\bar t)$ we find
\be
\label{f2}
f_2(x,\bar x) = \sum_i \int_1^x \frac{dt}{t} H_{w_i}(t) I_{w_i^\prime}(t,\bar x)
\ee
where, for a word $w$ made of the letters $0$ and $1$,
\be
I_{w}(x,\bar x) = \int_1^{\bar x} \frac{d\bar t}{\bar t - 1/x} H_w(\bar t) = (-1)^d \bigl( G(\tfrac{1}{x},w;\bar x) - G(\tfrac{1}{x},w;1)\bigr)\,.
\label{Iw}
\ee
We may now calculate the symbol of $f_2$. We note that
\be
d f_2(x,\bar x) = d \log x  \sum_i H_{w_i}(x) I_{w_i^\prime}(x,\bar x)
 - (x \leftrightarrow \bar x)\,.
\label{df2}
\ee
The symbol of $I_w$ is obtained recursively using
\be
\mathcal{S}(I_w(x,\bar x)) = \mathcal{S}(H_w(\bar x)) \otimes \frac{1-x \bar x}{1-x a_0} - (-1)^{a_0} \mathcal{S}(I_{w'}(x,\bar x)) \otimes \frac{x}{1-xa_0}\,,
\label{SIw}
\ee
where $w = a_0 w'$. When $w$ is the empty word $I(x,\bar x)$ is a logarithm,
\be
I(x,\bar x) = \log \frac{1-x \bar x}{1-x}\,.
\label{I1}
\ee
Using the relations (\ref{df2},\ref{SIw},\ref{I1}) we obtain the symbol of $f_2(x,\bar x)$. One finds 
that the result does not obey the initial entry condition (i.e. the first letters in the symbol are not only 
of the form $u=x \bar x$ or $v=(1-x)(1-\bar x)$). However, the inital entry condition can be uniquely 
restored by adding the symbols of single-variable functions in the form $\mathcal{S}(h_2(x)) - \mathcal{S}(h_2(\bar x))$. 
Inverting $x \leftrightarrow 1/x$ we may similarly treat $f_1(x,\bar x) = -f_2(1/x , 1/\bar x)$. 
Combining everything we obtain the symbol
\be
\mathcal{S}(\hat{f}(x,\bar x)) = \mathcal{S}(f_a(x,\bar x) + f_2(x,\bar x) + h_2(x) - h_2(\bar x) - f_2(1/x,1/\bar x) - h_2(1/x)+ h_2(1/\bar x))\,.
\ee
The symbol obtained this way agrees with that obtained by imposing the differential equation on an 
ansatz as described around eq. (\ref{newletters}).

Given that single-valuedness uniquely determines the solution of the differential equation 
(\ref{diffeq}) one might suspect that we can use this property to give an explicit representation of the 
function $h_2(x)$. Indeed this is the case. The integral formula (\ref{f2}) can, in principle, have 
discontinuities around any of the five divisors obtained by setting a letter from the set 
(\ref{newletters}) to zero.

Let us consider the discontinuity of $f_2(x,\bar x)$ at $ x =1/\bar x $,
\begin{align}
{\rm disc}_{x=1/\bar x} f_2(x,\bar x) &= \int_{1/\bar x}^x dt\,\,{\rm disc}_{t=1/\bar x} \int_1^{\bar x} d \bar t \frac{E_+(t,\bar t)}{1-t\bar t} \notag \\
&= \int_{1/\bar x}^x dt \,\,{\rm disc}_{\bar x =1/t} \int_1^{\bar x} d \bar t \frac{E_+(t,\bar t)}{1-t\bar t} \notag \\
&=- \int_{1/\bar x}^x \frac{dt}{t} (2 \pi i) E_+(t,1/t)\,.
\end{align}
The above expression vanishes due to the symmetry of $E_+$ under $x \leftrightarrow 1/x$ and the 
antisymmetry under $x \leftrightarrow \bar x$. The absence of such discontinuities is the reason 
that we split the original equation into two pieces, one for $f_a$ and one for $f_b$.

Now let us consider the discontinuity around $x=1$. We find
\begin{align}
{\rm disc}_{1-x}f_2(x,\bar x) &=- \int_1^x \frac{dt}{t} {\rm disc}_{1-t} \int_1^{\bar x} \frac{d\bar t}{\bar t - 1/t} E_+(t,\bar t) \notag \\
&= -(2 \pi i)\int_1^x \frac{dt}{t} E_+(t,1/t) + \int_1^x dt \int_1^{\bar x} d \bar t \, \frac{{\rm disc}_{1-t} E_+(t,\bar t)}{1-t\bar t} \,.
\end{align}
The first term above again vanishes due to the symmetries of $E_+$. The second term will cancel 
against the corresponding term involving ${\rm disc}_{1-\bar t} E_+(t,\bar t)$ in the integrand when 
we take the combination $[{\rm disc}_{1-x} - {\rm disc}_{1-\bar x}]f_2(x,\bar x)$.
The discontinuities at $x=1$ and $\bar x =1$ of $f_2$ therefore satisfy the single-valuedness 
conditions (\ref{SVconds}) since $E_+$ does. 

For the discontinuities at $x=0$  we find
\be
{\rm disc}_x f_2(x,\bar x) = \int_0^x dt\,\,{\rm disc}_t \int_1^{\bar x} d\bar t \frac{E_+(t,\bar t)}{1-t \bar t} = \int_0^x dt \int_1^{\bar x} \frac{d \bar t }{1-t\bar t}\,\,{\rm disc}_t E_+(t,\bar t)\,.
\ee
Now writing the $\bar t$ integral above as $\int_1^{\bar x} = \int_0^{\bar x} - \int_0^1$ and using $[{\rm disc}_t - {\rm disc}_{\bar t}]E_+(t,\bar t) =0$ we find
\begin{align}
[{\rm disc}_x - {\rm disc}_{\bar x}] f_2(x,\bar x) &= \biggl[-\int_0^x dt \int_0^1 \frac{d \bar t }{1-t\bar t}\,\,{\rm disc}_t E_+(t,\bar t)\biggr] + (x \leftrightarrow \bar x)\notag \\
&= {\rm disc}_x \biggl[-\int_0^x dt \int_0^1 \frac{d \bar t }{1-t\bar t} E_+(t,\bar t)\biggr] + (x \leftrightarrow \bar x)
\label{discx}
\end{align}
Thus $f_2(x,\bar x)$ is not single-valued by itself since the above combination of discontinuities 
(\ref{discx}) does not vanish. Note however that eq.~(\ref{discx}) is of the form $k(x) + k(\bar x)$, as 
necessary in order for it to be cancelled by adding a term of the form $h_2(x) - h_2(\bar x)$ to $f_2(x,\bar x)$.
We now construct such a function $h_2(x)$. 

Let
\be
h_2^0(x) = \int_0^x dt \int_0^1 \frac{d \bar t }{1-t\bar t} E_+(t,\bar t)\,.
\ee
Writing $E_+(t,\bar t) = \sum_i H_{w_i}(t) H_{w^\prime_i}(\bar t)$ we find
\be
h^0_2(x) = - \int_0^x \frac{dt}{t} \sum_i   H(w_i ; t) \int_0^1 \frac{d \bar t }{\bar t-1/t}\,\,H(w^\prime_i;\bar t)\,.
\ee
We can write
\be
 \int_0^1 \frac{d \bar t }{\bar t-1/t}\,\,H_{w^\prime_i}(\bar t) = (-1)^d G(1/t, w_i^\prime(0,1) ;1)\,,
 \ee
 where we have made explicit that $w_i^\prime$ is a word in the letters 0 and 1 and $d$ is the 
 number of 1 letters. 
 One can always rewrite this in terms of HPLs at argument $t$. Indeed we can recursively apply the 
 formula
\begin{align}
\label{recur1}
G(\tfrac{1}{t},a_2,a_3\ldots,a_n;1) = \int_0^t \frac{dr}{r-1} G(\tfrac{a_2}{r}, a_3,\ldots,a_n;1) - \int_0^t \frac{dr}{r} G(\tfrac{1}{r},a_3,\ldots,a_n;1)\,.
\end{align}
to achieve this. Note that $a_i \in \{ 0,1\}$ in the above formula. We also need
\be
G(\tfrac{1}{t},0_q;1) = (-1)^{q+1} H_{q+1}(t)\,.
\ee
Once this has been done, one can use standard HPL relations to calculate the products 
\be
H_{w_i}(t) G(1/t ,w^\prime_i ;1)
\ee
and perform the remaining integral from $0$ to $x$ w.r.t. $t$. We thus obtain a function $h^0_2$ 
whose discontinuity at $x=0$ is minus that of the $x$-dependent contribution to $[{\rm disc}_x - {\rm 
disc}_{\bar x}] f_2(x,\bar x)$.

In ref.~\cite{Drummond:2012bg} an explicit projection operator $\mathcal{F}$ was introduced which 
removes the discontinuity at $x=0$ of a linear combination of HPLs while preserving the 
discontinuity at $x=1$. The orthogonal projector $(1-\mathcal{F})$ removes the discontinuity at 
$x=1$ while preserving that at $x=0$. We define
\be
h_2(x) = (1 - \mathcal{F}) h_2^0(x)\,.
\ee
Explicitly we find
\begin{align}
h_2(x) &=\frac{151}{16} \zeta_{6} H_{2}+\frac{15}{2} \zeta_{3}^2 H_{2} -\frac{3}{2} \zeta_{2}  \zeta_{3} H_{2,0} -\frac{15}{4} \zeta_{5} H_{2,0} -\zeta_{2}  \zeta_{3} H_{2,1}+\frac{19}{4} \zeta_{4} H_{2,2} +2 \zeta_{3} H_{2,3} \notag \\
&-\zeta_{2}  H_{2,4} +\frac{21}{8} \zeta_{4} H_{2,0,0} +\frac{19}{4} \zeta_{4} H_{2,1,0} +\frac{17}{2} \zeta_{4} H_{2,1,1}+5 \zeta_{3} H_{2,1,2} -\zeta_{2}  H_{2,1,3}-\frac{3}{2} \zeta_{3} H_{2,2,0} \notag \\
&+\zeta_{3} H_{2,2,1}-\zeta_{2}  H_{2,2,2} +\frac{1}{2} \zeta_{2}  H_{2,3,0} -\zeta_{2}  H_{2,3,1}-\frac{3}{2} \zeta_{3} H_{2,0,0,0} -3 \zeta_{3} H_{2,1,0,0} -3 \zeta_{3} H_{2,1,1,0} \notag \\
 &+4 \zeta_{3} H_{2,1,1,1}+\zeta_{2}  H_{2,1,2,0} -2 \zeta_{2}  H_{2,1,2,1}+\frac{1}{2} H_{2,1,4,0} +\zeta_{2}  H_{2,2,1,0} +\frac{1}{2} H_{2,3,2,0} -\frac{1}{2} H_{2,4,0,0} \notag \\
&-H_{2,4,1,0} +\frac{1}{2} \zeta_{2}  H_{2,1,0,0,0} +\zeta_{2}  H_{2,1,1,0,0} +2 \zeta_{2}  H_{2,1,1,1,0} +H_{2,1,1,3,0} +H_{2,1,2,2,0} -H_{2,1,3,0,0}  \notag\\
&+2 H_{2,1,1,1,2,0} -H_{2,1,3,1,0} +H_{2,2,1,2,0} -\frac{1}{2} H_{2,2,2,0,0} -H_{2,2,2,1,0} +\frac{1}{2} H_{2,3,0,0,0} -H_{2,3,1,1,0} \notag \\
& -H_{2,1,1,2,0,0} +\frac{1}{2} H_{2,1,2,0,0,0} -H_{2,1,2,1,0,0} -2 H_{2,1,2,1,1,0} +H_{2,2,1,0,0,0} +H_{2,2,1,1,0,0} \notag \\
&+H_{2,1,1,1,0,0,0} \,.
\label{h2}
\end{align} 
Here the $H$ functions are all implicitly evaluated at argument $x$.

The contribution from $f_1(x,\bar x) = - f_2(1/x , 1/\bar x)$ is made single-valued by inversion on $x$. So we define
\be
h(x) = h_2(x) - h_2(1/x)\,.
\label{hfull}
\ee
Finally we deduce that the combination
\be
\hat{f}(x,\bar x) = f_a(x,\bar x) + f_1(x,\bar x) + f_2(x,\bar x) + h(x) - h(\bar x)
\label{fhatfinal}
\ee
is single-valued and obeys the differential equation (\ref{diffeq}) and hence describes the four-loop integral $I^{(4)}$ defined in equation (\ref{I4}). The equations (\ref{f2}), (\ref{h2}), (\ref{hfull}) and (\ref{fhatfinal}) therefore explicitly define the function $\hat {f}$.

\subsection{Expression in terms of multiple polylogarithms}

Now let us rewrite the integral form (\ref{f2}), (\ref{Iw}) for $f_2(x,\bar x)$ in terms of multiple polylogarithms. We use the following generalisation of relation (\ref{recur1}),
\begin{align}
\label{recur2}
G(\tfrac{1}{y},a_2,\ldots,a_n;z) =& \bigl(G(\tfrac{1}{z};y) - G(\tfrac{1}{a_2};y)\bigr)G(a_2,\ldots,a_n;z)\notag \\
& + \int_0^y \biggl(\frac{dt}{t-\tfrac{1}{a_2}} - \frac{dt}{t}\biggr) G(\tfrac{1}{t},a_3,\ldots,a_n;z)\,
\end{align}
to recursively rewrite the $G(\tfrac{1}{t},\ldots)$ appearing in the $I_{w_i^\prime}(t,\bar x)$ in eq.~(\ref{f2}) so that the $t$ appears as the final argument.
Note that in eq.~(\ref{recur2}), the two terms involving an explicit appearance of $1/a_2$ vanish in the case $a_2=0$.
The recursion begins with
\be
G(\tfrac{1}{y};z) = \log (1-yz) = G(\tfrac{1}{z};y)\,.
\ee
The recursion allows us to write the products $H_{w_i}(t) I_{w_i^\prime}(t,\bar x)$ as a sum of 
multiple polylogarithms of the form $G(w;t)$ where the weight vectors depend on $\bar x$. Then we can 
perform the final integration $dt/t$ to obtain an expression for $f_2$ in terms of multiple polylogs.

We may relate $f_1(x,\bar x)$ directly to $f_2(x,\bar x)$ since
\begin{align}
f_1(x,\bar x) &=  -\int_1^x \frac{dt}{t} \int_1^{\bar x} \frac{d\bar t}{\bar t} \frac{E_+(t,\bar t)}{1-t \bar t} = \int_1^x \frac{dt}{t^2} \int_1^{\bar x} \frac{d \bar t}{\bar t (\bar t - 1/t)} E_+(t,\bar t)\, \notag \\
&= \int_1^x \frac{dt}{t} \biggl[\int_1^{\bar x} \frac{d\bar t}{\bar t - 1/t} E_+(t,\bar t) - \int_1^{\bar x} \frac{d \bar t}{\bar t} E_+(t,\bar t)\biggr]\notag \\
&= f_2(x,\bar x) -  \sum_i [H_{0w_i}(x) - H_{0w_i}(1)][H_{0w_i^\prime}(\bar x) - H_{0w_i^\prime}(1)]\,.
\end{align}
For a practical scheme we express $E_+$ as a sum over $H_{w_i}(t) H_{w_i^\prime}(\bar t)$ and 
do the $\bar t$ integration. In any single term of the integrand of $f_2$, the recursion (\ref{recur2}) 
will lead to multiple polylogarithms of the type $G(\ldots,1/\bar x;t)$. Next, we take the 
shuffle product with the second polylogarithm and integrate over $t$. 

In this raw form our result is not manifestly antisymmetric under $x \leftrightarrow \bar x$.  Remarkably, in the sum over all terms only $G(0,1/\bar x, \ldots;x)$ remain. Upon rewriting
\begin{equation}
G\left(0,\tfrac{1}{\bar x},a_1,\ldots,a_n;x\right) \, = \, G(0;x) G\left(\tfrac{1}{\bar x},a_1,\ldots,a_n;x\right) - \int_0^x
\frac{dt}{t - \frac{1}{\bar x}} G(0;t) G(a_1,\ldots,a_n;t)
\end{equation}
we can use (\ref{recur2}) to swap $G(1/\bar x,\ldots;x)$ for  (a sum over) $G(\ldots,1/x;\bar x)$. Replacing the
original two-variable polylogarithms by 1/2 themselves and 1/2 the $x, \bar x$ swapped version, we can obtain a manifestly antisymmetric form. The shuffle algebra is needed to remove zeroes from the rightmost position of the weight vectors and to
bring the letters $1/x,1/\bar x$ to the left of all entries $1$. Finally we rescale to argument 1. 

In analogy to the notation introduced for the Hard integral let us write
\begin{equation}
G_{\hat 3,2,1} \, = \, G\left(0,0,\tfrac{1}{x \bar x},0,\tfrac{1}{x},\tfrac{1}{x};1\right)
\end{equation}
etc. 
Collecting terms, we find
\begin{align}
& I^{(4)}_{14;23}(x,\bar x) \, = \\
& -L_{2, 2, 4} + 2 L_{2, 3, 3} - L_{2, 4, 2} - 
 2 L_{2, 1, 1, 4} + 2 L_{2, 1, 2, 3} - 2 L_{2, 1, 3, 2} + 
 2 L_{2, 1, 4, 1} + 2 L_{2, 2, 1, 3}  - 2 L_{2, 2, 2, 2} \nonumber \\ &- 
 2 L_{2, 2, 3, 1} + 2 L_{2, 3, 1, 2} + 2 L_{2, 3, 2, 0} + 
 2 L_{2, 3, 2, 1} - 2 L_{2, 4, 1, 0} - 2 L_{2, 4, 1, 1} - 
 2 L_{3, 1, 3, 0}  + 2 L_{3, 3, 1, 0} \nonumber \\ & - 4 L_{2, 1, 1, 2, 2} + 
 4 L_{2, 1, 2, 2, 1} + 4 L_{2, 2, 1, 1, 2} - 
 4 L_{2, 2, 2, 1, 0} - 4 L_{2, 2, 2, 1, 1} - 
 4 L_{3, 1, 1, 2, 0} + 2 L_{3, 2, 1, 0, 0} \nonumber
 \end{align}
 \begin{align}
 & + 4 L_{3, 2, 1, 1, 0} + L_{0} \, \bigl(-H^-_{1, 2, 4} + 2 H^-_{1, 3, 3}
- H^-_{1, 4, 2} - 2 H^-_{1, 1, 1, 4} + 2 H^-_{1, 1, 2, 3} - 2 H^-_{1, 1, 3, 2} + 
   2 H^-_{1, 1, 4, 1} \nonumber \\ & + 2 H^-_{1, 2, 1, 3} - 2 H^-_{1, 2, 2, 2} - 
   2 H^-_{1, 2, 3, 1} + 2 H^-_{1, 3, 1, 2} + 2 H^-_{1, 3, 2, 1} - 
   2 H^-_{1, 4, 1, 1} - 4 H^-_{1, 1, 1, 2, 2} + 
   4 H^-_{1, 1, 2, 2, 1} \nonumber \\ & + 4 H^-_{1, 2, 1, 1, 2} - 
   4 H^-_{1, 2, 2, 1, 1}\bigr) \, +
4 H^-_{1, 2, 5} - 4 H^-_{1, 3, 4} - 4 H^-_{1, 4, 3} + 
 4 H^-_{1, 5, 2} + 8 H^-_{1, 1, 1, 5} - 4 H^-_{1, 1, 2, 4} \nonumber \\ & + 
 4 H^-_{1, 1, 4, 2} - 8 H^-_{1, 1, 5, 1} - 4 H^-_{1, 2, 1, 4} + 
 8 H^-_{1, 2, 3, 2} + 4 H^-_{1, 2, 4, 1} - 8 H^-_{1, 3, 1, 3} - 
 4 H^-_{1, 4, 1, 2} - 4 H^-_{1, 4, 2, 1} \nonumber \\  & + 8 H^-_{1, 5, 1, 1} + 
 8 H^-_{1, 1, 1, 2, 3} + 8 H^-_{1, 1, 1, 3, 2} - 
 8 H^-_{1, 1, 2, 3, 1} - 8 H^-_{1, 1, 3, 2, 1} - 
 8 H^-_{1, 2, 1, 1, 3} + 8 H^-_{1, 2, 3, 1, 1} \nonumber \\ &
- 8 H^-_{1, 3, 1, 1, 2} + 8 H^-_{1, 3, 2, 1, 1} +
\zeta_3 \, \bigl(8 \bar L_{2, 3} - 12 \bar L_{3, 2} - 12 \bar L_{2, 1, 2} + 
   12 \bar L_{2, 2, 1} - 12 \bar L_{3, 1, 0} - 16 \bar L_{3, 1, 1} \nonumber \\ &
- 12 L_{0} H^-_{1, 1, 2} + 12 L_{0} H^-_{1, 2, 1} +
   16 H^-_{1, 1, 3} - 16 H^-_{1, 3, 1} - 
   16 H^-_{1, 1, 1, 2} + 16 H^-_{1, 2, 1, 1}\bigr) \, +
2 \bar L_{3} \zeta_5 \nonumber \\ &
+ \zeta_3^2 \, \bigl(-24 L_{2} - 72 H^-_{2} - 48 H^-_{1, 1}\bigr) \, + \nonumber \\[2mm]
& G^+_{\hat 2} \, \bigl(-L_{4, 2} - L_{4, 0, 0} - 2 L_{4, 1, 0} - 
    2 L_{4, 1, 1} - 4 \bar L_{2, 1} \zeta_3 + 
    4 L_{0} \zeta_3 H^-_{2} + 4 L_{1} \zeta_3 H^-_{2} \nonumber \\ & + 
    3 \bar L_{2} H^-_{4} + 2 \bar L_{1, 1} H^-_{4} - 4 L_{1} H^-_{5} + 
    4 \zeta_3 H^-_{1, 2} - 2 \bar L_{0, 0} H^-_{1, 3} + 
    5 L_{0} H^-_{1, 4} - 2 L_{1} H^-_{1, 4} - 12 H^-_{1, 5} \nonumber \\ & + 
    12 \zeta_3 H^-_{2, 1} - \bar L_{0, 0} H^-_{2, 2} + 
    2 L_{0} H^-_{2, 3} - 2 L_{1} H^-_{2, 3} - 6 H^-_{2, 4} + 
    L_{0} H^-_{3, 2} - 2 L_{1} H^-_{3, 2} - 
    4 L_{0} H^-_{4, 1} \nonumber \\ & - 2 L_{1} H^-_{4, 1} + 2 H^-_{4, 2} + 
    16 H^-_{5, 1} + 16 \zeta_3 H^-_{1, 1, 1} - 
    2 \bar L_{0, 0} H^-_{1, 1, 2} + 6 L_{0} H^-_{1, 1, 3} - 
    12 H^-_{1, 1, 4} - 2 \bar L_{0, 0} H^-_{1, 2, 1} \nonumber \\  & + 
    6 L_{0} H^-_{1, 2, 2} - 8 H^-_{1, 2, 3} + 
    2 L_{0} H^-_{1, 3, 1} - 8 H^-_{1, 3, 2} + 4 H^-_{1, 4, 1} + 
    2 \bar L_{0, 0} H^-_{2, 1, 1} + 2 L_{0} H^-_{2, 1, 2} - 
    4 H^-_{2, 1, 3} \nonumber \\  & - 2 L_{0} H^-_{2, 2, 1} - 4 H^-_{2, 2, 2} + 
    4 H^-_{2, 3, 1} - 6 L_{0} H^-_{3, 1, 1} + 4 H^-_{3, 1, 2} + 
    12 H^-_{3, 2, 1} + 12 H^-_{4, 1, 1} \nonumber \\  & + 
    4 L_{0} H^-_{1, 1, 1, 2} - 8 H^-_{1, 1, 1, 3} - 
    8 H^-_{1, 1, 2, 2} - 4 L_{0} H^-_{1, 2, 1, 1} + 
    8 H^-_{1, 2, 2, 1} + 8 H^-_{1, 3, 1, 1} - L_{2} H^+_{4}\bigr) \, +
\nonumber \\[1mm]  &
 G^+_{\hat 3} \, \bigl(2 \bar L_{3, 2} + 4 \bar L_{3, 1, 0} + 4 \bar L_{3, 1, 1} - 
    8 \zeta_3 H^-_{2} - 6 \bar L_{2} H^-_{3} - 2 \bar L_{0, 0} H^-_{3} - 
    4 \bar L_{1, 1} H^-_{3} + 8 L_{0} H^-_{4} \nonumber \\  & + 6 L_{1} H^-_{4} - 
    20 H^-_{5} - 8 \zeta_3 H^-_{1, 1} + 8 L_{0} H^-_{1, 3} + 
    4 L_{1} H^-_{1, 3} - 20 H^-_{1, 4} + 6 L_{0} H^-_{2, 2} + 
    4 L_{1} H^-_{2, 2} \nonumber \\  & - 16 H^-_{2, 3} + 4 L_{0} H^-_{3, 1} + 
    4 L_{1} H^-_{3, 1} - 16 H^-_{3, 2} - 8 H^-_{4, 1} + 
    4 L_{0} H^-_{1, 1, 2} - 16 H^-_{1, 1, 3} + 
    4 L_{0} H^-_{1, 2, 1} \nonumber \\  & - 16 H^-_{1, 2, 2} - 8 H^-_{1, 3, 1} - 
    4 L_{0} H^-_{2, 1, 1} - 8 H^-_{2, 1, 2} + 8 H^-_{3, 1, 1} - 
    8 H^-_{1, 1, 1, 2} + 8 H^-_{1, 2, 1, 1} + 2 L_{2} H^+_{3}\bigr) \, +
\nonumber \\  &
 G^+_{\hat 2, 1} \, \bigl(2 \bar L_{3, 2} - \bar L_{4, 0} - 2 \bar L_{4, 1} + 
    2 \bar L_{3, 1, 0} - 12 L_{2} \zeta_3 + 16 \zeta_3 H^-_{2} - 
    4 \bar L_{2} H^-_{3} \nonumber \\  & - 2 \bar L_{0, 0} H^-_{3} + 8 L_{0} H^-_{4} + 
    8 L_{1} H^-_{4} - 20 H^-_{5} + 16 \zeta_3 H^-_{1, 1} + 
    4 L_{0} H^-_{1, 3} - 16 H^-_{1, 4} + 4 L_{0} H^-_{2, 2} \nonumber \\  & - 
    12 H^-_{2, 3} - 12 H^-_{3, 2} + 4 L_{0} H^-_{1, 1, 2} - 
    8 H^-_{1, 1, 3} - 8 H^-_{1, 2, 2} - 4 L_{0} H^-_{2, 1, 1} + 
    8 H^-_{2, 2, 1} + 8 H^-_{3, 1, 1}\bigr) \, + \nonumber \\[1mm]  &
 G^+_{\hat 4} \, \bigl(3 L_{0} H^-_{3} - 12 H^-_{4} + 3 L_{0} H^-_{1, 2} - 
    12 H^-_{1, 3} + 6 L_{0} H^-_{2, 1} - 12 H^-_{2, 2} - 
    12 H^-_{3, 1} + 6 L_{0} H^-_{1, 1, 1} \nonumber \\  & - 12 H^-_{1, 1, 2} - 
    12 H^-_{1, 2, 1}\bigr) \, +
 G^+_{\hat 3, 1} \, \bigl(2 L_{3, 0} + 4 L_{3, 1} - 8 \zeta_3 H^-_{1} + 
    2 L_{0} H^-_{3} - 4 L_{1} H^-_{3} - 8 H^-_{4} \nonumber \\  & + 
    4 L_{0} H^-_{1, 2} - 8 H^-_{1, 3} + 4 L_{0} H^-_{2, 1} - 
    8 H^-_{2, 2} - 8 H^-_{1, 1, 2} + 8 H^-_{2, 1, 1}\bigr) \, +
 G^+_{\hat 2, 2} \, \bigl(L_{4} + L_{3, 0} + 4 L_{3, 1} \nonumber \\  & - 
    16 \zeta_3 H^-_{1} - 4 L_{1} H^-_{3} + 4 H^-_{1, 3} + 
    4 H^-_{2, 2} + 8 H^-_{3, 1} - 4 L_{0} H^-_{1, 1, 1} + 
    8 H^-_{1, 2, 1} + 8 H^-_{2, 1, 1}\bigr) \, + \nonumber \\  &
 G^+_{\hat 2, 1, 1} \, \bigl(-2 L_{4} + 2 L_{3, 0} + 16 \zeta_3 H^-_{1} + 
    4 L_{0} H^-_{1, 2} - 8 H^-_{1, 3} - 8 H^-_{2, 2}\bigr) \, + \nonumber \\[1mm] &
 G^+_{\hat 5} \, \bigl(-4 H^-_{3} - 4 H^-_{1, 2} - 8 H^-_{2, 1} - 
    8 H^-_{1, 1, 1}\bigr) \, +
 G^+_{\hat 4, 1} \, \bigl(-6 \bar L_{2, 1} + 3 L_{0} H^-_{2} + 6 L_{1} H^-_{2} - 
    12 H^-_{3} \nonumber \\  & - 12 H^-_{1, 2} - 12 H^-_{2, 1}\bigr) \, +
 G^+_{\hat 3, 2} \, \bigl(-2 \bar L_{3} - 8 \bar L_{2, 1} + 2 L_{0} H^-_{2} + 
    8 L_{1} H^-_{2} - 4 H^-_{3} - 8 H^-_{1, 2} \nonumber \\  & - 8 H^-_{2, 1} + 
    8 H^-_{1, 1, 1}\bigr) \, +
 G^+_{\hat 3, 1, 1} \, \bigl(4 \bar L_{3} + 4 L_{0} H^-_{2} - 16 H^-_{3} - 
    8 H^-_{1, 2}\bigr) \, +
 G^+_{\hat 2, 3} \, \bigl(-4 \bar L_{3} - 8 \bar L_{2, 1} \nonumber \\  &
+ 2 L_{0} H^-_{2} + 8 L_{1} H^-_{2} - 4 H^-_{1, 2} - 8 H^-_{2, 1} + 
    8 H^-_{1, 1, 1}\bigr) \, +
 G^+_{\hat 2, 2, 1} \, \bigl(4 \bar L_{3} - 8 H^-_{3} - 8 H^-_{1, 2}\bigr) \, +
\nonumber \\  &
 G^+_{\hat 2, 1, 2} \, \bigl(4 \bar L_{3} + 4 \bar L_{2, 1} - 4 L_{1} H^-_{2} - 
    4 H^-_{3} + 8 H^-_{2, 1}\bigr) \, +
 G^+_{\hat 2, 1, 1, 1} \, \bigl(4 L_{0} H^-_{2} - 8 H^-_{3}\bigr) \, + \nonumber 
\end{align}
\begin{align}
%
& \bigl(10 G^+_{\hat 2, 4} + 8 G^+_{\hat 3, 3} + 3 G^+_{\hat 4, 2} - 8 G^+_{\hat 2, 1, 3} - 
   8 G^+_{\hat 2, 2, 2} - 4 G^+_{\hat 2, 3, 1} - 8 G^+_{\hat 3, 1, 2} - 
   8 G^+_{\hat 3, 2, 1} - 6 G^+_{\hat 4, 1, 1} \nonumber \\  & + 4 G^+_{\hat 2, 1, 1, 2} - 4 G^+_{\hat 2, 2, 1, 1}\bigr) \, L_{2} +
\bigl(-16 G^+_{\hat 2, 4} - 12 G^+_{\hat 3, 3} - 6 G^+_{\hat 4, 2} - 4 G^+_{\hat 5, 1} + 
   12 G^+_{\hat 2, 1, 3} \nonumber \\  & + 12 G^+_{\hat 2, 2, 2} + 8 G^+_{\hat 2, 3, 1} + 
   8 G^+_{\hat 3, 1, 2} + 8 G^+_{\hat 3, 2, 1} - 8 G^+_{\hat 2, 1, 1, 2} - 
   8 G^+_{\hat 2, 1, 2, 1} - 8 G^+_{\hat 3, 1, 1, 1}\bigr) \, H^-_{2} + \nonumber \\  &
\bigl(-16 G^+_{\hat 2, 4} - 16 G^+_{\hat 3, 3} - 12 G^+_{\hat 4, 2} - 8 G^+_{\hat 5, 1} + 
   8 G^+_{\hat 2, 1, 3} + 8 G^+_{\hat 2, 2, 2} + 8 G^+_{\hat 3, 1, 2}\bigr) \, H^-_{1, 1} + \nonumber \\[1mm]  &
\bigl(20 G^+_{\hat 2, 5} + 20 G^+_{\hat 3, 4} + 12 G^+_{\hat 4, 3} + 4 G^+_{\hat 5, 2} - 
   16 G^+_{\hat 2, 1, 4} - 12 G^+_{\hat 2, 2, 3} - 12 G^+_{\hat 2, 3, 2} - 
   16 G^+_{\hat 3, 1, 3} \nonumber \\  & - 16 G^+_{\hat 3, 2, 2} - 8 G^+_{\hat 3, 3, 1} - 
   12 G^+_{\hat 4, 1, 2} - 12 G^+_{\hat 4, 2, 1} - 8 G^+_{\hat 5, 1, 1} + 
   8 G^+_{\hat 2, 1, 1, 3} + 8 G^+_{\hat 2, 1, 2, 2} - 8 G^+_{\hat 2, 2, 2, 1} \nonumber \\  & - 
   8 G^+_{\hat 2, 3, 1, 1} + 8 G^+_{\hat 3, 1, 1, 2} - 
   8 G^+_{\hat 3, 2, 1, 1}\bigr) \, H^-_{1} + \nonumber \\[2 mm]
& G^-_{\hat 2, 1} \, \bigl(2 L_{3, 2} - L_{4, 0} - 2 L_{4, 1} + 
   L_{3, 0, 0} + 2 L_{3, 1, 0} - 12 \bar L_{2} \zeta_3 + 
   8 \zeta_3 H^+_{2} - 4 \bar L_{2} H^+_{3} \nonumber \\  & + 8 L_{1} H^+_{4} - 
   16 \zeta_3 H^+_{1, 1} - 4 H^+_{2, 3} + 4 L_{0} H^+_{3, 1} - 
   4 H^+_{3, 2} - 16 H^+_{4, 1} - 4 L_{0} H^+_{1, 1, 2} + 
   8 H^+_{1, 1, 3} \nonumber \\  & + 8 H^+_{1, 2, 2} + 4 L_{0} H^+_{2, 1, 1} - 
   8 H^+_{2, 2, 1} - 8 H^+_{3, 1, 1}\bigr) \, + \nonumber \\[1mm]  & 
G^-_{\hat 3, 1} \, \bigl(2 \bar L_{3, 0} + 4 \bar L_{3, 1} + 4 L_{1} \zeta_3 - 
   6 L_{0} H^+_{3} - 4 L_{1} H^+_{3} + 20 H^+_{4} - 
   4 L_{0} H^+_{1, 2} + 16 H^+_{1, 3} \nonumber \\  & - 4 L_{0} H^+_{2, 1} + 
   16 H^+_{2, 2} + 8 H^+_{3, 1} + 8 H^+_{1, 1, 2} - 8 H^+_{2, 1, 1}\bigr) \, +
G^-_{\hat 2, 2} \, \bigl(\bar L_{4} + \bar L_{3, 0} + 4 \bar L_{3, 1} + 8 L_{1} \zeta_3
\nonumber \\  & - 2 L_{0} H^+_{3} - 4 L_{1} H^+_{3} + 
   4 H^+_{4} + 4 H^+_{1, 3} + 4 H^+_{2, 2} + 
   4 L_{0} H^+_{1, 1, 1} - 8 H^+_{1, 2, 1} - 8 H^+_{2, 1, 1}\bigr) \, + \nonumber \\  &
G^-_{\hat 2, 1, 1} \, \bigl(-2 \bar L_{4} + 2 \bar L_{3, 0} - 8 L_{1} \zeta_3 - 
   4 L_{0} H^+_{3} + 16 H^+_{4} - 4 L_{0} H^+_{1, 2} + 
   8 H^+_{1, 3} + 8 H^+_{2, 2}\bigr) \, + \nonumber \\[1mm]  &
G^-_{\hat 4, 1} \, \bigl(-3 L_{2, 0} - 6 L_{2, 1} + 3 L_{0} H^+_{2} + 
   6 L_{1} H^+_{2}\bigr) \, +
G^-_{\hat 3, 2} \, \bigl(-2 L_{3} - 2 L_{2, 0} - 8 L_{2, 1} + \nonumber \\  &
   2 L_{0} H^+_{2} + 8 L_{1} H^+_{2} - 8 H^+_{1, 2} - 
   8 H^+_{2, 1} - 8 H^+_{1, 1, 1}\bigr) \, +
G^-_{\hat 3, 1, 1} \, \bigl(4 L_{3} - 2 L_{2, 0} + 8 H^+_{1, 2}\bigr) \, +
\nonumber \\  &
G^-_{\hat 2, 3} \, \bigl(-4 L_{3} - 2 L_{2, 0} - 8 L_{2, 1} - 
   L_{0, 0, 0} - 8 \zeta_3 + 2 L_{0} H^+_{2} + 
   8 L_{1} H^+_{2} - 12 H^+_{1, 2} \nonumber \\  & - 8 H^+_{2, 1} - 
   8 H^+_{1, 1, 1}\bigr) \, +
G^-_{\hat 2, 2, 1} \, \bigl(4 L_{3} + 2 L_{0, 0, 0} + 16 \zeta_3 + 
   8 H^+_{1, 2}\bigr) \, +
G^-_{\hat 2, 1, 2} \, \bigl(4 L_{3} + L_{2, 0} \nonumber \\  & + 4 L_{2, 1} + 
   L_{0, 0, 0} + 8 \zeta_3 - 2 L_{0} H^+_{2} - 
   4 L_{1} H^+_{2} + 8 H^+_{1, 2}\bigr) \, +
G^-_{\hat 2, 1, 1, 1} \, \bigl(-2 L_{2, 0} - 2 L_{0, 0, 0} - 16 \zeta_3\bigr) \, +
\nonumber \\[1mm]
& G^-_{\hat 5, 1} \, \bigl(4 H^+_{2} + 8 H^+_{1, 1}\bigr) \, +
G^-_{\hat 4, 2} \, \bigl(3 \bar L_{2} + 12 H^+_{1, 1}\bigr) \, +
G^-_{\hat 4, 1, 1} \, \bigl(-6 \bar L_{2} + 12 H^+_{2}\bigr) \, + 
G^-_{\hat 3, 3} \, \bigl(8 \bar L_{2} + 2 \bar L_{0, 0} \nonumber \\ & - 4 H^+_{2} + 16 H^+_{1, 1}\bigr) \, +
G^-_{\hat 3, 2, 1} \, \bigl(-8 \bar L_{2} - 4 \bar L_{0, 0} + 8 H^+_{2}\bigr) \, +
G^-_{\hat 3, 1, 2} \, \bigl(-8 \bar L_{2} - 2 \bar L_{0, 0}  + 8 H^+_{2} \nonumber \\ & - 8 H^+_{1, 1}\bigr) \, +
 G^-_{\hat 3, 1, 1, 1} \, \bigl(4 \bar L_{0, 0} + 8 H^+_{2}\bigr) \, +
G^-_{\hat 2, 4} \, \bigl(10 \bar L_{2} + 4 \bar L_{0, 0} - 4 H^+_{2} + 16 H^+_{1, 1}\bigr) \, + \nonumber \\ &
G^-_{\hat 2, 3, 1} \, \bigl(-4 \bar L_{2} - 4 \bar L_{0, 0}\bigr) \, +
G^-_{\hat 2, 2, 2} \, \bigl(-8 \bar L_{2} - 4 \bar L_{0, 0} + 4 H^+_{2} - 8 H^+_{1, 1}\bigr) \, +
 G^-_{\hat 2, 2, 1, 1} \, \bigl(-4 \bar L_{2} + 8 H^+_{2}\bigr) \, + \nonumber \\ &
G^-_{\hat 2, 1, 3} \, \bigl(-8 \bar L_{2} - 4 \bar L_{0, 0} + 4 H^+_{2} - 8 H^+_{1, 1}\bigr) \, +
 G^-_{\hat 2, 1, 2, 1} \, \bigl(4 \bar L_{0, 0} + 8 H^+_{2}\bigr) \, +
G^-_{\hat 2, 1, 1, 2} \, \bigl(4 \bar L_{2} + 4 \bar L_{0, 0}\bigr) \,
+ \nonumber \\[1mm]  &
\bigl(-10 G^-_{\hat 2, 5} - 8 G^-_{\hat 3, 4} - 3 G^-_{\hat 4, 3} + 
   10 G^-_{\hat 2, 1, 4} + 8 G^-_{\hat 2, 2, 3} + 8 G^-_{\hat 2, 3, 2} + 
   4 G^-_{\hat 2, 4, 1} + 8 G^-_{\hat 3, 1, 3} \nonumber \\  & + 8 G^-_{\hat 3, 2, 2} + 
   8 G^-_{\hat 3, 3, 1} + 3 G^-_{\hat 4, 1, 2} + 6 G^-_{\hat 4, 2, 1} - 
   8 G^-_{\hat 2, 1, 1, 3} - 8 G^-_{\hat 2, 1, 2, 2} - 4 G^-_{\hat 2, 1, 3, 1} - 
   4 G^-_{\hat 2, 2, 1, 2} \nonumber \\  & + 4 G^-_{\hat 2, 3, 1, 1} -
   8 G^-_{\hat 3, 1, 1, 2} - 
   8 G^-_{\hat 3, 1, 2, 1} - 6 G^-_{\hat 4, 1, 1, 1} + 
   4 G^-_{\hat 2, 1, 1, 1, 2} - 4 G^-_{\hat 2, 1, 2, 1, 1}\bigr) \, L_{0} +
\nonumber \\  &
\bigl(-10 G^-_{\hat 2, 5} - 10 G^-_{\hat 3, 4} - 6 G^-_{\hat 4, 3} - 2 G^-_{\hat 5, 2} + 
   8 G^-_{\hat 2, 1, 4} + 6 G^-_{\hat 2, 2, 3} + 6 G^-_{\hat 2, 3, 2} + 
   8 G^-_{\hat 3, 1, 3} \nonumber \\  & + 8 G^-_{\hat 3, 2, 2} + 4 G^-_{\hat 3, 3, 1} + 
   6 G^-_{\hat 4, 1, 2} + 6 G^-_{\hat 4, 2, 1} + 4 G^-_{\hat 5, 1, 1} - 
   4 G^-_{\hat 2, 1, 1, 3} - 4 G^-_{\hat 2, 1, 2, 2} + 4 G^-_{\hat 2, 2, 2, 1}
\nonumber \\  & + 
   4 G^-_{\hat 2, 3, 1, 1} - 4 G^-_{\hat 3, 1, 1, 2} + 
   4 G^-_{\hat 3, 2, 1, 1}\bigr) \, L_{1} \, + \nonumber
\end{align}
\begin{align}
& 20 G^-_{\hat 2, 6} + 20 G^-_{\hat 3, 5} + 12 G^-_{\hat 4, 4} + 4 G^-_{\hat 5, 3} - 
 20 G^-_{\hat 2, 1, 5} - 16 G^-_{\hat 2, 2, 4} - 12 G^-_{\hat 2, 3, 3} - 
 12 G^-_{\hat 2, 4, 2} \nonumber \\  & - 20 G^-_{\hat 3, 1, 4} - 16 G^-_{\hat 3, 2, 3} - 
 16 G^-_{\hat 3, 3, 2} - 8 G^-_{\hat 3, 4, 1} - 12 G^-_{\hat 4, 1, 3} - 
 12 G^-_{\hat 4, 2, 2} - 12 G^-_{\hat 4, 3, 1} - 4 G^-_{\hat 5, 1, 2} \nonumber \\  & - 
 8 G^-_{\hat 5, 2, 1} + 16 G^-_{\hat 2, 1, 1, 4} + 12 G^-_{\hat 2, 1, 2, 3} + 
 12 G^-_{\hat 2, 1, 3, 2} + 8 G^-_{\hat 2, 2, 1, 3} + 8 G^-_{\hat 2, 2, 2, 2} - 
 8 G^-_{\hat 2, 3, 2, 1} - 8 G^-_{\hat 2, 4, 1, 1} \nonumber \\  & + 16 G^-_{\hat 3, 1, 1, 3} + 
 16 G^-_{\hat 3, 1, 2, 2} + 8 G^-_{\hat 3, 1, 3, 1} + 8 G^-_{\hat 3, 2, 1, 2} - 
 8 G^-_{\hat 3, 3, 1, 1} + 12 G^-_{\hat 4, 1, 1, 2} + 12 G^-_{\hat 4, 1, 2, 1} + 
 8 G^-_{\hat 5, 1, 1, 1} \nonumber \\  & - 8 G^-_{\hat 2, 1, 1, 1, 3} - 
 8 G^-_{\hat 2, 1, 1, 2, 2} + 8 G^-_{\hat 2, 1, 2, 2, 1} + 
 8 G^-_{\hat 2, 1, 3, 1, 1} - 8 G^-_{\hat 3, 1, 1, 1, 2} + 
 8 G^-_{\hat 3, 1, 2, 1, 1} \nonumber
\end{align}

\subsection{Numerical consistency tests for $I^{(4)}$}

In order to check the correctness of the result from the previous section, we evaluated $I^{(4)}$ 
numerically and compared it to a direct numerical evaluation of the coordinate space integral using 
{\tt FIESTA}. In detail, we evaluate the conformally-invariant function $f(u,v) = x_{13}^2 x_{24}
^2\,I^{(4)}(x_1,x_2,x_3,x_4)$ by first applying a conformal transformation to send $x_4$ to infinity, 
the integral takes the simplified form,
\begin{equation}
\lim_{x_4\to \infty} x_{13}^2 x_{24}^2\,I^{(4)}_{14;23} = \frac{1}{\pi^8}\int \frac{d^4x_5 d^4x_6 d^4x_7  d^4 x_8\; x_{13}^2}{x_{15}^2 x_{18}^2 x_{25}^2 x_{26}^2 x_{37}^2 x_{38}^2 x_{56}^2 x_{67}^2 x_{78}^2}\, ,
\end{equation}
and then using the remaining freedom to fix $x_{13}^2=1$ so that $u=x_{12}^2$ and $v=x_{23}^2$. 
In comparison with the two 3-loop integrals, the extra loop in this case yields a moderately more 
cumbersome numerical evaluation. As such, we modify the setup for the 3-loop examples slightly 
and only perform $5\times 10^5$ integral evaluations. We nevertheless obtain about 5 digits of 
precision, and excellent agreement with the analytic function at 40 different points. See
Table~\ref{tab:4loop} for an illustrative sample of points.
\begin{table}[!t]
\begin{center}
\begin{tabular}{ccccc}
\hline\hline
$u$&$v$& Analytic & {\tt FIESTA} & $\delta$\\
\hline
0.1 & 0.2 & 156.733& 156.733 & 4.9e-7 \\
0.2 & 0.3 & 116.962& 116.962 & 5.9e-8 \\
0.3 & 0.1 & 110.366& 110.366 & 2.8e-7 \\
0.4 & 0.5 & 84.2632& 84.2632 & 1.4e-7 \\
0.5 & 0.6 & 75.2575& 75.2575 & 1.4e-7 \\
0.6 & 0.2 & 78.3720& 78.3720 & 3.7e-8 \\
0.7 & 0.3 & 70.7417& 70.7417 & 6.8e-8 \\
0.8 & 0.9 & 58.6362& 58.6363 & 1.4e-7 \\
0.9 & 0.5 & 60.1295& 60.1295 & 1.1e-7 \\
\hline\hline
 \end{tabular}
 \caption{\label{tab:4loop}Numerical comparison of the analytic result for $x_{13}^2 x_{24}^2\, I^{(4)}(x_1,x_2,x_3,x_4)$ against {\tt FIESTA} for several values of the conformal cross ratios.}
  \end{center}
 \end{table}

\section{Conclusions}
\label{sec:Conclude}

Recent years have seen a lot of advances in the analytic computation of Feynman integrals contributing to 
the perturbative expansion of physical observables. In particular, a more solid understanding of the 
mathematics underlying the leading singularities and the classes of functions that appear at low loop 
orders have opened up new ways of evaluating multi-scale multi-loop Feynman integrals analytically.

In this paper we applied some of these new mathematical techniques to the computation of the two so far unknown 
integrals appearing in the three-loop four-point stress-tensor correlator in ${\cal N} = 4$ SYM, and 
even a first integral occurring in the planar four-loop contribution to the same function. The 
computation was made possible by postulating that these integrals can be written as a sum over all
 the leading singularities (defined as the residues at the global poles of the loop integrand), each leading singularity being multiplied by a 
pure transcendental function that can be written as a $\mathbb{Q}$-linear combination of single-valued multiple polylogarithms in one complex variable. After a suitable choice was made for the 
entries that can appear in the symbols of these functions, the coefficients can easily be fixed by matching to some 
asymptotic expansions of the integrals in the limit where one of the cross ratios vanishes. In all 
cases we were able to integrate the symbols obtained form this procedure to a unique 
polylogarithmic function, thus completing the analytic computation of the the three-loop four-point 
stress-tensor correlator in ${\cal N} = 4$ SYM. While for the Easy integral the space of 
polylogarithmic function is completely classified in the mathematical literature, new classes of 
multiple polylogarithms appear in the analytic results for the Hard integral and the four-loop integral
we considered. 

One might wonder, given that the Hard integral function $H^{(a)}$ involves genuine two-variable 
functions, whether there could have been a similar contribution to the Easy integral, compatible with 
all asymptotic limits. Indeed there does exist a symbol of a single-valued function, not expressible in 
terms of SVHPLs alone, which evades all constraints from the asymptotic limits. In other words the 
function is power suppressed in all limits, possibly up to terms proportional to zeta values. However, 
the evidence we have presented  (in particular the numerical checks) strongly suggests that such a contribution 
is absent and therefore the Easy integral is expressible in terms of SVHPLs only.

We emphasise that the techniques we used for the computation are not limited to the rather special 
setting of the ${\cal N}=4$ model. First,  by sending a point to infinity a conformal four-point integral 
becomes a near generic three-point integral. Such integrals appear as master integrals for 
phenomenologically relevant processes, like for example the quantum corrections to the decay of a heavy particle into two 
massive particles. Second,
the conformal integrals we calculated have the structure $\sum \, R_i \, F_i$ (so residue times pure 
function) that is also observed for integrals contributing to on-shell amplitudes. However, we believe 
that this is in fact a common feature of large classes of Feynman integrals (if not all) and one 
purpose of this work is to advocate our combination of techniques as a means of solving many other diagrams. 

Further increasing the loop-order or the number of points might eventually hamper our prospects of success. Indeed, beyond problems of merely combinatorial nature there are also more fundamental issues, for example to what extent multiple polylogarithms exhaust the function spaces. It is anticipated in ref.~\cite{CaronHuot,Hamed} that elliptic integrals will eventually appear in higher-point on-shell amplitudes. Via the correlator/amplitude duality this observation will eventually carry over to our setting. Nevertheless, some papers~\cite{Hamed,mason} also hint at a more direct albeit related way of evaluating loop-integrals by casting them into a `$d\log$-form', which should have a counterpart for off-shell correlators.

\section*{Acknowledgements}
\label{acknowledge}

We acknowledge stimulating discussions with Simon Caron-Huot.~We are grateful to Alexander Smirnov for the possibility to use his {\tt c++} version of {\tt FIRE}.  PH would also like to acknowledge many inspirational discussions  with Hugh Osborn and Francis Dolan (to whom this work is dedicated) from 2005-6 in which we were attempting to find the three-loop correlator from the corresponding twist two anomalous dimensions using many similar ideas to the current work.  CD is supported by the ERC grant `IterQCD'. BE is supported by the DFG `eigene Stelle' 32302603.  PH is supported by STFC through the Consolidated
Grant number ST/J000426/1. 
JP is supported in part by the US Department of Energy under contract
DE--AC02--76SF00515.
The work of VS was supported by the Alexander
von Humboldt Foundation (Humboldt Forschungspreis) and by the Russian Foundation for Basic Research through
grant 11-02-01196.

\appendix
\section{Asymptotic expansions of the Easy and Hard integrals}
\label{app:asymptotics}
In this appendix
we collect the asymptotic expansions of the different orientations of the Easy and 
Hard integrals in terms of harmonic polylogarithms. The results for $E_{14;23}$ and $H_{12;34}$ 
were already presented in Section~\ref{sec:asymptotic_expansion}. The results for the other 
orientations are given below.
\begin{eqnarray}\label{app1eq:1}
x_{13}^2 x_{24}^2\,E_{12;34} & = & \log^3u\,\Big[-\frac{1}{3 x^2} \Big(2 H_{1,2}+ H_{1,1,1}\Big)+\frac{1}{3 x}\Big(H_{1,2}+H_{1,1,1}\Big)\Big]\\
\nonumber&+&\log^2 u\,\Big[\frac{2}{x^2} \Big(2 H_{2,2}+H_{2,1,1}+2 H_{1,3}+H_{1,1,2}\Big)\\
\nonumber&-&\frac{1}{2 x}\Big(-4 H_{2,2}-3 H_{2,1,1}-4 H_{1,3}-H_{1,2,1}-4 H_{1,1,2}\Big)\Big] \\
\nonumber&+&\log u\,\Big[ \frac{1}{x^2}\Big(-16 H_{3,2}-8 H_{3,1,1}-16 H_{2,3}-8 H_{2,1,2}-8 H_{1,4}+4 \
H_{1,3,1}\\
\nonumber&-&4 H_{1,2,2}-H_{1,2,1,1}-4 H_{1,1,3}+2 \
H_{1,1,2,1}-H_{1,1,1,2}\\
\nonumber&+&\frac{1}{x}\Big( 8 H_{3,2}+5 H_{3,1,1}+8 H_{2,3}+H_{2,2,1}+6 H_{2,1,2}+4 \
H_{1,4}-H_{1,3,1}\\
\nonumber&+&5 H_{1,2,2}+H_{1,2,1,1}+4 H_{1,1,3}-2 \
H_{1,1,2,1}+H_{1,1,1,2}\Big)\Big] \nonumber
\end{eqnarray}
\begin{eqnarray}
\phantom{x_{13}^2 x_{24}^2\,E_{12;34} \ \ }&+& \frac{1}{x^2}\Big( 4 \zeta _3 H_{1,2}+2 \zeta _3 H_{1,1,1}+32 H_{4,2}+16 H_{4,1,1}+32 \
H_{3,3}+16 H_{3,1,2}
\nonumber \\
&+&16 H_{2,4}-8 H_{2,3,1}+8 H_{2,2,2}+2 \
H_{2,2,1,1}+8 H_{2,1,3}-4 H_{2,1,2,1}+2 H_{2,1,1,2}
\nonumber \\
&-&8 H_{1,4,1}+4 \
H_{1,3,2}+4 H_{1,3,1,1}+4 H_{1,2,3}-2 H_{1,2,2,1}+2 H_{1,2,1,2}-4 \
H_{1,1,3,1} \nonumber \\
\nonumber&+&H_{1,1,2,1,1}-H_{1,1,1,2,1}\Big)
+\frac{1}{x}\Big(-4 \zeta _3 H_{2,1}-6 \zeta _3 H_{1,2}-2 \zeta _3 H_{1,1,1}-16 \
H_{4,2}\\
\nonumber&-&10 H_{4,1,1}-16 H_{3,3}-10 H_{3,1,2}-8 H_{2,4}+4 H_{2,3,1}-8 \
H_{2,2,2}-2 H_{2,2,1,1}\\
\nonumber&-&6 H_{2,1,3}+4 H_{2,1,2,1}-2 H_{2,1,1,2}+2 \
H_{1,4,1}-6 H_{1,3,2}-4 H_{1,3,1,1}-6 H_{1,2,3}\\
\nonumber&+&2 H_{1,2,2,1}-2
H_{1,2,1,2}+4 H_{1,1,3,1}-H_{1,1,2,1,1}+H_{1,1,1,2,1}-8 \zeta _3 \
H_3\\
\nonumber&+& 20 \zeta _5 H_1\Big)+\mathcal{O}(u)\,,
\end{eqnarray}
\begin{eqnarray}
x_{13}^2 x_{24}^2\,E_{13;24} & = &
\frac{\log u}{x}\Big(H_{2,2,1}-H_{2,1,2}+H_{1,3,1}-H_{1,2,1,1}-H_{1,1,3}+H_{1,1,2,1}-6 \
\zeta _3 H_2\Big)  \nonumber\\
\nonumber&+&\frac{1}{x}\Big(4 \zeta _3 H_{2,1}-2 \zeta _3 H_{1,2}-2 H_{3,2,1}+2 H_{3,1,2}-2 H_{2,3,1}+H_{2,2,1,1}+2 H_{2,1,3}\\
\nonumber&-&2 H_{2,1,2,1}+H_{2,1,1,2}-4 H_{1,4,1}+3 H_{1,3,1,1}+H_{1,2,1,2}+4 H_{1,1,4}-2 H_{1,1,3,1}\\
&-&H_{1,1,2,2}-H_{1,1,2,1,1}-H_{1,1,1,3}+H_{1,1,1,2,1}+12 \zeta _3 H_3\Big)+\mathcal{O}(u) \,, \label{app1eq:1b}
\end{eqnarray}
\begin{eqnarray}
x_{13}^4 x_{24}^4\,H_{13;24} & = & \log^3u\,\Big[\frac{1}{3x^2}\Big(2 H_{2,1}-H_{1,2}-H_{1,1,1}\Big)+\frac{1}{3 (1-x) x} \Big(H_{2,1}-H_3\Big)\Big]\label{app1eq:2}\\
\nonumber&+&\log^2u\,\Big[\frac{1}{x^2}\Big(-4 H_{3,1}-2 H_{2,2}+2 H_{1,3}+2 H_{1,2,1}+2 H_{1,1,2}\Big)\\
\nonumber&+&
\frac{1}{(1-x) x}\Big(-2 H_{3,1}- H_{2,2}- H_{2,1,1}- H_{1,3}+ H_{1,2,1}+4 H_4\Big)\Big]\\
\nonumber&+&\log u\,\Big[\frac{1}{x^2}\Big(16 H_{3,2}+8 H_{3,1,1}+8 H_{2,3}-8 H_{2,2,1}-4 H_{1,4}-12 H_{1,3,1}\\
\nonumber&-&4 H_{1,2,2}+2 H_{1,2,1,1}-4 H_{1,1,3}-2 H_{1,1,1,2}\Big)
+\frac{1}{(1-x) x} \Big(4 H_{4,1}+4 H_{3,2}\\
\nonumber&+&6 H_{3,1,1}+4 H_{2,3}+2 H_{2,1,2}+8 H_{1,4}-4 H_{1,3,1}-2 H_{1,2,2}-2 H_{1,2,1,1}\\
\nonumber&-&2 H_{1,1,3}+2 H_{1,1,2,1}-20 H_5\Big)\Big]+ \frac{1}{x^2}\Big(32 \zeta _3 H_{2,1}-16 \zeta _3 H_{1,2}-16 \zeta _3 H_{1,1,1}\\
\nonumber&+&64 H_{5,1}-32 H_{4,2}-32 H_{4,1,1}-24 H_{3,3}+16 H_{3,2,1}-16 H_{3,1,2}-24 H_{2,4}\\
\nonumber&+&40 H_{2,3,1}-4 H_{2,2,1,1}-8 H_{2,1,3}+4 H_{2,1,1,2}+40 H_{1,4,1}+4 H_{1,3,2}\\
\nonumber&-&8 H_{1,3,1,1}-4 H_{1,2,3}+4 H_{1,2,2,1}-4 H_{1,2,1,2}+8 H_{1,1,1,3}\Big)\\
\nonumber&+&\frac{1}{(1-x) x} \Big(16 \zeta _3 H_{2,1}-4 H_{4,2}-12 H_{4,1,1}-4 H_{3,3}-12 H_{3,2,1}-8 H_{3,1,2}\\
\nonumber&-&12 H_{2,4}+4 H_{2,3,1}+2 H_{2,2,1,1}-4 H_{2,1,2,1}+2 H_{2,1,1,2}-20 H_{1,5}+4 H_{1,4,1}\\
\nonumber&+&4 H_{1,3,2}+6 H_{1,3,1,1}+4 H_{1,2,3}+2 H_{1,2,1,2}+8 H_{1,1,4}-4 H_{1,1,3,1}-2 H_{1,1,2,2}\\
\nonumber&-&2 H_{1,1,2,1,1}-2 H_{1,1,1,3}+2 H_{1,1,1,2,1}-16 \zeta _3 H_3+40 H_6\Big)+\mathcal{O}(u)\,,
\end{eqnarray}
\begin{eqnarray}
x_{13}^4 x_{24}^4 \,H_{14;23} & = & \frac{\log^3u}{3x}\,\Big[\frac{1}{x}\Big(2 H_{2,1}-H_{1,2}+2 H_{1,1,1}\Big)-2 H_{2,1}-H_{1,2}-2 H_{1,1,1}-H_3\Big] \nonumber \\
& + & \log^2u\,\Big[-\frac{2}{x^2} \Big(2 H_{3,1}+H_{2,2}-H_{1,3}+2 H_{1,2,1}+2 H_{1,1,2}\Big) \label{app1eq:2b}\\
\nonumber&+&\frac{4}{x} \Big(H_{3,1}+H_{2,2}+H_{1,3}+H_{1,2,1}+H_{1,1,2}+H_4\Big)\Big] \\ 
\nonumber & + & \log u\,\Big[\frac{4}{x^2} \Big(4 H_{3,2}-4 H_{3,1,1}+2 H_{2,3}+4 H_{2,1,2}-H_{1,4}+2 H_{1,3,1}+4 H_{1,2,2}\\
\nonumber&-&2 H_{1,2,1,1}+4 H_{1,1,3}+2 H_{1,1,1,2}\Big)
+\frac{4}{x} \Big(2 H_{4,1}+4 H_{3,2}-2 H_{3,1,1}+4 H_{2,3}\\
\nonumber&+&2 H_{2,1,2}+5 H_{1,4}+2 H_{1,3,1}+4 H_{1,2,2}-2 H_{1,2,1,1}+4 H_{1,1,3}+2 H_{1,1,1,2}\\
\nonumber&+&5 H_5\Big)\Big]+\frac{8}{x^2} \Big(4 \zeta_3 H_{2,1}-2 \zeta_3 H_{1,2}+4 \zeta_3 H_{1,1,1}+8 H_{5,1}-4 H_{4,2}+8 H_{4,1,1}\\
\nonumber&-&6 H_{3,3}+4 H_{3,2,1}-4 H_{3,1,2}-3 H_{2,4}+2 H_{2,3,1}-4 H_{2,2,2}+2 H_{2,2,1,1}\\
\nonumber&-&6 H_{2,1,3}-2 H_{2,1,1,2}+2 H_{1,4,1}-3 H_{1,3,2}+4 H_{1,3,1,1}-5 H_{1,2,3}+2 H_{1,2,2,1}\\
\nonumber&-&2 H_{1,2,1,2}-4 H_{1,1,4}-2 H_{1,1,2,2}-2 H_{1,1,1,3}\Big)+\frac{8}{x}\Big(-4 \zeta _3 H_{2,1}-2 \zeta _3 H_{1,2}\\
\nonumber&-&4 \zeta _3 H_{1,1,1}+3 H_{4,2}-2 H_{4,1,1}+3 H_{3,3}-2 H_{3,2,1}+4 H_{2,4}+2 H_{2,2,2}\\
\nonumber&+&2 H_{2,1,3}+5 H_{1,5}+3 H_{1,3,2}-2 H_{1,3,1,1}+3 H_{1,2,3}-2 H_{1,2,2,1}+4 H_{1,1,4}\\
\nonumber&+&2 H_{1,1,2,2}+2 H_{1,1,1,3}-2 \zeta _3 H_3+5 H_6\Big)+\mathcal{O}(u)\,,
\end{eqnarray}

\section{An integral formula for the Hard integral}

We want to find an integral formula for pure functions which involve $x-\bar{x}$ in the symbol as well as $x,\bar{x},1-x,1-\bar{x}$. We are interested in single-valued functions, i.e. ones obeying the constraints on the discontinuities,
\be
[{\rm disc}_x - {\rm disc}_{\bar x}] f(x, \bar x) =0\, , \qquad [{\rm disc}_{1-x} - {\rm disc}_{1 - \bar x}] f(x, \bar x) =0\,.
\ee
and with no other discontinuities.

It will be sufficient for us to consider functions whose symbols have final letters drawn from a restricted set of letters,
\be
\cS(F) = \cS(X) \otimes \frac{x}{\bar x} + \cS(Y) \otimes \frac{1-x}{1-\bar x} + \cS(Z) \otimes (x- \bar x)\,.
\label{SF}
\ee
where $X,Y,Z$ are single-valued functions of $x, \bar x$.

We will suppose also that the function $F$ obeys $F(x,x) = 0$, as required to remove the poles at $x=\bar x$ present in the leading singularities of the conformal integrals. We therefore take $Z(x,x)=0$ also.
If $F$ has a definite parity under $x \leftrightarrow \bar x$ then $X$ and $Y$ have the opposite parity while $Z$ has the same parity.

The functions $X,Y$ and $Z$ are not independent of each other. Integrability (i.e. $d^2 F =0$) imposes the following restrictions,
\be
d X \wedge d \log \frac{x}{\bar x} + d Y \wedge d \log \frac{1-x}{1-\bar x} + d Z \wedge d \log(x-\bar x) = 0\,.
\label{int}
\ee
We may then define the derivative of $F$ w.r.t $x$ to be
\be
\partial_x F(x,\bar x) = \frac{X}{x} - \frac{Y}{1-x} + \frac{Z}{x-\bar x}\,,
\ee
so that
\be
F(x,\bar x) = \int_{\bar x}^x dt \biggl[ \frac{X(t,\bar x)}{t} - \frac{Y(t,\bar x)}{1-t} + \frac{Z(t ,\bar x)}{t-\bar x} \biggr]\,.
\label{intdx}
\ee

A trivial example is the Bloch-Wigner dilogarithm function, defined via,
\be
F_2(x,\bar x) = \log x \bar x \, (H_1(x) - H_1(\bar x)) - 2(H_2(x) - H_2(\bar x))\,.
\ee
It has a symbol of the form (\ref{SF}) where
\be
X_1 = \log (1-x)(1-\bar x), \qquad Y_1 = -\log x \bar x\, \qquad Z_1=0\,.
\ee
Thus we can write the integral formula (\ref{intdx}) for $F_2$.

\subsection{Limits}

We want to be able to calculate the limits of the functions to compare with the asymptotic expressions obtained in Section~\ref{sec:asymptotic_expansion}.
The formula (\ref{intdx}) allows us to calculate the limit $\bar x \rightarrow 0$ (which means dropping any power suppressed terms in this limit). 
We may commute the limit and integration
\be
\lim_{\bar x \rightarrow 0}F(x,\bar x) = \int_{\bar x}^x dt \lim_{\bar x \rightarrow 0} \biggl[ \frac{X(t,\bar x)}{t} -\frac{Y(t,\bar x)}{1-t} + \frac{Z(t,\bar x)}{t}\biggr]\,.
\ee
In the second and third terms one may also set the lower limit of integration to zero. directly. In the first one should take care that contributions from $X(t,\bar x)$ which do not vanish as $t \rightarrow 0$ produce extra logarithms of $\bar x$, beyond those explicitly appearing in the limit of $X$, as the lower limit approaches zero.

\subsection{First non-trivial example (weight three)}

The first example of a single-valued function whose symbol involves $x-\bar{x}$ is at weight three \cite{Chavez:2012kn}. There is exactly one such function at this weight, i.e. all single-valued functions can be written in terms of this one and single-valued functions constructed from single-variable HPLs with arguments $x$ and $\bar x$ only. It obeys $F_3(x,\bar{x}) = - F_3(\bar{x},x)$. The symbol takes the form (\ref{SF}) with
\begin{align}
X_2 &= - \log (x \bar x) (H_1(x) + H_1(\bar x)) + \frac{1}{2}(H_1(x) + H_1(\bar x))^2 \, , \notag \\
Y_2 &= -\frac{1}{2} \log^2 (x \bar x) + \log (x \bar x) (H_1(x) + H_1(\bar x))\, , \notag \\
Z_2 &= 2\log x \bar x \, (H_1(x) - H_1(\bar x)) - 4(H_2(x) - H_2(\bar x))\,.
\end{align}
Note that $X_2,Y_2$ and $Z_2$ are single-valued and that $Z_2$ is proportional to the Bloch-Wigner dilogarithm (it is the only antisymmetric weight-two single-valued function so it had to be). They obey the integrability condition (\ref{int}) so we can write the integral formula (\ref{intdx}) to define the function $F_3$.

We have constructed a single-valued function with a given symbol, but in fact this function is uniquely defined since there is no antisymmetric function of weight one which is single-valued which could be multiplied by $\zeta_2$ and added to our result. Moreover, since it is antisymmetric in $x$ and $\bar x$, we cannot add a constant term proportional to $\zeta_3$.

Looking at the limit $\bar x \rightarrow 0$ we find, following the discussion above, 
\begin{align}
\lim_{\bar x \rightarrow 0} F_3(x,\bar x) = &\frac{1}{2} \log^2 \bar x  H_{1}(x) + \log \bar x \bigl(H_{2}(x) +H_{1, 0}(x) -H_{1, 1}(x)\bigr) \notag \\
&- 3 H_{3}(x)  - H_{1, 2}(x)+ H_{2, 0}(x) + 
 H_{2, 1}(x) + H_{1, 0, 0}(x) - H_{1, 1, 0}(x)
\end{align}
Starting from the original symbol for $F_3$ and taking the limit $\bar x \rightarrow 0$ we see that the above formula indeed correctly captures the limit.

\subsection{Weight five example}
\label{sect-weight5ex}

We now give an example directly analogous to the weight-three example above but at weight five. The example we are interested in is symmetric $F_5(x, \bar x) = F_5(\bar x, x)$. It has a symbol of the canonical form (\ref{SF}) with 
\begin{align}
X_4(x,\bar x) &= (\mathcal{L}_{0,0,1,1} - \mathcal{L}_{1,1,0,0} - \mathcal{L}_{0,1,1,1}+\mathcal{L}_{1,1,1,0})\,, \notag \\
Y_4(x,\bar x) &= (\mathcal{L}_{0,0,0,1} - \mathcal{L}_{1,0,0,0} - \mathcal{L}_{0,0,1,1} + \mathcal{L}_{1,1,0,0})\,, \notag \\
Z_4(x,\bar x) &= (\mathcal{L}_{0,0,1,1} + \mathcal{L}_{1,1,0,0} - \mathcal{L}_{0,1,1,0} - \mathcal{L}_{1,0,0,1})\,.
\end{align}

The above functions are single-valued and obey the integrability condition and therefore define a single-valued function of two variables of weight five via the integral formula.

Taking the limit $\bar x \rightarrow 0$ we find
\begin{align}
\lim_{\bar x \rightarrow 0} F_5(x,\bar x) =  
 & H_{1, 1} \bar{H}_{0, 0, 0} + 
(H_{1, 1, 0} -  H_{1, 1, 1} )\bar{H}_{0, 0} \notag \\
&+(-   H_{3, 1} +  H_{2, 1, 1} + 
  H_{1, 1, 0, 0}  -  H_{1, 1, 1, 0})\bar{H}_{0}  \notag \\
  &- H_{1, 4} -  H_{2, 3} + 2 H_{4, 1} +  H_{1, 3, 1} -  H_{3, 1, 0} -  H_{3, 1, 1} \notag \\
  &+  H_{2, 1, 1, 0} + 
  H_{1, 1, 0, 0, 0} -  H_{1, 1, 1, 0, 0}  + 2 H_{1, 1} \zeta_3\,.
\end{align}
This formula correctly captures the limit taken directly on the symbol of $F_5$. This weight-five function plays a role in the construction of the Hard integral.

\subsection{The function $H^{(a)}$ from the Hard integral}

The function $H^{(a)}$ from the Hard integral is a weight-six symmetric function obeying the condition $H^{(a)}(x,x)=0$. The symbol of $H^{(a)}$ is known but is not of the form (\ref{SF}). However, we can use shuffle relations to rewrite the symbol in terms of logarithms of $u$ and $v$ and functions which end with our preferred set of letters. We find the symbol can be represented by a function of the form
\begin{align}
&H^{(a)}(1-x,1-\bar x) \notag \\
&\qquad= \bigl(2 H_{0,0}(u) + 4 H_{0}(u)H_{0}(v) + 8 H_{0,0}(v)\bigr) \bigl(\mathcal{L}_{0,0,1,1} +  \mathcal{L}_{1,1,0,0} -  \mathcal{L}_{0,1,1,0} - \mathcal{L}_{1,0,0,1}\bigr) \notag \\
& \qquad \quad - 8F_5 \bigl(H_0(u) + 2 H_0(v)\bigr) + F_6\,.
\label{Hsdef}
\end{align}
Here $F_5$ is the weight-five function defined in Section~\ref{sect-weight5ex}. The function $F_6$ is now one whose symbol is of the form (\ref{SF}), where the functions $X_5,Y_5$ and $Z_5$ take the form
\begin{align}
X_5 &= 20 \mathcal{L}_{0,0,0,1,1} + 12 \mathcal{L}_{0,0,1,1,0} - 32 \mathcal{L}_{0,0,1,1,1} - 8 \mathcal{L}_{0,1,0,1,1} - 12 \mathcal{L}_{0,1,1,0,0} - 8 \mathcal{L}_{0,1,1,0,1} \notag \\
&+16 \mathcal{L}_{0,1,1,1,1} - 8\mathcal{L}_{1,0,0,1,1} + 8 \mathcal{L}_{1,0,1,1,0} - 20 \mathcal{L}_{1,1,0,0,0} + 8 \mathcal{L}_{1,1,0,0,1} + 8 \mathcal{L}_{1,1,0,1,0} \notag \\
&+ 32 \mathcal{L}_{1,1,1,0,0} - 16 \mathcal{L}_{1,1,1,1,0} - 16\mathcal{L}_{1,1} \zeta_3\,, \\
Y_5 &= 20 \mathcal{L}_{0,0,0,0,1} - 32 \mathcal{L}_{0,0,0,1,1} - 8 \mathcal{L}_{0,0,1,1,0} + 16 \mathcal{L}_{0,0,1,1,1} - 8 \mathcal{L}_{0,1,0,0,1} + \mathcal{L}_{0,1,1,0,0} \notag \\
&- 20 \mathcal{L}_{1,0,0,0,0} + 8 \mathcal{L}_{1,0,0,1,0} + 16 \mathcal{L}_{1,0,0,1,1} + 8 \mathcal{L}_{1,0,1,0,0} + 32 \mathcal{L}_{1,1,0,0,0} - 16 \mathcal{L}_{1,1,0,0,1}  \notag \\
&- 16 \mathcal{L}_{1,1,1,0,0} - 16 \mathcal{L}_{1,0} \zeta_3 + 64 \mathcal{L}_{1,1} \zeta_3\,. \\
Z_5 &= 32F_5\,.
\end{align}
Note that the $\zeta_3$ terms have been chosen in such  a way the the functions $X_5,Y_5$ and $Z_5$ obey the integrability condition (\ref{int}). The integral formula for $F_6$ based on the above functions will give a single-valued function with the correct symbol, i.e. one such that $H^{(a)}$ defined in eq.~(\ref{Hsdef}) has the correct symbol and is single-valued.

We recall that the Hard integral takes the form
\be
H_{14;23} = \frac{1}{x_{13}^4 x_{24}^4} \biggl[  \frac{H^{(a)}(1-x,1-\bar x)}{(x-\bar x)^2} + \frac{H^{(b)} (1-x,1-\bar x)}{(1-x \bar x)(x-\bar x)} \biggr]\,.
\ee
Calculating the limit $\bar x \rightarrow 0$ we find that $H^{(a)}$ reproduces the terms proportional to $1/x^2$ in the limit exactly, including the zeta terms. Note that in this limit the contributions of $H^{(a)}$ and $H^{(b)}$ are distinguishable since the harmonic polylogarithms come with different powers of $x$. Since there are no functions of weight four or lower which are symmetric in $x$ and $\bar x$ and which vanish at $x=\bar x$ and which vanish in the limit $\bar x \rightarrow 0$, we conclude that $H^{(a)}$ defined in eq.~(\ref{Hsdef}) is indeed the function.
Comparing numerically with the formula obtained in Section~\ref{sec:Hard} we indeed find agreement to at least five significant figures.

\section{A symbol-level solution of the four-loop differential equation}
In this appendix we sketch an alternative approach to the evaluation of the four-loop integral. Mor 
precisely, we will show how the function $I^{(4)}$ can be determined using symbols and the 
coproduct on multiple polylogarithms. We start from the differential equation~\eqref{diffeq}, which 
we recall here for convenience,
\beq
 \partial_x \partial_{\bar x} \hat{f}(x,\bar x) =  -\frac{1}{(1-x \bar x)x \bar x} E_1(x,\bar x) - \frac{1}{(1-x \bar x)} E_2(x,\bar x)\,,
\eeq
where we used the abbreviations $E_1(x,\bar x) = E(1-x,1-\bar x)$ and $E_2(x,\bar x) = E(1-1/x,
1-1/\bar x)$. We now act with the symbol map $\cS$ on the differential equation, and we get
\beq\label{eq:symb_diff_eq}
 \partial_x \partial_{\bar x} \cS[\hat{f}(x,\bar x)] =  -\frac{1}{(1-x \bar x)x \bar x} \cS[E_1(x,\bar x)] - \frac{1}{(1-x \bar x)} \cS[E_2(x,\bar x)]\,,
\eeq
where the differential operators act on tensors only in the last entry, e.g.,
\beq
\partial_x[a_1\otimes\ldots\otimes a_n] = [\partial_x\log a_n]\,a_1\otimes\ldots\otimes a_{n-1}\,,
\eeq
and similarly for $\partial_{\bar x}$. It is easy to see that the tensor 
\beq
S_1 = \cS[E_1(x,\bar x)]\otimes\left(1-\frac{1}{x\bar x}\right)\otimes(x\bar x) + \cS[E_2(x,\bar x)]\otimes\left(1-{x\bar x}\right)\otimes(x\bar x) 
\eeq
solves the equation~\eqref{eq:symb_diff_eq}. However, $S_1$ is not integrable in the pair of entries 
(6,7), and so $S_1$ is not yet the symbol of a solution of the differential equation. In order to obtain 
an integrable solution, we need to add a solution to the homogeneous equation associated to
eq.~\eqref{eq:symb_diff_eq}. The homogeneous solution can easily be obtained by writing down the 
most general tensor $S_2$ with entries drawn from the set $\{x,\bar x,1-x,1-\bar x,1-x\bar x\}$ that 
has the correct symmetries and satisfies the first entry condition and 
\beq
  \partial_x \partial_{\bar x} S_2= 0\,.
  \eeq
  In addition, we may assume that $S_2$ satisfies the integrability condition in all factors of the 
  tensor product except for the pair of entries $(6,7)$, because $S_1$ satisfies this condition as well. 
  The symbol of the solution of the differential equation is then given by $S_1+S_2$, subject to the 
  constraint that the sum is integrable. It turns out that there is a unique solution, which can be 
  written in the schematic form
  \beq\bsp\label{eq:form_symbol_4_loop}
  \cS[\hat f(x,\bar x)] &\,= s_1^-\otimes u\otimes u 
+ s^-_2 \otimes v \otimes u + s^-_3 \otimes\frac{1-x}{1-\bar x}\otimes\frac{x}{\bar x}
+ s^+_4 \otimes\frac{x}{\bar x}\otimes u \\
  &\,+ s^+_5 \otimes u \otimes\frac{x}{\bar x}
+s^+_6\otimes\frac{1-x}{1-\bar x}\otimes u +s^+_7\otimes v\otimes\frac{x}{\bar x}
+s^-_8\otimes\frac{x}{\bar x}\otimes\frac{x}{\bar x}\\
&\,+s^-_9\otimes(1-u)\otimes u\,,
  \esp\eeq
  where $s_i^\pm$ are (integrable) tensor that have all their entries drawn from the set $\{x,\bar x,1-x,1-\bar x\}$ and the superscript refers to the parity under an exchange of $x$ and $\bar x$.
  
The form~\eqref{eq:form_symbol_4_loop} of the symbol of $\hat f(x,\bar x)$ allows us to make the 
following more refined ansatz: as the $s_i^\pm$ are symbols of SVHPLs, and using the fact that the 
symbol is the maximal iteration of the coproduct, we conclude that there are linear combinations $f^
\pm_i(x,\bar x)$ of SVHPLs of weight six (including products of zeta values and SVHPLs of lower 
weight) such that $\cS[f^\pm_i(x,\bar x)] = s_i^\pm$ and 
  \beq\bsp\label{eq:coproduct_4_loop}
\Delta_{6,1,1}[\hat f(x,\bar x)] &\,=  f_1^-(x,\bar x)\otimes \log u\otimes \log u 
+ f^-_2(x,\bar x) \otimes \log v \otimes \log u\\
&\, + f^-_3(x,\bar x) \otimes\log\frac{1-x}{1-\bar x}\otimes\log\frac{x}{\bar x}
+ f^+_4(x,\bar x) \otimes\log \frac{x}{\bar x}\otimes \log u \\
  &\,+ f^+_5(x,\bar x) \otimes \log u \otimes\log \frac{x}{\bar x}
+f^+_6(x,\bar x)\otimes\log\frac{1-x}{1-\bar x}\otimes \log u \\
  &\,+f^+_7(x,\bar x)\otimes \log v\otimes\log\frac{x}{\bar x}
+f^-_8(x,\bar x)\otimes\log\frac{x}{\bar x}\otimes\log\frac{x}{\bar x}\\
&\,+f^-_9(x,\bar x)\otimes\log(1-u)\otimes u\,.
  \esp\eeq
  The coefficients of the terms proportional to zeta values and SVHPLs of lower weight (which were 
  not captured by the symbol) can easy be fixed by appealing to the differential equation, written in 
  the form\footnote{We stress that differential operators act in the last factor of the coproduct, just 
  like for the symbol.}
  \beq
  (\mathrm{id}\otimes\partial_x\otimes\partial_{\bar x})\Delta_{6,1,1}[\hat f(x,\bar x)] = 
  -\frac{1}{(1-x \bar x)x \bar x} E_1(x,\bar x)\otimes1\otimes1 - \frac{1}{(1-x \bar x)} E_2(x,\bar x)\otimes1\otimes1\,.
\eeq
The expression~\eqref{eq:coproduct_4_loop} has the advantage that it captures more information 
about the function $\hat f(x,\bar x)$ than the symbol alone. In particular, we can use
eq.~\eqref{eq:coproduct_4_loop} to derive an iterated integral representation for $\hat f(x,\bar x)$ with 
respect to $x$ only. To see how this works, first note that there must be functions $A^\pm(x,\bar x)$, 
that are respectively even and odd under an exchange of $x$ and $\bar x$, such that
\beq\label{eq:coproduct_7_1}
\Delta_{7,1}[\hat f(x,\bar x)] = A^-(x,\bar x)\otimes\log u + A^+(x,\bar x)\otimes\log \frac{x}{\bar x}\,.
\eeq
with
\beq\bsp\label{eq:cop_6_1}
\Delta_{6,1}[A^-(x,\bar x)] &\,= 
f_1^-(x,\bar x)\otimes \log u 
+ f^-_2(x,\bar x) \otimes \log v 
+ f^+_4(x,\bar x) \otimes\log \frac{x}{\bar x}\\
&\,+f^+_6(x,\bar x)\otimes\log\frac{1-x}{1-\bar x}
+f^-_9(x,\bar x)\otimes\log(1-u)\,,\\
\Delta_{6,1}[A^+(x,\bar x)] &\,= f^-_3(x,\bar x) \otimes\log\frac{1-x}{1-\bar x}
  + f^+_5(x,\bar x) \otimes \log u\\
&\,  +f^+_7(x,\bar x)\otimes \log v
+f^-_8(x,\bar x)\otimes\log\frac{x}{\bar x}\,.
  \esp\eeq
  
The (6,1) component of the coproduct of $A^+(x,\bar x)$ does not involve $\log(1-u)$, and so it can 
entirely be expressed in terms of SVHPLs. We can thus easily obtain the result for $A^+(x,\bar x)$ 
by writing down the most general linear combination of SVHPLs of weight seven that are even 
under an exchange of $x$ and $\bar x$ and fix the coefficients by requiring the (6,1) component of 
the coproduct of the linear combination to agree with eq.~\eqref{eq:cop_6_1}. In this way we can fix
$A^+(x,\bar x)$ up to zeta values of weight seven (which are integration constants of the original 
differential equation).

The coproduct of $A^-(x,\bar x)$, however, does involve $\log(1-u)$, and so it cannot be expressed 
in terms of SVHPLs alone. We can nevertheless derive a first-order differential equation for $A^-(x,
\bar x)$. We find
\beq\bsp
\partial_xA^-(x,\bar x) &\,= \frac{1}{x}\left[f^-_1(x,\bar x)+f^+_4(x,\bar x)\right]
-\frac{1}{1-x}\left[f^-_2(x,\bar x)+f^+_6(x,\bar x)\right]\\
&\,-\frac{\bar x}{1-x\bar x}f_9^-(x,\bar x)\\
&\,\equiv K(x,\bar x)\,.
\esp\eeq
The solution to this equation is
\beq
A^-(x,\bar x) = h(\bar x) + \int_{\bar x}^xdt\,K(t,\bar x)\,,
\eeq
where $h(\bar x)$ is an arbitrary function of $\bar x$. The integral can easily be performed in terms 
of multiple polylogarithms. Antisymmetry of $A^-(x,\bar x)$ under an exchange of $x$ and $\bar x$ requires $h(\bar x)$ to vanish identically, because
\beq
A^-(x,\bar x) = h(\bar x) + \int_{\bar x}^xdt\,\partial_tA^-(t,\bar x) = h(\bar x) + A^-(x,\bar x) - A^-(\bar x,\bar x)= h(\bar x) + A^-(x,\bar x)\,.
\eeq
We thus obtain a unique solution for $A^-(x,\bar x)$.

Having obtained the analytic expressions for $A^\pm(x,\bar x)$ (up to the integration constants in $A^+(x,\bar x)$), we can easily obtain a first-order differential equation for $\hat f(x,\bar x)$,
\beq\label{eq:first_order_diff_eq}
\partial_x\hat f(x,\bar x) = \frac{1}{x}[A^-(x,\bar x)  + A^+(x,\bar x)]\,.
\eeq
The solution reads
\beq
\hat f(x,\bar x) = \int_{\bar x}^x\frac{dt}{t}[A^-(t,\bar x)  + A^+(t,\bar x)]\,.
\eeq
The integral can again easily be performed in terms of multiple polylogarithms and the 
antisymmetry of $\hat f(x,\bar x)$ under an exchange of $x$ and $\bar x$ again excludes any 
arbitrary function of $\bar x$ only. The solution to eq.~\eqref{eq:first_order_diff_eq} is however not 
yet unique, because of the integration constants in $A^+(x,\bar x)$, and we are left with three free 
coefficients of the form,
\beq
(c_1\,\zeta_7+c_2\, \zeta_5\,\zeta_2+c_3\,\zeta_4\,\zeta_3)\log\frac{x}{\bar x}\,.
\eeq
The free coefficients can be fixed using the requirement that $\hat f(x,\bar x)$ be single-valued (see 
the discussion in Section~\ref{sec:4loops}). Alternatively, they can be fixed by requiring that $\hat 
f(x,\bar x)$ be odd under inversion of $(x,\bar x)$ and vanish at $x=\bar x$. We checked that the 
resulting function agrees analytically with the result derived in Section~\ref{sec:4loops}.


\end{document}